\documentclass[twocolumn]{aastex}
\usepackage{spr-astr-addons}
\usepackage{url}

\begin{document}
\noindent{\it Accepted for publication in Astrophysics and Space Science}
\begin{center}
\title{\textbf{ NONAXISYMMETRIC INSTABILITIES IN SELF-GRAVITATING DISKS. 
II  LINEAR AND QUASI-LINEAR ANALYSES }}

\author{ Kathryn Z. Hadley, Paul Fernandez, James N. Imamura,
Erik Keever, Rebecka Tumblin, \& William Dumas }

\affil{ Institute of Theoretical Science, University of Oregon,
Eugene, OR 97403-1229 }
\end{center}
\begin{abstract}
We studied global nonaxisymmetric hydrodynamic 
instabilities in an extensive collection of 
hot, self-gravitating polytropic disk systems, systems
that covered a wide expanse of the parameter space relevant to 
protostellar and protoplanetary systems. We examined
equilibrium disk models
varying three parameters: the ratio of the
inner to outer equatorial radii, 
the ratio of star mass to disk mass,
and the rotation law exponent $q$.
We took the polytropic index $n$ = 1.5 and examined the exponents $q =$ 1.5 and 2, and 
the transitional one $q$ = 1.75.
For each of these sets of parameters, we 
examined models with inner to outer radius 
ratios from 0.1 to 0.75, and star mass to disk mass ratios from 0
to 10$^3$.
We numerically calculated the growth rates and oscillation frequencies 
of low-order nonaxisymmetric disk modes,
modes with azimuthal dependence $\propto$ e$^{im\phi}$. 
Low-$m$ modes are found to dominate with the character and 
strength of instability strongly dependent on disk self-gravity.
Representatives of each mode type are examined in detail, and torques and mass 
transport rates are calculated.

\end{abstract}
\section {INTRODUCTION} \label{sec_intro}

Star formation takes place in Giant Molecular clouds 
where collapse of small embedded cloud cores is 
triggered by external mechanisms such as 
shock waves or stellar winds. For nonrotating cloud
cores, our understanding of the star formation process is 
well in-hand. However, it is 
clear that in general, rotation must be taken into account.
Observations show that molecular
cloud cores typically have specific angular momenta
$\sim$ $10^{21-22}$ 
cm$^2$ s$^{-1}$ ({\it e.g.,} see Tohline 2002). 
Clouds with such high specific angular momenta 
cannot collapse directly into stars. Only a few percent of the
cloud matter, that nearest the rotation axis, goes directly into the 
formation of the central object; the rest forms a massive 
circumstellar disk ({\it e.g.,} see Shu, Adams, \& Lizano 1987, 
Tohline 2002). Star formation thus hinges on how mass
from the disk finds its way onto the nascent star.
Molecular viscosity as a mechanism is ineffective at 
transport of angular momentum in astrophysical disks.
Other mechanisms are needed to enhance
transport above that which results from 
binary particle collisions (Armitage 2011). Hydrodynamic 
and/or magnetohydrodynamic nonaxisymmetric 
instabilities have been proposed as ways to supply 
the needed dissipation, either directly through wave transport 
or indirectly as mechanisms 
that generate turbulence which supplies the needed effective
viscosity (Balbus \& Hawley 1998).
We model the redistribution of angular momentum 
through global nonaxisymmetric hydrodynamic 
instabilities in this and follow-up papers.
Recent reviews of the fluid 
mechanics involved in young stellar objects by Shariff (2009) 
and Armitage (2011) include summaries of 
observed characteristics of various classes of objects, 
as well as discussions of various mechanisms involved,
focusing on magnetic field 
effects, radiation transport and turbulence.

The stability of nonmagnetic disks has been of interest since the 
late nineteenth century when Dyson (1893) first investigated what he 
called {\it anchor rings}.  As with many systems in physics, the stability 
analyses began with simplified models, adding increasing complexity over 
time. Serious attempts at the stability analysis of non-self-gravitating 
thick disks began with Papaloizou \& Pringle (1984, 1985) who 
studied isentropic disks with power
law differential rotation. They made the important discovery that 
disks may be dynamically unstable to global nonaxisymmetric modes with 
azimuthal dependence given by $e^{im\phi}$.  For the special cases 
of a thin cylindrical shell and a thin isothermal ring, a threshold of 
instability was found for low-$m$ modes and slender tori such that disks 
were found to be unstable for a range of angular momentum profiles. For
disks with power law angular rotation distributions with 
exponent $q$,
Kelvin-Helmholtz-like instabilities were found to dominate disks for low 
$q$ while sonic instability dominates systems near a constant specific 
angular momentum profile, $q = 2$. These sonic instabilities 
were later named P 
modes. Papaloizou \& Pringle (1987) subsequently performed 
work on higher order 
modes, studying modes trapped near the 
inner and outer disk boundaries by an 
evanescent region around corotation, modes for which 
the fluid speed equals the speed of the perturbation. 
Kojima (1986, 1989) further analyzed non-self-gravitating isentropic thick 
disks for $q$ $=$ 2 and $n$ $=$ 0, 1.5, and 3.0, where $n$ is the 
polytropic index. Kojima found the disks were unstable for almost all cases 
calculated and that the growth rate decreased for either sufficiently large 
or small radial widths, and also decreased with decreasing $q$.  
The growth rates 
showed little dependence on compressibility, with only small differences 
between his $n$ $=$ 1.5 and $n$ $=$ 3.0 calculations.

The effect of self-gravity was first included in the analytic and 
numerical investigations of long 
wavelength modes found in slender, incompressible tori 
by Goldreich, Goodman \& Narayan (1986). Their theory 
used a thin ribbon approximation to investigate the two-dimensional
incompressible limit of the narrow torus. 
They showed that two new modes emerged, 
one with corotation at the density maximum, called the J mode (for the Jeans 
instability) and a second with corotation outside the ribbon, called the I mode 
(intermediate between P and J modes). Goodman \& Narayan (1988) 
further investigated I modes and J modes adding self-gravity to their 
calculations of 3D slender incompressible tori with $q$ $=$ 2 
and two-dimensional slender incompressible tori with varying $q$.  
They found that I and J modes were strongly influenced by 
self-gravity showing character different from the P modes.

Papaloizou \& Lin (1989) used a variational principle approach 
to study thin (flat) self-gravitating disks. They found modes which fell 
into three categories determined by the distribution of vortensity (see also 
Papaloizou \& Savonije 1991). One kind of mode is associated with extrema in 
vortensity, corresponding to a disk where corotation is located at the radius 
of the maximum density. A second mode depicts modes generated by the gradient 
of vortensity on the disk boundaries, corresponding to the existence of the 
corotation radius outside the disk. A third mode is associated with internal 
variations in the vortensity gradient. These modes show corotation inside the 
disk, but not necessarily at the density maximum.

An important development in the study of disks occurred 
when  Adams, Ruden, \& Shu (1989)
showed that the {\it indirect} stellar potential could couple the star and 
disk, and drive one-armed spiral modes, $m$ = 1 modes in disks. Symmetry 
arguments showed that multi-armed modes with $m \ge 2$ could not drive 
the central star off the disk center of mass and so could not contribute to the
indirect stellar potential. Adams, Ruden \& Shu (1989) found that $m = 1$ 
modes were unstable for high mass disks ($M_*/M_d$ $\approx$ 1), attributing 
instability to SLING amplification 
(however, see Heemskerk, Papaloizou, \& Savonije
1991). Noh, Vishniac \& Cochran (1992) 
studied $m = 1$ modes in Keplerian ($q = 1.5$) 
disks for high and low disk masses with emphasis on sensitivity to the outer 
disk boundary conditions. They found that low mass disks, down to 
$M_*/M_d \approx 2.0$, were unstable to $m = 1$ modes only when a reflecting 
outer boundary existed, with growth rates increasing rapidly with 
an increase in disk mass and that there were two types of $m = 1$ modes 
(see also Hadley \& Imamura 2011).

Mathematically simple systems, such as infinitesimally thin disks, 
self-gravitating annuli and tori with constant mass density and circular 
cross-sections, have been studied extensively. Fully 3D, self-gravitating 
disks have received much less attention. Self-gravitating polytropic disks were 
analyzed by Eriguchi \& Hachisu (1983) and Hachisu \& Eriguchi (1985a, 1985b). 
Tohline \& Hachisu (1990) performed nonlinear calculations for $n$ $=$ 1.5, 
varying $q$, for extremely small mass stars, $10^{-9}$ $<$ $M_*/M_d$ $<$ $10^{-6}$, 
making these disks fully self-gravitating. Their analysis included eight models 
but was extended in a second paper, Woodward, Tohline \& Hachisu (1994) where 
a more extensive study was performed, this time including models where the 
star-to-disk ratio was much larger. Hadley \& Imamura (2011), 
performed linear stability analyses on self-gravitating toroids,  
$M_*/M_d$ $=$ 0.0. In addition, they modeled the early nonlinear evolution 
using a quasi-linear theory and, for selected models, fully nonlinear 
techniques. In our present work, we perform an extensive study of 
nonaxisymmetric global instabilities in thick, self-gravitating star-disk 
systems creating a large catalog of star/disk systems covering most of the 
parameter space relevant to protostellar and protoplanetary systems. 
We consider
star/disk systems for $n = 1.5$, $q = 1.5, 1.75$ and $2$, for star masses of 
$0.0 \le M_*/M_d \le 10^3$ and inner to outer edge aspect ratios of 
$0.1 < r_-/r_+ < 0.75$. We discuss how the trends found in the 
non-self-gravitating disks and thin disks systems carry 
over to self-gravitating 
thick disks, as well as how the extra degree of freedom leads to new behavior. 
We perform quasi-linear analysis and compare our linear and quasi-linear 
modeling results with nonlinear simulations in Paper III of this series 
(Hadley {\it et al.} 2014).

The remainder of our paper is organized as follows. Section 2 introduces our 
mathematical methods and concepts. Section 3 presents our results.  Section 
4 contains discussion and applications with comparison of our results with 
those of previous studies. Section 5 contains a summary of our results and conclusions.

\section { EQUILIBRIUM DISKS } \label{sec_equil}

\subsection{ Equilibrium Disk Equations } \label{sec_eq_eqns}

The inviscid, adiabatic, hydrodynamic equations are
\begin{equation}
\partial_t \rho + {\bf \nabla\cdot}(\rho{\bf v}) = 0
\end{equation}
and 
\begin{equation}
\rho (\partial_t+ {\bf v\cdot\nabla} ) {\bf v} 
= -{\bf \nabla} P - \rho {\bf \nabla} \Phi_g
\end{equation}
where $P$ is the pressure, $\rho$ is the mass density, ${\bf v}$ is the 
%
%
velocity, and $\Phi_g$ is the total gravitational potential
composed of the disk self-gravitational 
potential, $\Phi_d$, and the stellar potential, $\Phi_*$. The disk 
potential is found from solution of the Poisson equation
\begin{equation}
\nabla^2\Phi_d = 4\pi{\rm G}\rho
\end{equation}
where G is the gravitational constant. The 
stellar potential is found from the potential formula for a point mass 
\begin{equation}
\Phi_* = -\frac{{\rm G}M_*} {|{\bf r}-{\bf r_*}|},
\end{equation}
where ${\bf r}$ is the field point and ${\bf r_*}$ is the position
of the star. In equilibrium, $r_*$ = 0, but for the perturbed flow,
we determine $r_*$ using the stellar equation of motion given by
\begin{equation}
\frac{d\vec{v}_*}{dt} = -\nabla \Phi_{d}.
\end{equation}
The perturbed form of the stellar equation of motion is
given in equation \ref{linear_stareq}.

We calculate equilibrium 
models using the self-consistent field method (SCF, Hachisu [1986]) under the 
following assumptions:
(i) axial symmetry;
(ii) the fluid rotates on cylinders; and
(iii) the disk has mirror symmetry across the equatorial plane.
For isentropic fluids the relationship between $P$ and $\rho$  
is $P = K \rho^{1+1/n}$ where P is pressure,  
$\rho$ is mass density,  $n$ is the polytropic index, and K is the 
polytropic constant. We investigate models with $n = 1.5$. 
The velocity field is defined using a power law angular velocity distribution 
\begin{equation}
\Omega(\varpi) = \Omega_{\circ}\left( \frac {\varpi} {r_{\circ}} \right)^{-q}\label{Omega}
\end{equation}
where $r_{\circ}$ is the radius of the density maximum, $\Omega_{\circ}$ is the 
frequency of the fluid at $r_{\circ}$, and $\varpi$ is the cylindrical radial 
coordinate. Keplerian rotation refers to $q = 1.5$, and the axisymmetrically neutrally-stable case of constant specific angular momentum is $q = 2$. 
Unless otherwise noted, all quantities are presented in polytropic units 
where G $=$ K $=$ $M_d$ $=$ 1, G is the gravitational constant, and 
$M_d$ is the the disk mass. Conversion between polytropic units and physical 
units can be made using transformations found in Williams \& Tohline (1987).

The mass continuity equation is identically satisfied and the momentum 
conservation equation may be integrated once for axisymmetric disks that rotate 
on cylinders. In cylindrical coordinates we find 
\begin{equation}
\frac{\gamma}{\gamma-1}{\rm K}\rho^{1/n}-\frac{{\rm G}M_*}{r}+\Phi_d
-\int\Omega_{\circ}^2r_{\circ}^{2q}\varpi^{1-2q}{\rm d}\varpi-C = 0
\end{equation}
where $\gamma$ = 1+1/$n$, $M_*$ is the mass of the star,
$C$ is the integration constant whose value is determined by 
the boundary conditions.  Defining dimensionless variables
\begin{equation}
\begin{split}
\psi=\frac{\rho}{\rho_{\circ}},\;\;
\theta=\frac{\varpi}{r_{\circ}},\;\;
\xi=\frac{r}{r_{\circ}},\;\; 
\varphi_d = \frac {\Phi_d} {\Phi_{\circ}},\;\; \\
{\rm and}\;\;
\varphi_c = - 
\int \theta^{1-2q}{\rm d}\theta
\end{split}
\end{equation}
where $\varphi_c$ is the centrifugal potential, we arrive at
\begin{equation}
\frac {\gamma} {\gamma-1} \psi^{1/n}-
\left(\frac {M_*} {M_d}\right)\frac{1}{\xi}
+\varphi_d
+h_{\circ}^2\varphi_c-C^{\prime} = 0 .  \label{equilib_eqn}
\end{equation}
After setting
\begin{equation}
\frac {\Phi_{\circ}} {{\rm K} \rho_{\circ}^{1/n}} = 1.\;\;
h_{\circ}^2 
= \frac {\Omega_{\circ}^2r_{\circ}^2} { {\rm K}\rho_{\circ}^{1/n} }
,\;\;
C^{\prime} = \frac {C} { {\rm K}\rho^{1/n} }
\end{equation}
and using the Poisson equation to show that 
\begin{equation}
\Phi_{\circ} = {\rm G} \rho_{\circ} r_{\circ}^2 \rightarrow
\frac { {\rm G}M_*} { {\rm K}\rho_{\circ}^{1/n}r_{\circ} } = \frac{M_*}{M_d} .
\end{equation}

We solve Equation \ref{equilib_eqn} as follows: (i) values for $\psi$ are guessed; (ii) 
$\varphi_d$ and $\varphi_c$ are calculated; (iii) using $\varphi_d$ 
and $\varphi_c$, Equation \ref{equilib_eqn} is inverted for $\psi$; (iv) if the guessed 
and calculated $\psi$ agree to within a predetermined tolerance, the 
calculation is stopped. If they do not, the guess for $\psi$ is improved 
and steps (ii) and (iii) repeated. This is continued until the guessed and 
calculated $\psi$ are consistent. We use a global test for convergence by 
monitoring the changes in the constants $h_{\circ}^2$ and $C^{\prime}$ 
from iteration to iteration. We quantified the accuracy 
of our result using the virial theorem,
2 (T+T$_{th}$)+W $=$ 0, where T is the rotational kinetic energy, 
T$_{th}$ is the kinetic energy in thermal motion, and W is the
gravitational energy. In practice, this quantity does not equal zero.
In general, our models had
\begin{equation}
{\rm VT} = |\frac {T+T_{th}+0.5 W}{-0.5W}| < 10^{-3}-10^{-4},
\end{equation}
However, for small $M_*/M_d$ and large $r_-/r_+$, VT could be as large as 0.01.

\subsection{ Equilibrium Disk Properties } \label{sec_eq_props}

\begin{figure}
\begin{center}
\includegraphics[trim=0 2in 0 1in,clip,width=4in]{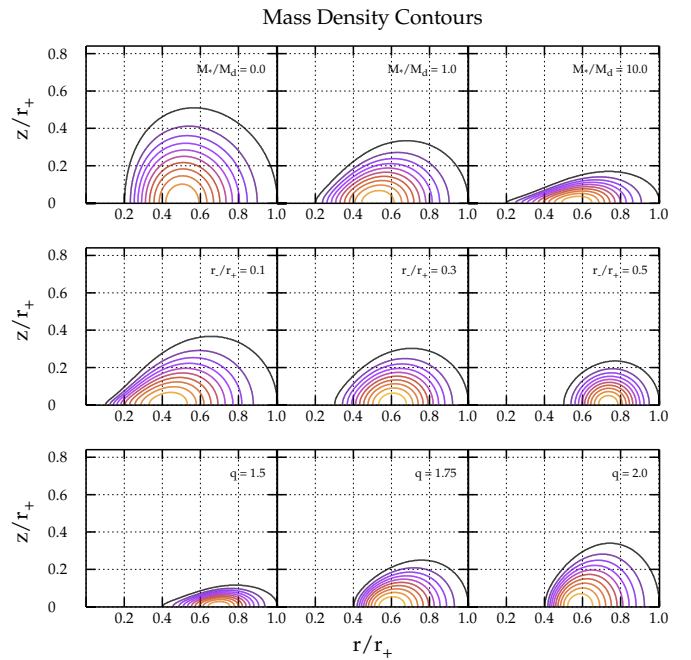}
\end{center}
\caption{
Top row: Mass density contours for models of varying values of $M_*/M_d$
with $q$ = 1.5 and $r_-/r_+$ = 0.20. Contours trace ten divisions between
the arbitrary low density of $10^{-30}$ and the max density.
Middle row: Contours for models with $q=1.5$ and $M_*/M_d = 1$, 
sweeping $r_-/r_+$.
Bottom row: Contours for models with $M_*/M_d=25$ and $r_-/r_+=.4$, 
sweeping $q$.
} 
\label{aEquiContours}
\end{figure}

An extensive library of equilibrium disk models was compiled  
covering the parameter space occupied by typical protostellar 
and protoplanetary disks. Disk models were grouped into families 
defined by $n$ and $q$, where family members were parameterized by $r_-/r_+$, 
the ratio of the inner and outer radii of the disk, and $M_*/M_d$, the ratio 
of the star mass disk mass. We modeled $q$ $=$ 1.5, 1.75 
and 2 disks for $r_-/r_+$ = 0.05 to 0.75, and $M_*/M_d$ = 0 to 
10$^3$. Density contours 
in meridional slices for representative models 
are shown in Figure \ref{aEquiContours} to 
qualitatively illustrate the effects of varying $M_*/M_d$, $q$ and $r_-/r_+$.
Large star masses, small inner radii and shallow rotation curves all have the 
effect of flattening the disk. Small stars, large inner radii and steep 
rotation curves have the opposite effect leading to values of $h/r \approx 1$
where $h$ is the disk thickness at radius $r$. 
The latter is where we expect thin-disk approximations to break down.

We investigate disk stability based on local and global
macroscopic properties of the equilibrium disks. Some examples 
of parameters include 
the well-known Toomre Q parameter which indicates local instability
in thin disks if Q falls below unity, where
\begin{equation}
Q=\frac{c_s\kappa}{\pi{\rm G}\Sigma},
\end{equation}
$c_s$ is the local sound speed, and $\kappa$ is the epicyclic frequency,
\begin{equation}
\kappa^2 = (4-2q)\Omega^2
\end{equation}
for power law rotation (Toomre 1964).
For nonaxisymmetric instability, a corresponding necessary and 
sufficient condition does not exist but it has been suggested 
that systems which have $Q \lesssim 1.5-1.7$ anywhere
in the disk are unstable to nonaxisymmetric instability
({\it e.g.}, Durisen {\it et al.} 2007).  
For $q$ = 2, Q = 0. 
For $q$ = 1.5, disks with small
$M_*/M_d$ show large regions where $Q < 1.5-1.7$ while for large
$M_*/M_d$, $q = 1.5$ disks show $Q > 1.5-1.7$ everywhere. An additional 
effect of $Q$ arises in regions where $Q < 1$. In such 
regions, traveling waves damp.
For $q$ $\approx$ 2, the disk has $Q \approx 0$ because $\kappa^2$ $\approx$ 0. 
Qualitatively similar behavior between $q = 1.5$ and $1.75$ is seen. The region 
where $Q < 1$ begins wide, decreases in width monotonically with increasing 
$M_*/M_d$ and finally disappears.
On grounds of the Q parameter, disks near $q = 2$ are likely to be much more
unstable than $q$ = 1.5 and 1.75 disks. For $q = 1.5$ and $1.75$, systems 
with large $M_*/M_d$ are likely to be stable based on $Q$.

Another parameter commonly used in the analysis of equilibrium disks is 
the ratio of the rotational kinetic energy to the absolute value of 
the gravitational potential energy, $T/|W|$. $T/|W|$ is particularly 
useful in the analysis of gravito-rotation driven nonaxisymmetric
modes in star-like objects.  
$q = 1.5$ disks, Kepler-like disks, show higher values of $T/|W|$ for large $M_*/M_d$ than do 
$q = 2$ disks. $T/|W| \rightarrow 0.5$ for the largest $M_*/M_d$ 
and $r_-/r_+$ for all $q$ and varying $q$ and $r_-/r_+$ has little effect on
$T/|W|$ for large $M_*/M_d$. $T/|W|$ also varies differently
across parameter space for different $q$ making it less useful
as a universal stability indicator for star/disk systems than has been 
found for toroids (Hadley \& Imamura 2011) and {\it spheroidal} objects
({\it e.g.,} Tassoul 1978).

Two local measures of the importance of self-gravity,
\begin{equation}
p^2 = \frac{4\pi{\rm G\rho_{\circ}}}{\Omega_{\circ}^2},
\end{equation}
and
\begin{equation}
\eta=\frac{\Omega_k^2}{\Omega_{\circ}^2}
\end{equation}
are useful indicators of 
stability in thin-disks and ICTs ({\it e.g.,} Christodoulou \& Narayan 1992,
Andalib, Tohline, \& Christodoulou 1997). Here,
$\Omega_k$ is the Keplerian angular velocity given by $\Omega_k$
= $\sqrt{GM_*/r_{\circ}^3}$.
Although $p$ and $\eta$ show different behavior and also vary with
$q$, individually, each
may be a useful indicator for the stability properties of 
thick, self-gravitating disk systems.

\section{ NONAXISYMMETRIC DISK INSTABILITIES } \label{sec_instabilities}

A foundation is presented for the understanding of nonaxisymmetric disk modes,
beginning with the linearized evolution equations. This is followed by detailed
descriptions of the character of the several types of modes observed (J, E/P, I, m=1),
followed by laying out, in ($r_-/r_+,M_*/M_d$) space for given $q$, the eigenvalues
and character of the dominant low-m modes. The section concludes with
discussion and calculation of the angular momentum transport properties of
representative modes from all classes, with particular attention to $m=1$.

\subsection{ Linearized Evolution Equations } \label{sec_in_lineq}

Linearly unstable modes are found from solution of an Initial Value Problem 
(IVP) formulated from the hydrodynamics equations evaluated using Eulerian 
perturbations of the form, 
\begin{equation}
A = A_{\circ}+\delta A(\varpi,z,t) e^{im\phi},
\end{equation}
where $A_{\circ}$ is the equilibrium solution and $\delta A$ is the perturbed 
amplitude in the meridional plane. For all complex perturbed quantities, the
physical solution corresponds to the real part.

From the perturbations and the hydrodynamic 
equations, a set of linearized evolution equations is formed 
(Hadley \& Imamura 2011). In cylindrical coordinates, 
the linearized hydrodynamic equations are
\begin{eqnarray}
\partial_t \delta\rho
&=&
-im\Omega\delta\rho-
\rho_{\circ} \frac {\delta v_{\varpi}} {\varpi} - 
\delta v_{\varpi}\partial_{\varpi}
\rho_{\circ} - \delta v_{z} \partial_z\rho_{\circ} \nonumber \\
&& -\rho_{\circ} \left( \partial_{\varpi}
 \delta v_{\varpi}
 +\frac {im} {\varpi} \delta v_{\phi}
 +\partial_z \delta v_{z} \right) \\
\partial_t \delta v_{\varpi}
&=&
-im\Omega 
\delta v_{\varpi}
+ 2\Omega \delta v_{\phi}
-\frac {\gamma P_{\circ}} {\rho_{\circ}^2} \partial_{\varpi}
\delta\rho \nonumber \\
&& -(\gamma-2) \frac {\delta\rho} {\rho_{\circ}^2}
\partial_{\varpi} P_{\circ}
-\partial_{\varpi}\delta\Phi_g, \\
\partial_t\delta v_{\phi}
&=&
-im\Omega 
\frac {\delta v_{\varpi}} {\varpi} \partial_{\varpi}(\Omega\varpi^2)
-\frac{im}{\varpi}\frac{\gamma P_o}{\rho_o^2}\delta\rho \nonumber \\
&& -\frac{im}{\varpi}\delta\Phi_g, \\
\partial_t\delta v_{z}&=&-im\Omega \delta v_{z}
-\frac{\gamma P_o}{\rho_o^2}
\partial_z\delta
\rho-(\gamma-2)\frac{\delta\rho}{\rho_o^2}
\partial_zP_o \nonumber \\
&& -\partial_z\delta\Phi_g.
\end{eqnarray}

The perturbed disk gravitational potential $\delta\Phi_d$ 
is found by solving the linearized Poisson equation,
\begin{eqnarray}
\nabla^2(\delta\Phi_de^{im\phi}) = 4\pi{\rm G}\delta\rho e^{im\phi}
\end{eqnarray}
for the azimuthal mode {\it m} under consideration.
The star follows a spiral trajectory giving rise to the 
{\it indirect potential},
\begin{equation}
\delta\Phi{*} = -\frac { {\rm GM}_*}{r}\left(\frac{{\bf \delta_*\cdot r}}{r^2}
\right)
\end{equation}
where $\delta_*$ is the perturbed location of the star.  
Both $\delta\Phi_d$ and $\delta\Phi_*$
are included in the evolution equations. 
We find ${\bf \delta_*}$ by solving
the linearized stellar equation of motion,
\begin{eqnarray}
\partial_{tt}{\bf \delta}_*  &=&
\pi{\rm G}({\bf \hat{x}+i\hat{y}})\int\frac{\delta\rho}{r^3}
{\rm \varpi^2d\varpi dz} \nonumber \\
&& +\pi{\rm G}{\bf \delta_* }\int\frac{\rho_{\circ}}{r^3}
\left(3(\frac{\varpi^2}{r^2})-2\right){\rm \varpi^2d\varpi dz},  \label{linear_stareq}
\end{eqnarray}
simultaneously with the linearized evolution equations.
The Cartesian coordinates of the star are related as
$\delta_{*,y} = i\delta_{*,x}$ so that the perturbed 
stellar potential is
\begin{equation}
\delta\Phi_{*} = -\frac { {\rm GM}_*}{r}\left(\frac{\delta_{*,x}^R
+i\delta_{*,x}^I}{r^2}\right)
\varpi e^{i\phi}
\end{equation}
where $\delta_{*,x}^R$ and $\delta_{*,x}^I$ are the real and imaginary
parts of the perturbed $\hat{x}$ component of the stellar position.

Equilibrium values for $\rho$ and ${\bf v}$ are used as the background for 
solution of the IVP equations. The IVP is evolved on the same grid as the 
equilibrium models. Spatial derivatives are given in finite difference 
form and time derivatives left continuous. The perturbed solutions are 
advanced in time using a fourth order Runge-Kutta method. The numerical code 
is described in detail in Hadley \& Imamura (2011). We usually used grid 
sizes of $n_{\varpi}\times n_{z} =  512 \times 512$ although for disks with 
large $r_-/r_+$ and $q$ $=$ 1.5, we sometimes used grids of size 
$1024 \times 1024$.
Convergence tests were run with resolutions of $256^2$, $512^2$ and $1024^2$, 
with eigenvalues typically agreeing with each other to
within a few percent. Less agreement was found near transition regions,
boundaries between mode types, which tended to shift slightly in $r_-/r_+$ for a
given $M_*/M_d$.

Boundary conditions consist of mirror symmetry about the 
equatorial plane. Perturbed velocities in the ${\varpi}$ and z directions are set to 
zero on the surface of the disk, while the mass density perturbation is 
set to zero gradient. 
The gravitational potential is solved in the Coulomb 
gauge and computed at the outer grid boundaries using a sum over spherical harmonics up to $l=16$.

We find that our results agree with the early-time behavior of full nonlinear
simulations made using the CHYMERA code
(Hadley {\it et al.} 2014). 
In tests, we agreed with the work of Blaes (1985), who found 
analytic
eigenvalues for infinitely slender $q=2$ non-self-gravitating tori within about
10\%. The evolution equation coefficients at the disk outer boundary are weakly
singular, and likely introduce some problems for our fixed grid code 
causing the discrepancy.

The evolution of the perturbed disk was followed by monitoring 
$|\delta\rho|/\rho_{\circ}$ at three points in the disk midplane to ensure that 
instability is global in nature. A model is 
deemed dynamically stable if it shows no global 
growth after 30 - 40 $\tau_{\circ}$ where $\tau_{\circ}$ is the rotation period 
at the radius of the density maximum.  We monitor growth until time dependence
has settled into stable exponential behavior, 
\begin{equation}
f(t) = f_0 e^{-i \omega_m t} \label{harmonicpert}
\end{equation}
for any perturbed quantity 
$f$. We choose these and other sign conventions to 
be consistent with previous workers in the field ({\it e.g.,} see Kojima 1986).
Growth rates and oscillation frequencies are determined from least squares fits 
to the logarithm of the amplitude and the phase, respectively. For our 
definition, the real part of $\omega$ refers to the frequency of the 
perturbation while the imaginary part refers to the growth rate. Prograde modes 
have ${\cal R}(\omega) < 0$. For our choice of the form for the azimuthal 
eigenfunction, normalized eigenvalues are defined as		
\begin{equation}
y_1(m) = -\left(\frac{{\cal R}(\omega_m)}{\Omega_{\circ}}+m\right)\;\;{\rm and}\;\;
y_2(m) = \frac{{\cal I}(\omega_m)}{\Omega_{\circ}}. \label{define_y}
\end{equation}
Here ${\cal R}$(z) and ${\cal I}$(z) take the real and imaginary parts
of a complex variable.

The corotation radius, $r_{co}$, is where the pattern frequency of 
the mode equals the orbital frequency of the fluid.
 The co-rotation radius acts as a resonance point
in the disk where periodic forcing may amplify the density perturbation
(see Goldreich \& Tremaine 1979). If there is no real component in the
density perturbation, a singularity may arise at the corotation resonance point
(see eq. \ref{harmonicpert}-\ref{define_y}).  The $y_1(m)$ are
defined so that if $y_1(m) < 0$, $r_{co} >r_{\circ}$ and if 
$y_1(m) > 0$, $r_{co} < r_{\circ}$.  For power-law $\Omega$,
$r_{co}$ is given by:
\begin{equation}
\frac{r_{co}}{r_{\circ}} = \left(\frac{y_1(m)}{m}+1\right)^{-1/q}
\end{equation}

The inner and outer Lindblad resonances $r_{ilr}$ and $r_{olr}$,
are located where the real part of $\omega_m$
equals $\pm \kappa$.
For power-law $\Omega$, $r_{ilr}$ and $r_{olr}$
are related to $r_{co}$ by:
\begin{equation}
\frac{r_{lr}}{r_{co}}= \left(1\pm\frac{\sqrt{4-2q}}{m}\right)^{1/q}
\end{equation}

The vortensity is defined as 
\begin{equation}
{\bf \Lambda} = \frac {\bf \nabla \times v}{\Sigma} 
\end{equation}
where $\Sigma$ is the column density. 
For axial disks that rotate on cylinders
with power law angular velocity, the only nonzero component 
of ${\bf \Lambda}$ is 
\begin{equation}
\Lambda_z = (2-q)\left(\frac{\Omega}{\Sigma}\right) .
\end{equation}
Vortensity modes arise when corotation falls at extrema in $\Lambda_z$ 
(Papaloizou \& Savonije 1991).

For analysis purposes, we calculate the work done locally by the perturbed 
kinetic energy, and the perturbed enthalpy which 
also accounts for the perturbation in 
the acoustic energy (see Kojima 1989). The perturbed 
kinetic energy and acoustic energy 
are designated as $E_k$ and $E_h$, respectively and, 
for a polytrope, are given by:
\begin{equation}
{\rm E}_k = \frac{1}{2}\rho\langle\delta v_{\varpi}^2+\delta v_{\phi}^2+
\delta v_z^2\rangle
\end{equation}
\begin{equation}
{\rm E}_h = \frac{1}{2}\gamma\frac{P}{\rho^2}\langle\delta\rho^2\rangle
\end{equation}
Here, the brackets represent time-averaged perturbed quantities.
The total energy of the mode is the sum of the two. The time rate of  
change of the perturbed energy may be broken down into the time rate of 
change of the stresses:
\begin{equation}
\partial_t{\sigma}_{R}  = 
-\rho_{\circ}\varpi\partial_{\varpi}(\delta v_{\phi}\delta v_{\varpi})
\end{equation}
\begin{equation}
\partial_t{\sigma}_{G} =
-\rho_{\circ}(\delta{\bf v\cdot\nabla})(\delta\Phi_d+\delta\Phi_*)
\end{equation}
\begin{equation}
\partial_t{\sigma}_{h}  = 
-{\bf \nabla\cdot}(\delta P\delta v)
\end{equation}
where $\sigma_{R}$ is the 
Reynolds stress, $\sigma_{G}$ is the gravitational
stress, and $\sigma_{h}$ is the acoustic stress.

\subsection{ Classification Of Nonaxisymmetric Disk Modes } \label{sec_in_class}

Modes are identified from their 
morphological and dynamical properties. We use characteristics including
the winding of their arms, the regions in the disk where 
they have the largest amplitude, locations of corotation
and vortensity extrema, and the self-gravity parameter $p$ to classify modes.
We have also examined the ratio of gravitational and Reynolds stress 
rates integrated over the disk,
\begin{equation}
{\cal R} = \frac {\int{\partial_t\sigma}_{G}{\rm d}^3{x}}
{\int{\partial_t\sigma}_{R}{\rm d}^3{x}}
\end{equation}
and found that it tracks mode type almost exactly as $p$ does for $m \ge 2$,
confirming the relative importance of self-gravity.

No one characteristic is sufficient or necessary to define 
mode type. In the following subsections we present representative 
models that characterize each mode type. Tables 1 to 4
summarize the properties of the modes presented in the 
figures and discussed in the following sections.
The azimuthal mode numbers $m$ with the
highest growth rate in the amplitude of the density perturbation
are illustrated and discussed in Sections \ref{sec_in_regime} and \ref{sec_Msummary}.
The reader may find it helpful to reference this section to 
understand which azimuthal mode number dominates respective 
modes (J, E/P, I, m=1) discussed.

\subsubsection{J Modes} \label{sec_in_J}

J modes, the Jeans-like modes, are driven by self-gravity. 
They dominate systems with narrow, $r_-/r_+$ $>$ 0.30,
and high mass disks, $M_*/M_d$ $<$ 0.2, where 
the self-gravity parameter $p \gtrsim$ 7.5.
For given $M_*/M_d$, J modes with successively
higher $m$ dominate as $r_-/r_+$ increases (see Figs \ref{q15y_isocontours}-\ref{q20y_isocontours}).

\begin{figure*}
\begin{center}
\includegraphics[width=3.0in]{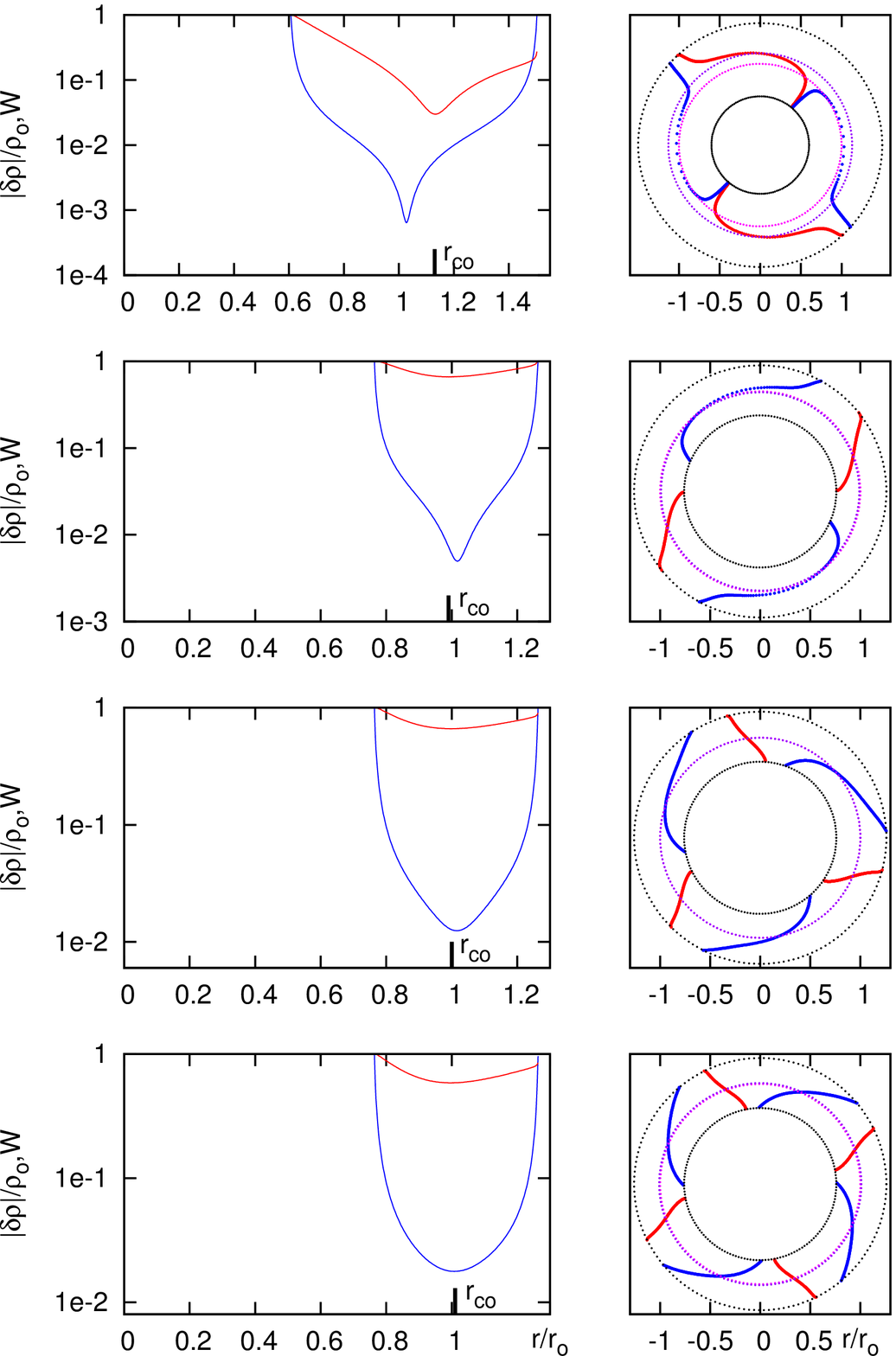}
\includegraphics[width=3.0in]{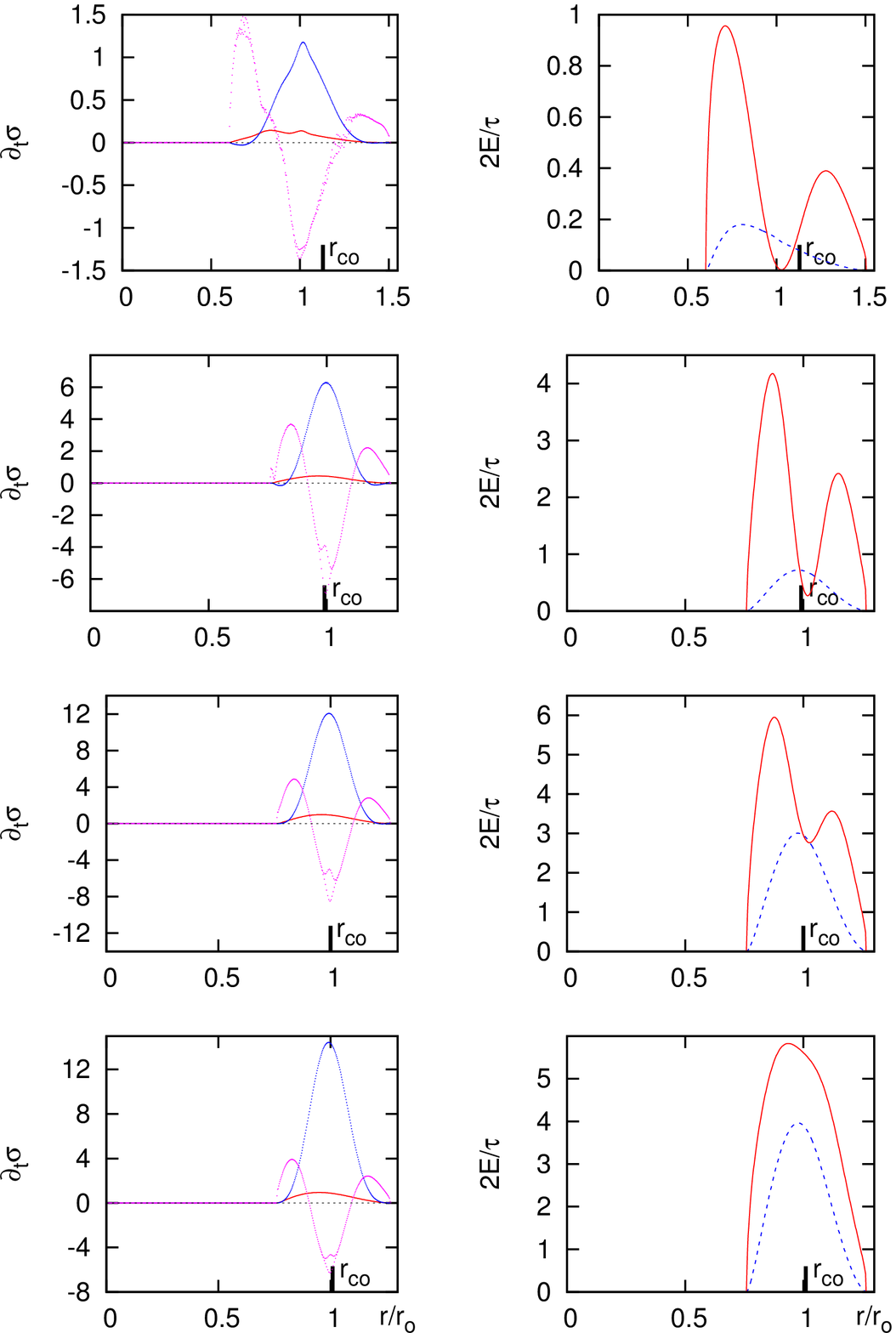}
\end{center}
\caption{
$m$ = 2, 3, and 4 J modes for $q$ = 1.5 and $M_*/M_d$ = 0.01 systems with 
$r_-/r_+$ = 0.4 and 0.6, models J1, J2a, J2b, and J2c from top-to-bottom.
We show $\delta\rho$ and ${\cal W}$ amplitudes and phases,
$\partial_t\sigma$, and $\delta$J. For the 
eigenfunctions, the blue curve is for $\delta\rho/\rho_{\circ}$ and the
red curve for ${\cal W}$. For the 
$\partial_t\sigma$, the Reynolds stress is
the red curve, the gravitational stress the blue curve, and the acoustic
stress rate the magenta curve. For the perturbed energies, the kinetic energy is
the blue curve and the enthalpy the red curve.
For the first column, the ratios of the unnormalized maximum values for $|\delta\rho|$/${|\cal W|}$ are 
2687, 9629, 4954 and 4256,
respectively, from top-to-bottom.
}
\label{Jmode_plots}
\end{figure*}

In Table 1, we outline the parameters of four representative 
star disk systems which exhibit J modes. The properties of typical J 
modes are illustrated 
in Figure \ref{Jmode_plots}. We show constant phase loci for 
$\delta\rho/\rho_{\circ}$ and ${\cal W}$, amplitudes
of $\delta\rho/\rho_{\circ}$ and ${\cal W}$ subject to an 
arbitrary scaling factor, work integrals, and 
stress rates in the disk midplane for $q$ = 1.5 disks with 
$M_*/M_d$ = 0.01 and large $r_-/r_+$. Here
\begin{equation}
{\cal W} = \gamma\frac{P_{\circ}}{\rho_{\circ}^2}\delta\rho+\delta\Phi
\end{equation}
is an alternative 
eigenfunction formed from the sum of the perturbed enthalpy and perturbed 
gravitational potential
(Papaloizou \& Pringle 1984). ${\cal W}$ is a more natural eigenfunction for
disks than is $\delta\rho$ ({\it e.g.,} Hadley \& Imamura 2011). This 
is apparent in the location of corotation which more closely tracks 
minima in ${|\cal W|}$ than minima in $|\delta\rho|$.   
 The benefit of tracking the ${\cal W}$ eigenfunction is consistent 
with the corotation singularity at threshold being removed 
by ${\cal W}$ = 0 rather than by $\delta\rho$ = 0. This behavior is also clear 
for the $m$ $\ge$ 2 I and J modes. Note that in Figure 
\ref{Jmode_plots} and in the rest of the paper, the amplitudes
of $\delta\rho/\rho_{\circ}$ and ${\cal W}$ are normalized using their 
respective maximum values in the disk midplane. 
We have included the ratios of the unnormalized maximum values for
$|\delta\rho/\rho_{\circ}|$ / ${|\cal W|}$ in the figure captions.

\begin{table*}[t]
\centering
Table 1: Representative J Modes
\vskip 0.1in
\begin{tabular}{*{10}{c}}
\hline\hline
 &
$r_-/r_+$ &
$r_+/r_{\circ}$ &
$r_{\circ}$ &
$\tau_{\circ}$ &
$J$ &
$m$ &
$y_1,y_2$ &
$r_{ilr}/r_{\circ},r_{olr}/r_{\circ}$ &
$r_{co}/r_{\circ}$ \\
\hline
J1 &0.402 &1.51 &6.47 &187& 1.43 &2 &-0.344,0.110
&0.714,1.49 &1.13 \\
J2a &0.600 &1.27 & 14.8 & 560 & 2.49 &2 &0.0297,0.840
&\nodata,1.30 &0.990 \\
J2b& " & " & " & " & " & 3 &-0.0271,1.57
&0.768,1.22 &1.01 \\
J2c & " &"  & " & " & " & 4 & -0.0844,1.85
&0.837,1.18 &1.01 \\
\hline
\end{tabular}
\label{Jmode_table}

\caption{Examples of J modes and their properties, giving (left to right)
 the radial aspect ratio of the disk, the disk outer radius, radius of 
density max, characteristic 
rotation time at density maximum $\tau_0$ in polytrope units, total angular 
momentum, azimuthal mode number $m$, oscillation frequencies and growth rates of
perturbed density, the radii of the inner and outer Lindblad resonances and the corotation radius.}

\end{table*}

J modes do not show large variation 
in their properties. Dominant azimuthal wavenumbers for these models 
are illustrated in Figure \ref{y1y2_small_plot}, Section \ref{sec_in_regime}.     
The constant phase loci for $\delta\rho/\rho_{\circ}$  
are barlike for $r < r_{\circ}$, and show trailing spiral arms that
extend for $\sim\pi/m$ radians for $r > r_{\circ}$ (Figure 
\ref{Jmode_plots}, column 2).
The arms shown by ${\cal W}$ lead those of $\delta\rho$ 
indicating that the gravitational perturbation 
leads the enthalpy perturbation. The work integrals 
(Figure \ref{Jmode_plots}, column 4)
for the $m$ = 2 modes show two distinct peaks in $E_h$, the first 
peak larger than the second. $E_k$ 
has one peak which is much lower in amplitude than the $E_h$ 
peaks, near the center of the disk for large $r_-/r_+$ 
models and skewed toward the inner edge for models with small 
$r_-/r_+$. For the $m$ = 3 mode, the minimum between the peaks does 
not go to zero, and the $m$ = 4 mode has only one peak 
each for $E_k$ and $E_h$, with the $E_k$ at higher amplitude. 
As for the stress rates (Figure \ref{Jmode_plots}, column 3)
 ${\partial_t\sigma}_{h}$ dominates near the inner
edge with higher amplitude for the $r_-/r_+$ = 0.40 
model than for the $r_-/r_+$ = 0.60 model. In the 
central part of the disk, ${\partial_t\sigma}_{G}$ 
is positive while ${\partial_t\sigma}_{h}$ 
is negative. For the $m$ = 2 mode, 
the amplitudes of ${\partial_t\sigma}_{G}$  and
${\partial_t\sigma}_{h}$ are roughly comparable, while for the $m > 2$
mode, the amplitude of ${\partial_t\sigma}_{G}$ is higher than that 
of ${\partial_t\sigma}_{h}$. ${\partial_t\sigma}_{R}$
is positive, but significantly lower in amplitude than 
either ${\partial_t\sigma}_{G}$ or ${\partial_t\sigma}_{h}$, 
and is skewed toward the 
inner edge in the $r_-/r_+$ 
= 0.60 model. The stress rates for the $m$ = 2, 3 and 4 are 
similar to each other. 

\subsubsection{P \& Edge Modes} \label{sec_in_PE}

P and edge modes are driven by coupling of inertial waves across 
corotation. For both, corotation and the minimum in ${|\cal W|}$ sit 
at $\sim$ $r_{\circ}$. P and edge modes are dominant in
the region in ($r_-/r_+,M_*/M_d$)-space to the far right and below 
the J mode corner (see Figures \ref{eigenvalues_m12a}
and \ref{eigenvalues_m12b}). Dominant azimuthal wavenumbers for these models
are illustrated in Figure \ref{y2_plots}, Section \ref{sec_in_regime}. 
We identify three characteristic behaviors.
For given $M_*/M_d$ and increasing
$r_-/r_+$, P and edge modes show: (i) bars near the 
inner edge of the disk 
with short forward phase shifts at $r_{\circ}$ which switch to 
long trailing arms outside $r_{\circ}$, sometimes winding around 
the disk repeatedly; (ii) a similar mode with central bars 
and short leading phase shifts but with short trailing 
arms outside $r_{\circ}$; and (iii) another mode with smoothly winding leading 
spiral arms. The instabilities with large winding number are referred 
to as edge modes. Edge modes are associated with the low $r_-/r_+$ 
and/or high $M_*/M_d$. 
The modes in (ii) and (iii) are associated with the humps shown in the 
NSG $y_2$ plot discussed in $\S\ref{sec_in_hiM}$.

\begin{table*}[t]
\centering
Table 2: Representative P and Edge $m$ = 2 Modes
\vskip 0.1in
\begin{tabular}{*{8}{c}}
\hline\hline
 &
$r_-/r_+$ &
$r_+/r_{\circ}$ &
$r_{\circ}$ &
$\tau_{\circ}$ &
$J$ &
$y_1,y_2$ &
$r_{co}/r_{\circ}$ \\
\hline
P1 &0.452 &1.60 &0.254 &0.0808 &5.05 &-0.0649,0.0956
& 1.02 \\
P2 &0.500 &1.49&0.403 &0.161 &6.35 &-0.227,0.151
& 1.06 \\
P3 &0.600 &1.33 &1.09 &0.712 &10.4 &-0.152,0.212
& 1.04 \\
P4 &0.700 &1.21& 3.37 &3.88 &18.4 &-0.0736,0.126
& 1.02 \\
E1 &0.101 &5.52 & 6.13$\times10^{-3}$ &3.03$\times10^{-4}$ & 0.786 &0.428,0.067
& 0.908 \\
E2 &0.202 &2.99 & 0.0229 & 2.18$\times10^{-3}$ &1.51 &0.177,0.0833
& 0.959 \\
E3 &0.402 &1.74 &0.159 &0.04 &3.99 &-0.0524,0.141
& 1.01 \\
\hline
\end{tabular}
\label{PEmode_table}
\end{table*}

\begin{figure*}
\begin{center}
\includegraphics[width=3.0in]{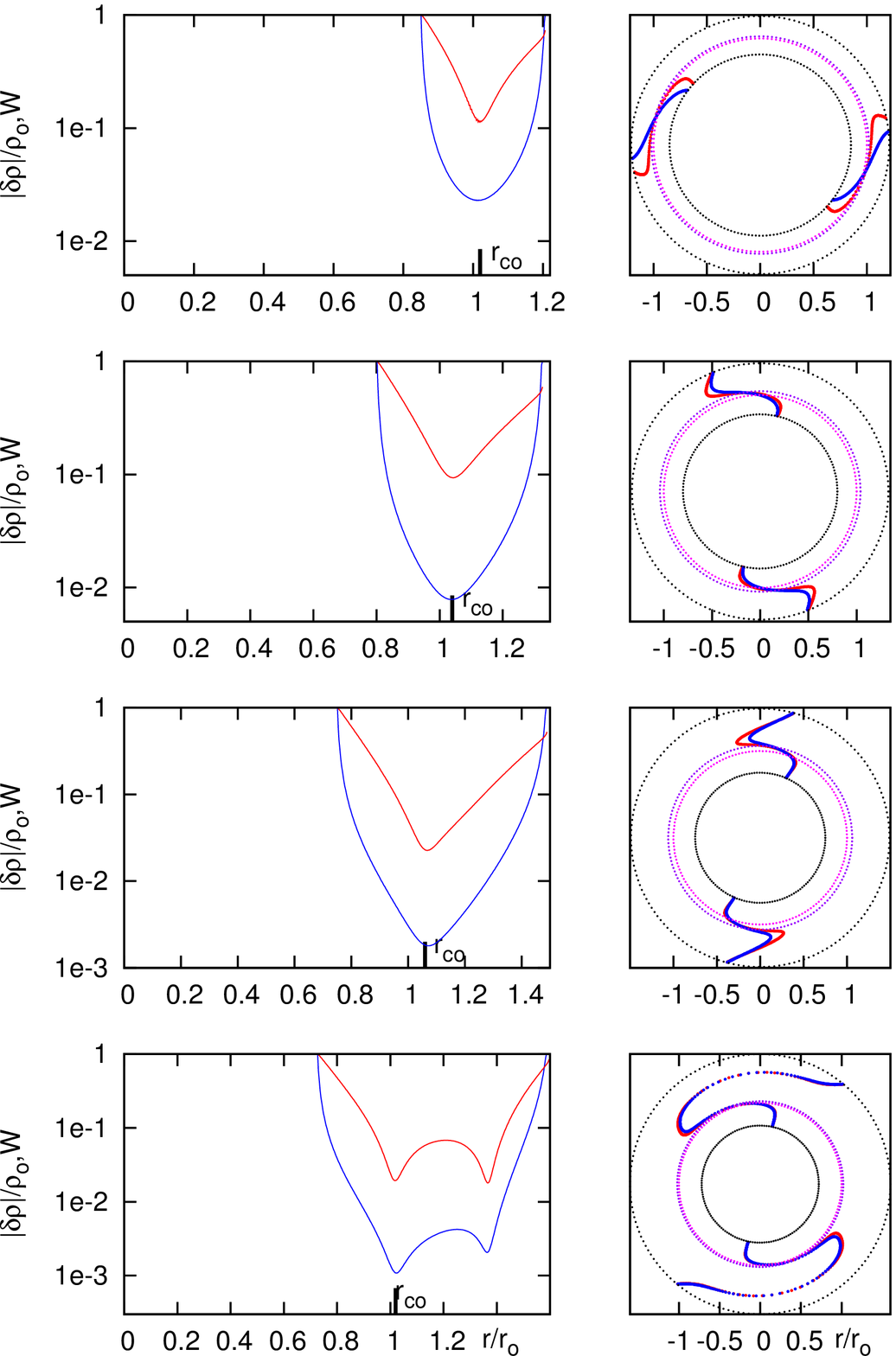}
\includegraphics[width=3.0in]{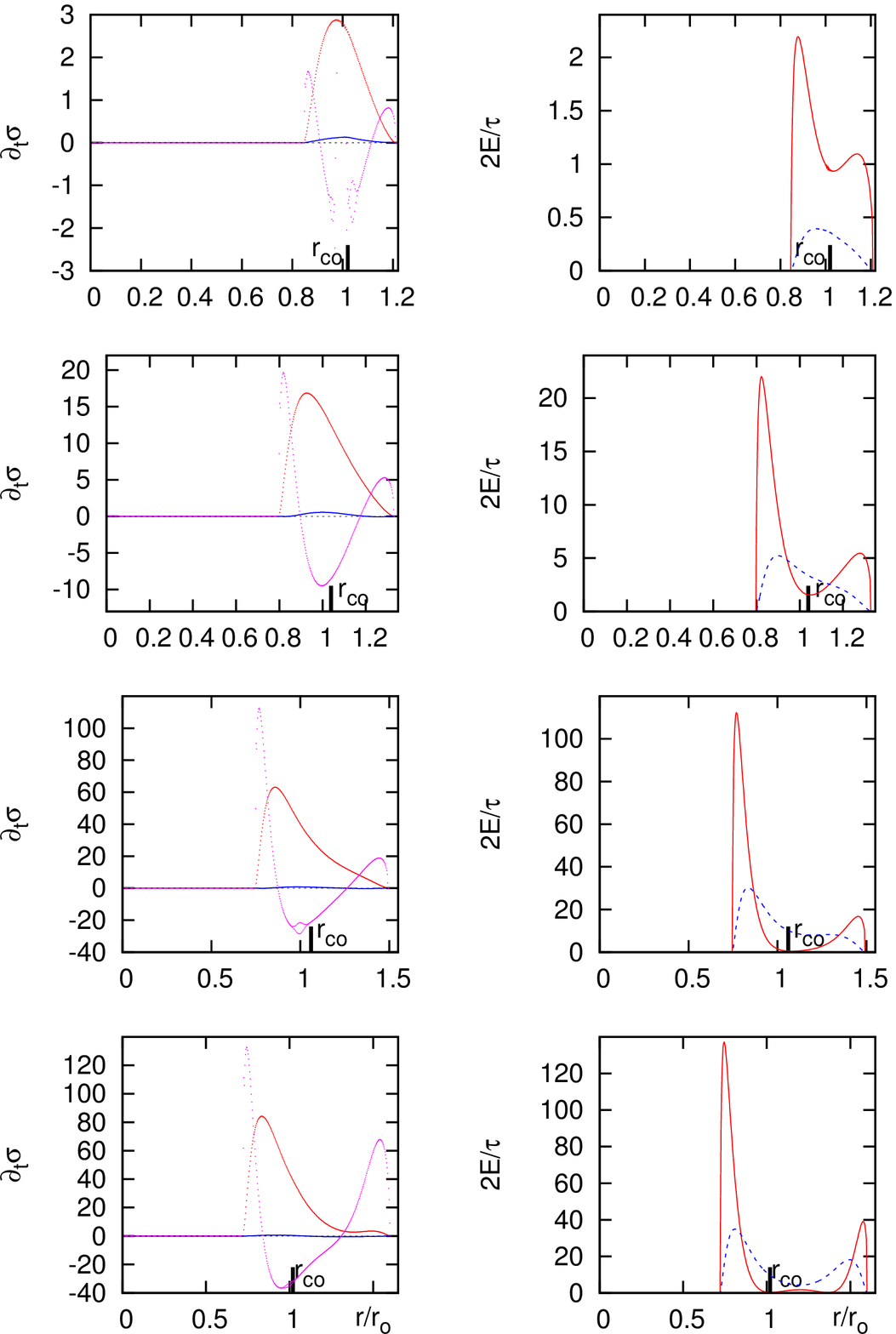}
\end{center}
\caption{
Barlike P modes, $m$ = 2, for disks with $q$ = 2 and 
$M_*/M_d$ = 10$^2$ for $r_-/r_+$ $>$ 0.45, models P4, P3, P2, and P1 from
top-to-bottom. 
We show $\delta\rho$ and ${\cal W}$ amplitudes and phases,
$\partial_t\sigma$, and $\delta$J. For the 
eigenfunctions, the blue curve is for $\delta\rho/\rho_{\circ}$ and the
red curve for ${\cal W}$. For the 
$\partial_t\sigma$, the Reynolds stress is
the red curve, the gravitational stress the blue curve, and the acoustic
stress the magenta curve. For the perturbed energies, the kinetic energy is
the blue curve and the enthalpy the red curve.
For the first column, the ratios of the unnormalized maximum values for $|\delta\rho|$/${|\cal W|}$ =
99.43, 17.70, 4.74 and 4.48,
respectively, from top-to-bottom.
}
\label{Pmode_plots}
\end{figure*}

\begin{figure*}
\begin{center}
\includegraphics[width=3.0in]{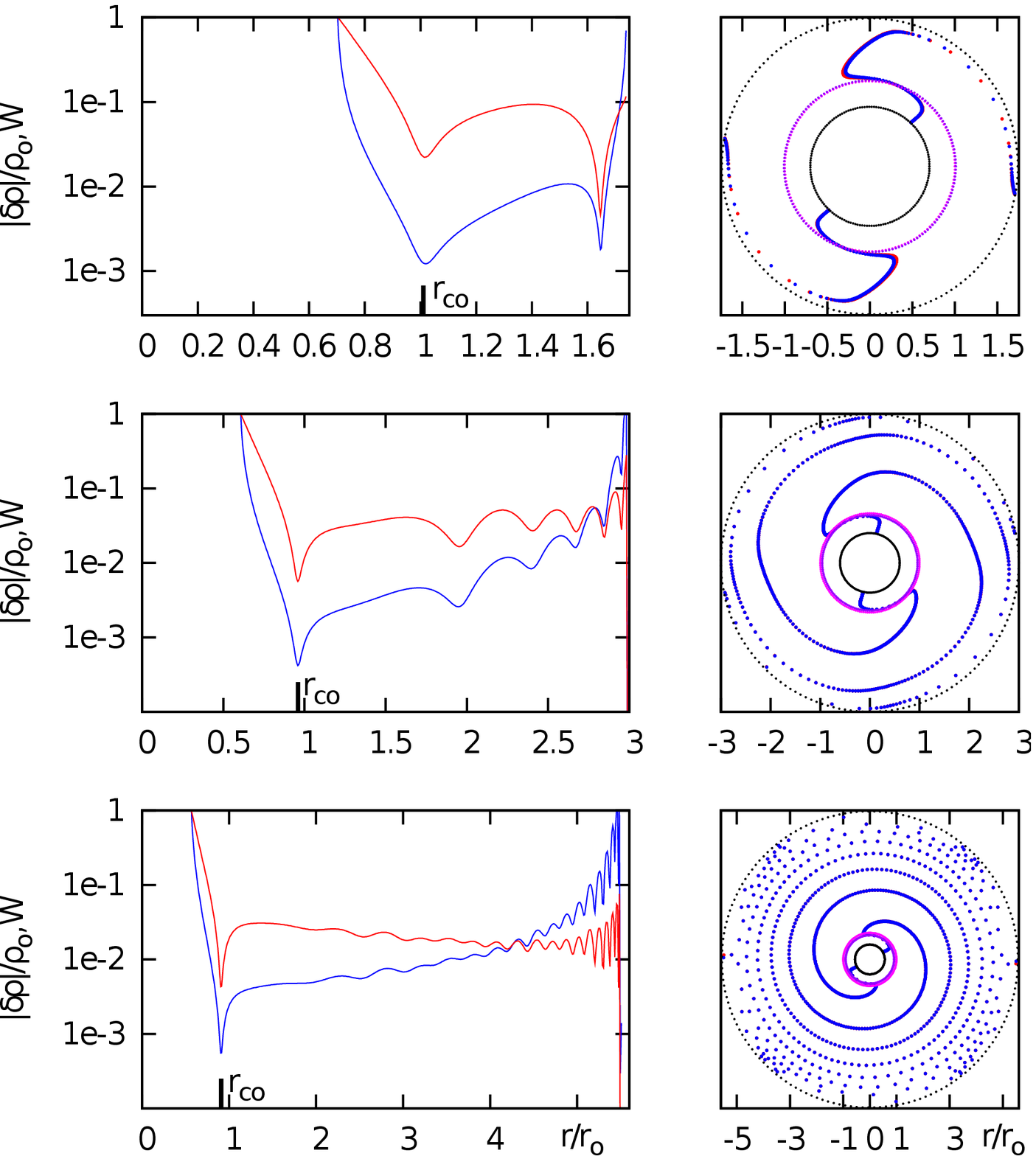}
\includegraphics[width=3.0in]{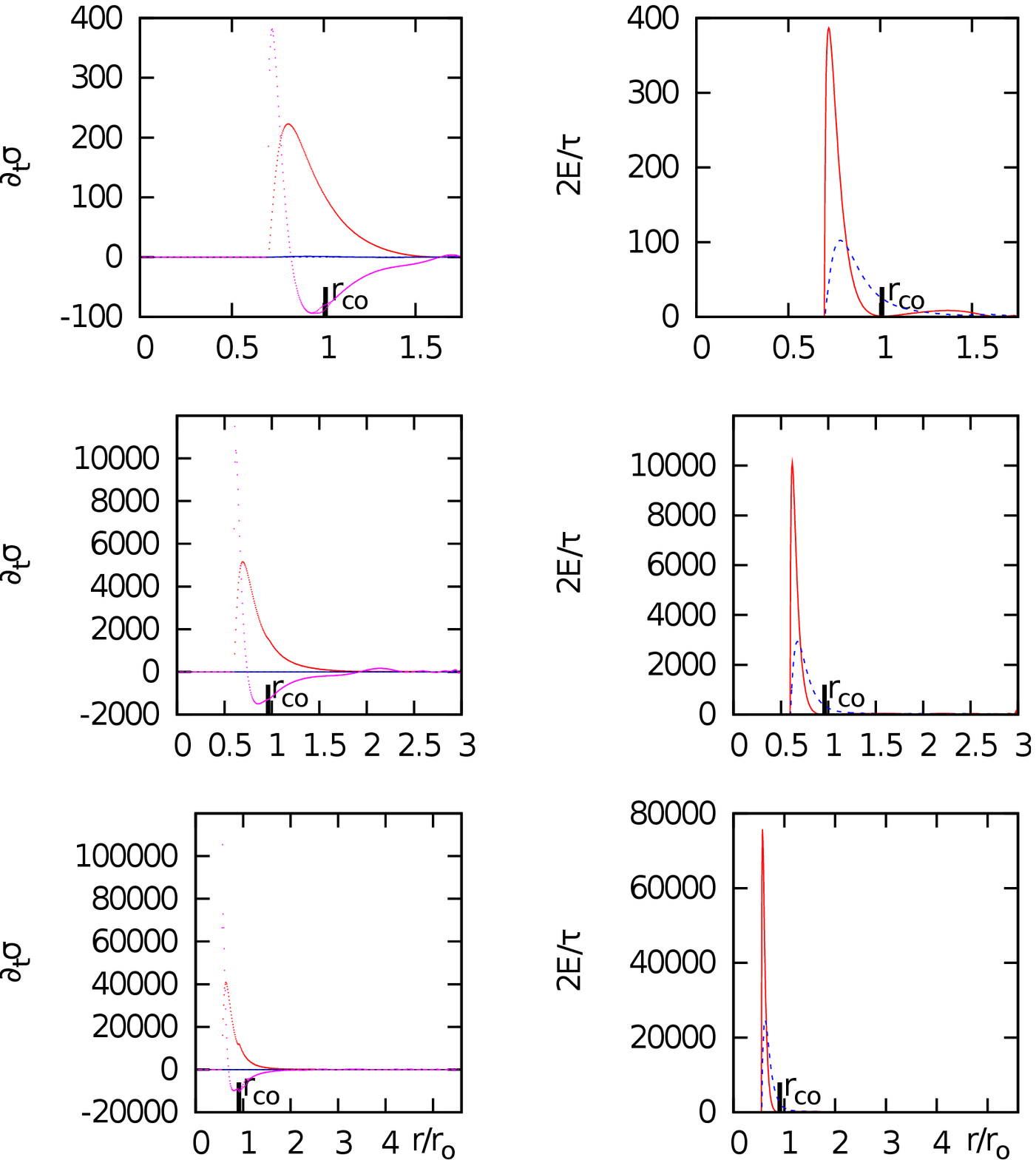}
\end{center}
\caption{
Barlike edge modes, $m$ = 2, for disks with $q$ = 2 and 
$M_*/M_d$ = 10$^2$ for $r_-/r_+$ $<$ 0.45,
models E3, E2, and E1 from top-to-bottom.
We show $\delta\rho$ and ${\cal W}$ amplitudes and phases,
$\partial_t\sigma$, and $\delta$J. For the 
eigenfunctions, the blue curve is for $\delta\rho/\rho_{\circ}$ and the
red curve for ${\cal W}$. For the $\partial_t\sigma$, the 
Reynolds stress is the red
curve, the gravitational stress the blue curve, and the acoustic
stress the magenta curve. For the perturbed energies, the kinetic energy is
the blue curve and the enthalpy the red curve.
For the first column, the ratios of the unnormalized maximum values for $|\delta\rho|$/${|\cal W|}$ are
0.32, 0.14 and 0.0093,
respectively, from top-to-bottom.
}
\label{Edgemode_plots}
\end{figure*}

The characteristic behavior of edge modes and P modes 
is illustrated in 
Figures \ref{Pmode_plots} and \ref{Edgemode_plots} where 
a sequence of figures of $m$ = 2 modes for $q$ = 2,
$M_*/M_d$ = 10$^2$, and
$r_-/r_+$ = 0.70 to 0.10 systems are shown (see also Table 2). 
These models strongly resemble corresponding NSG disks. 
Threshold behavior between edge and P modes for NSG disks 
occurs at $r_-/r_+$ = 0.50, where models with $r_-/r_+$ $<$ 0.50 exhibit
corotation radii slightly inside $r_{\circ}$ and greater winding of the
trailing arms, and models with $r_-/r_+$ $>$ 0.50 exhibit
corotation radii slightly outside $r_{\circ}$ and shorter trailing arms.
For self-gravitating models, the threshold between edge and P modes is
not as clearly defined. At transition, corotation moves
across $r_{\circ}$ but does not coincide with the 
extension of the arms. For example, $r_{co}$ = 1.01$r_{\circ}$ for the 
model at $r_-/r_+$ = 0.45, even though it has extended arms. 
The first forward phase shift falls near $r_{\circ}$, with a rapid switch to 
a trailing arm. As $r_-/r_+$ decreases from 0.60, $r_{\circ}$ moves further 
inside the geometric center of the disk, leaving more room for the 
trailing arm to extend. 
Note that there is a 
change in $y_1$, as seen in Figure \ref{eigenvalues_m12b}  ($y_1$ plotted for 
$q$ = 2, $m$ = 2, 
$M_*/M_d$ {\it vs.} $r_-/r_+$) in that models exhibiting P mode behavior 
have $y_1$ $<$ 0, while edge modes have $y_1$ $>$ 0. In edge modes, the number 
of wraps increases with decreasing $r_-/r_+$. The corresponding eigenfunctions 
in Figure \ref{Edgemode_plots} indicate that a second 
minimum appears when the trailing 
arm extends beyond the inner bar. For low $r_-/r_+$,
the winding of the arms increases.  
The minima occur with more 
frequency toward the outer edge of the disk. 
The P mode work integral plot has 
two peaks in $E_h$, one lying close to the inner and one closer
to the outer edge of the disk.
As $r_-/r_+$ decreases, a broad valley develops between them. $E_k$ has 
a peak which lies inside the inner $E_h$ peak, with a shoulder 
across the central region, going to zero at the outer edge 
of the disk. The edge mode work integral plot also has a 
narrow $E_h$ peak near the inner edge which contains the peak in $E_k$, 
but both have very low amplitude except near the inner 
edge. The work integrals become oscillatory for smaller $r_-/r_+$. 
The stress plots show 
acoustic flux dominating near the inner and outer edges while 
Reynolds stress dominates the inner disk region, carrying 
opposite sign. Stress due to self-gravity is negligible for
the P and edge modes.

\subsubsection{I$^-$ \& I$^+$ Modes} \label{sec_in_I}

The I modes, modes with properties intermediate between 
those of the J and P modes  
were discovered by Goodman \& Narayan (1986). 
There are two types of I modes, fast
I$^-$ modes with corotation inside $r_{\circ}$, sometimes inside
$r_-$, and slow I$^+$ 
modes with corotation outside $r_{\circ}$, sometimes outside 
$r_+$. Dominant azimuthal wavenumbers for I modes
lie in parameter space illustrated in Figures \ref{y1y2_small_plot} 
and \ref{y2_plots}, Section \ref{sec_in_regime}. The I mode region 
is seen in Figure \ref{eigenvalues_m12a} Section \ref{sec_in_regime} as 
a wide arc extending to the right and below 
the J mode corner previously defined, ending in stable models curving 
from ($r_-/r_+,M_*/M_d$) = (0.70,10.0) to ($r_-/r_+,M_*/M_d$) 
= (0.20,0.01). 
Equilibrium disks which show I modes consistently have $p \lesssim 6-7.5$
and $p \gtrsim 3$ for $q=2$ or $\gtrsim 2$ for $q=1.5$, both increasing with increasing $m$. 
Density contours are nearly concentric for small $M_*/M_d$ 
disks, becoming somewhat flatter for high $M_*/M_d$ disks. The threshold $r_-/r_+$ 
above which the mode is type I$^-$ roughly $.5$ for $q=1.5$, $m=2$ and
is an increasing function of all of $m$, $q$ and $M_*/M_d$ (see Figs 
\ref{q15y_isocontours}-\ref{q20y_isocontours}, dashed lines).

\begin{table*}[t]
\centering
Table 3: Representative I$^-$ and I$^+$ $m$ = 2 Modes
\vskip 0.1in
\begin{tabular}{*{11}{c}}
\hline\hline
 &
$M_*/M_d$ &
$q$ &
$r_-/r_+$ &
$r_+/r_{\circ}$ &
$r_{\circ}$ &
$\tau_{\circ}$ &
$J$ &
$y_1,y_2$ &
$r_{ilr}/r_{\circ},r_{olr}/r_{\circ}$ &
$r_{co}/r_{\circ}$ \\
\hline
I1 &0.1 &1.5 &0.600 &1.27 & 14.9 &511 &2.75 &0.905,0.504
&\nodata,1.02 &0.780 \\
I2 &5   &1.5 &0.600 &1.26 &16.2 &176 &9.41 &0.784,0.432
&\nodata,1.05 & 0.802 \\
I3 &7   &2  &0.600 &1.30 &7.68 &49.8 &7.47 &-0.747,0.169
& \nodata & 1.26 \\
I4 &0.2 &1.5  &0.402 &1.50 &6.60 &152 &1.84 &-0.993,0.601
&0.996,\nodata & $>$ $r_+$\\
I5 &7  &1.5  &0.500 &1.36 &11.6 &91.6 &9.32 &-0.737,0.274
&0.856,\nodata&1.36 \\
\hline
\end{tabular}
\label{Imode_table}
\end{table*}

\begin{figure*}
\begin{center}
\includegraphics[width=3.0in]{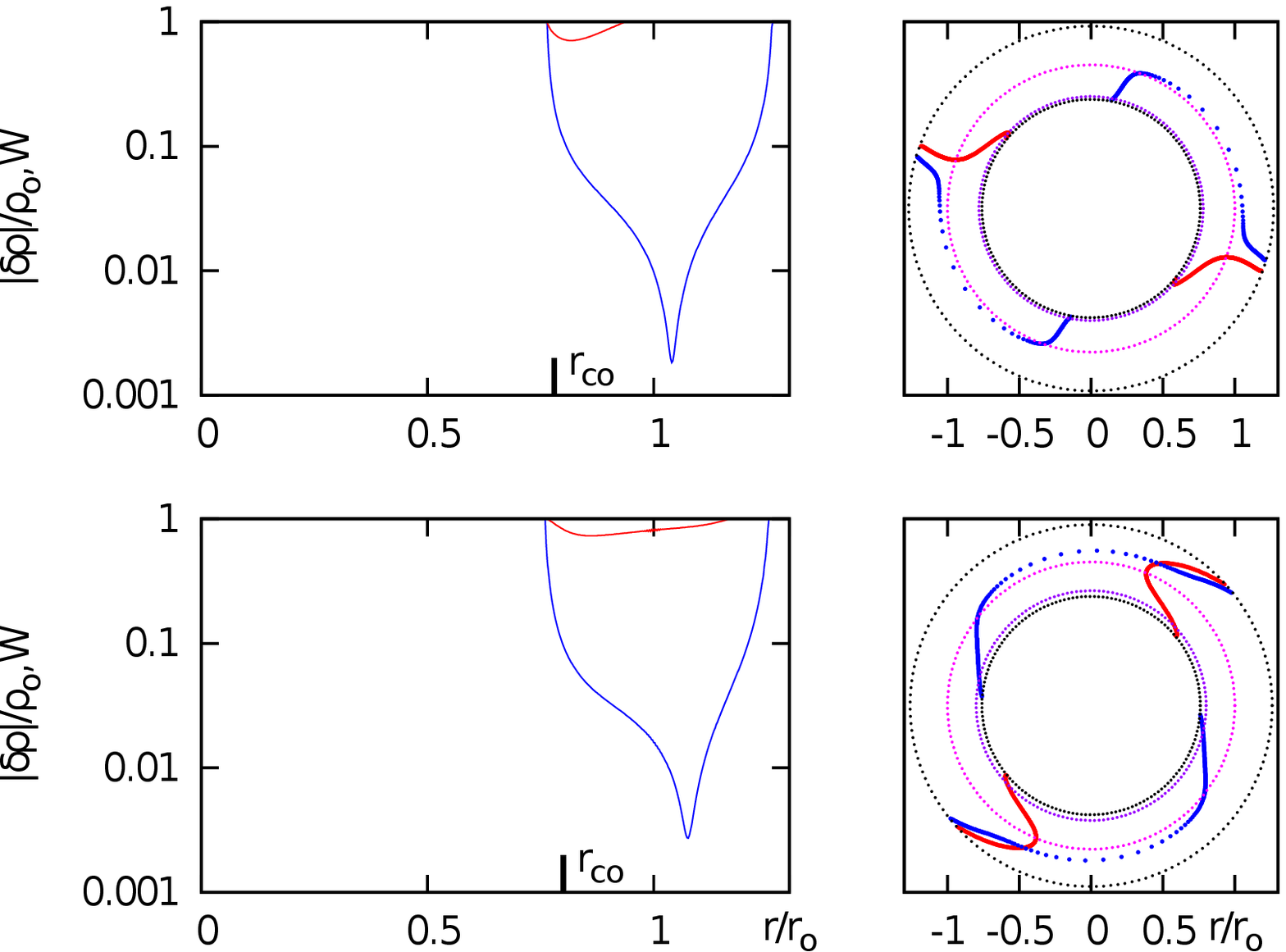}
\includegraphics[width=3.0in]{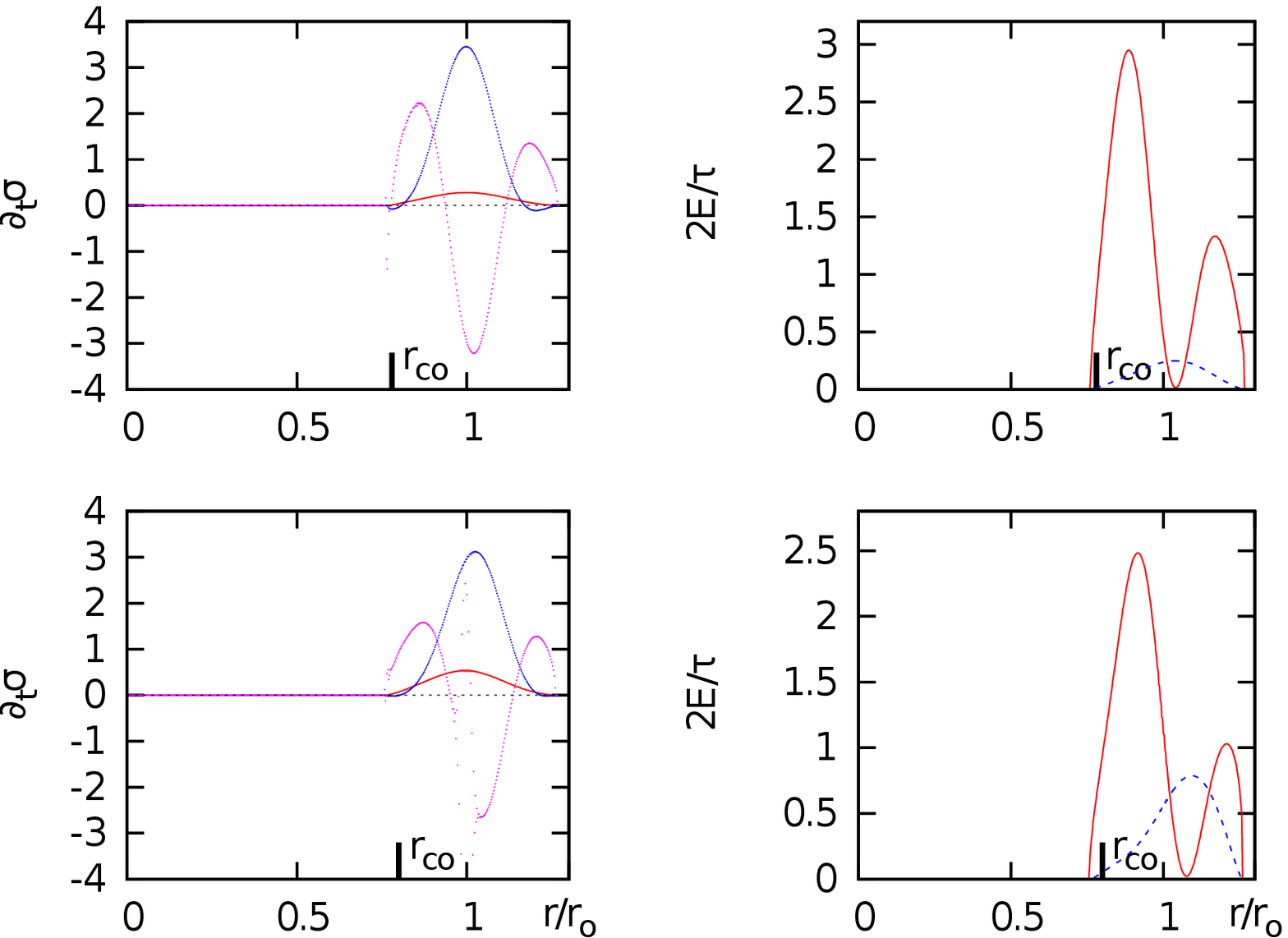}
\end{center}
\caption{
Barlike I$^-$ modes, $m$ = 2 modes,
with corotation inside r$_{\circ}$ and 
sometimes inside $r_-$ for disks with $q$ = 1.5 and $M_*/M_d$ = 
0.1 and 5 for $r_-/r_+$= 0.6, models 
I1 and I2 from top-to-bottom.
We show eigenfunctions, $\delta\rho$ and ${\cal W}$ 
amplitudes and phases, $\partial_t\sigma$, and $\delta$J. For the 
eigenfunctions, the blue curve is for $\delta\rho/\rho_{\circ}$ and the
red curve for ${\cal W}$. For the 
$\partial_t\sigma$, the Reynolds stress is
the red curve, the gravitational stress the blue curve, and the acoustic
stress the magenta curve. For the perturbed energies, the kinetic energy is
the blue curve and the enthalpy the red curve.
For the first column, the ratios of the unnormalized maximum values for $|\delta\rho|$/${|\cal W|}$ are
7685 and 5221,
respectively, from top-to-bottom.
}
\label{Iminusmode_plots}
\end{figure*}

Figure \ref{Iminusmode_plots} depicts and Table 3 presents 
relevant properties for typical 
I$^-$ modes. Corotation falls well inside $r_{\circ}$. $|{\cal W}|$ shows a 
minimum near $r_{co}$ while $|\delta\rho/\rho_{\circ}|$
shows a minimum near $r_{\circ}$.
The phase plots show a central bar and an outer 
bar connected by a trailing $\pi/m$ phase shift 
in $\delta\rho/\rho_{\circ}$
slightly outside $r_{\circ}$. ${\cal W}$ is out of phase 
with $\delta\rho/\rho_{\circ}$
at the inner edge, with a short leading arm that switches 
to trailing at $r_{co}$, coming into phase with $\delta\rho/\rho_{\circ}$
at the outer edge. Models where $M_*/M_d < 1$ exhibit 
bars near the inner and outer edges of the disk, while in 
higher $M_*/M_d$ models, the bars become less perpendicular to the 
disk edges with $\delta\rho/\rho_{\circ}$ and 
${\cal W}$ in phase and trailing at 
the outer edge. For the $M_*/M_d$ = 5 model pictured here,
$\delta\rho/\rho_{\circ}$
and ${\cal W}$ come into phase at $r/r_{\circ}$ = 1.09. These trends are also 
seen in $q$ = 1.75 and 2, in that the $M_*/M_d$ $<$ 1 models exhibit 
bars and the bars are less perpendicular in higher $M_*/M_d$ 
models. $\delta\rho/\rho_{\circ}$ 
amplitudes are similar to J modes, but the ${\cal W}$
amplitudes of I$^-$ modes typically have a dip near the inner 
edge of the disk. The work integrals
for these models are similar in character to those of 
the J modes. The $M_*/M_d$ = 5 model shows stronger 
dominance of $E_k$ in the disk, especially outside $r_{\circ}$. 
Stress plots are generally similar to those seen in the J modes. 

\begin{figure*}
\begin{center}
\includegraphics[width=3.0in]{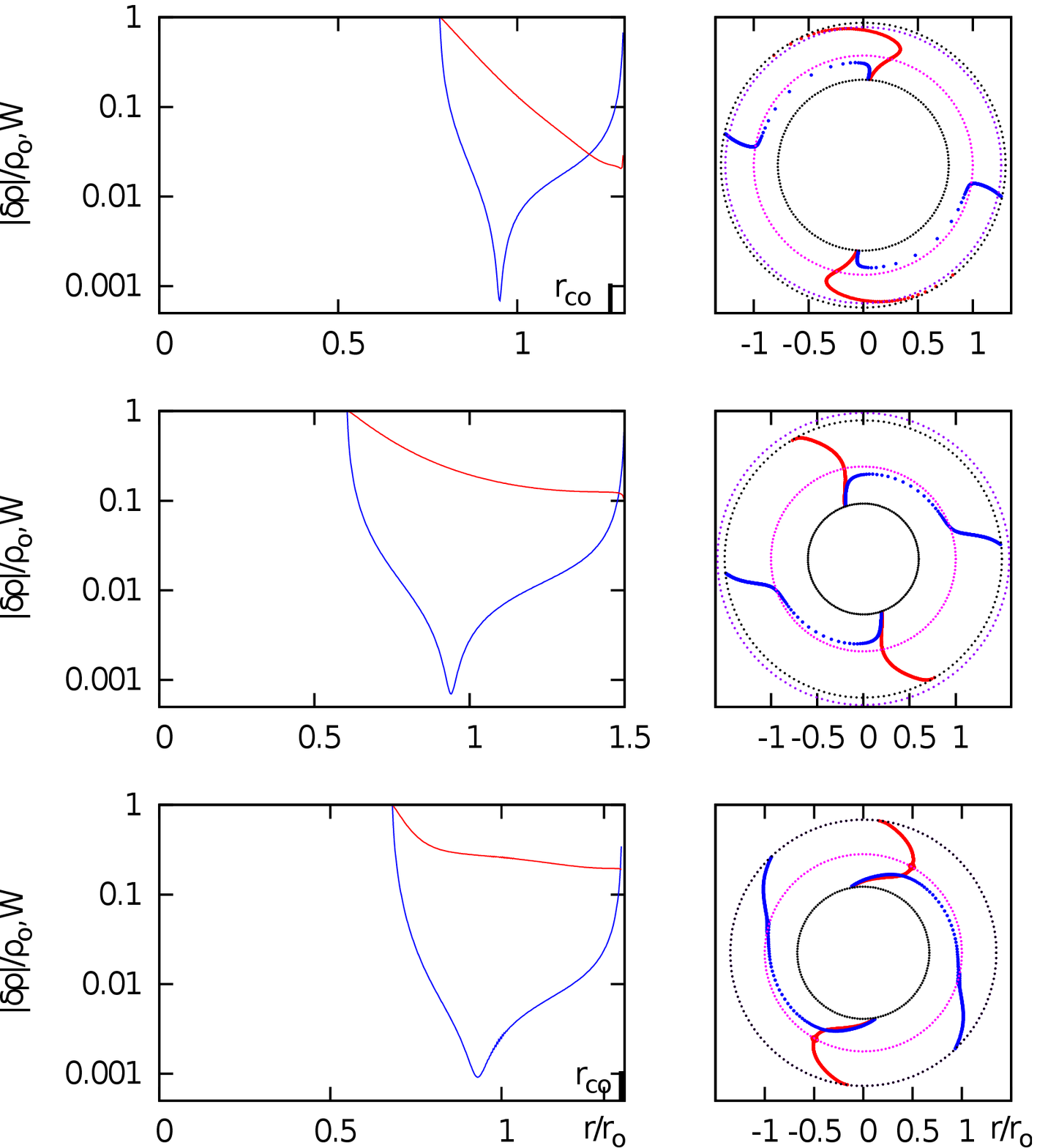}
\includegraphics[width=3.0in]{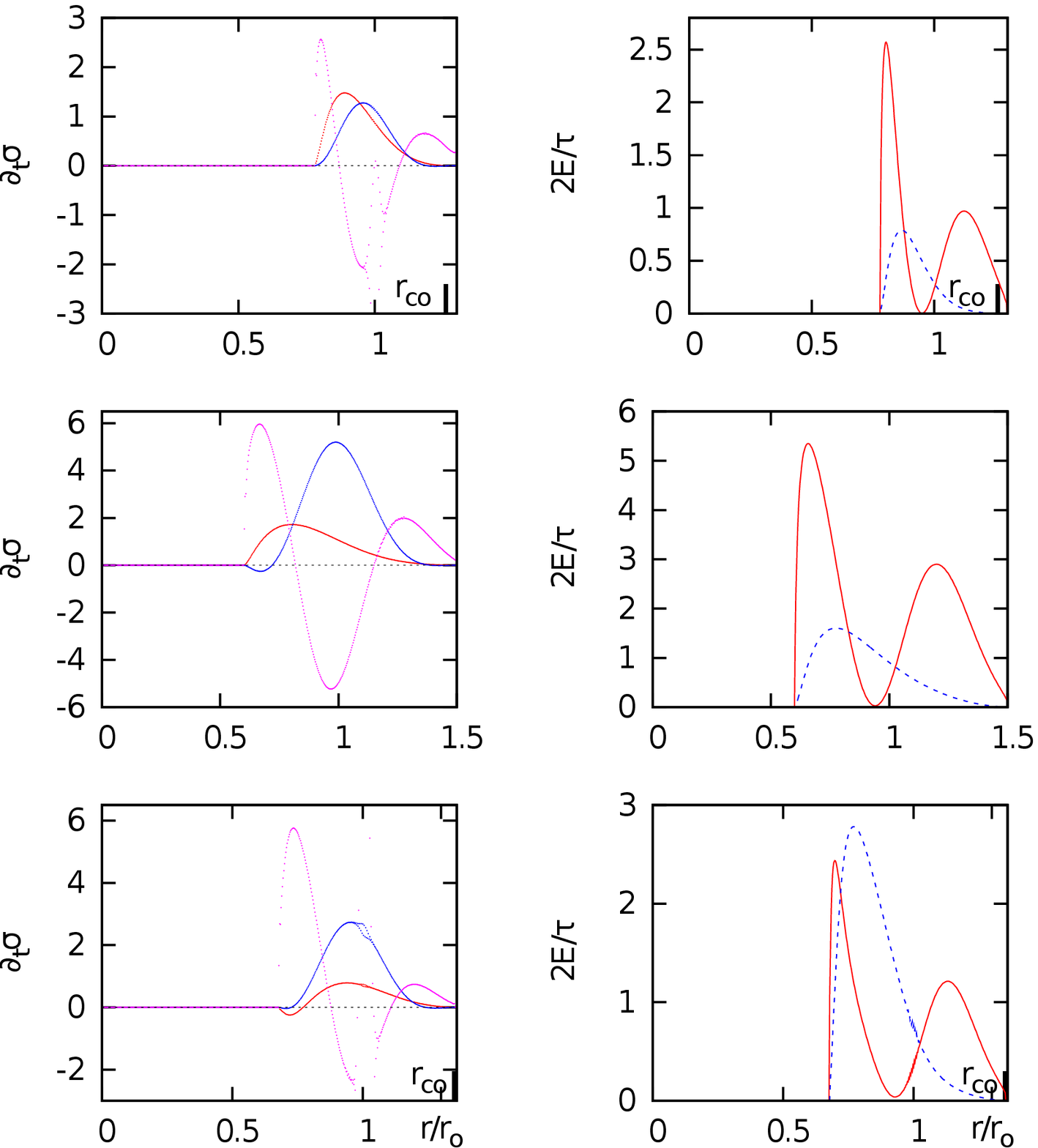}
\end{center}
\caption{
Barlike I$^+$ modes, $m$ = 2 modes,
with corotation outside r$_{\circ}$ and 
sometimes outside $r_+$ for disks with 
$q$ = 2, $M_*/M_d$ = 
7, and $r_-/r_+$= 0.6, model
I3, 
$q$ = 1.5, $M_*/M_d$ = 
0.2, and $r_-/r_+$= 0.402, model
I4, and 
$q$ = 1.5, $M_*/M_d$ = 
7, and $r_-/r_+$= 0.5, model
I5, from top-to-bottom.
We show $\delta\rho$ and ${\cal W}$ amplitudes and phases,
$\partial_t\sigma$, and $\delta$J. For the 
eigenfunctions, the blue curve is for $\delta\rho/\rho_{\circ}$ and the
red curve for ${\cal W}$. For the 
$\partial_t\sigma$, the Reynolds stress is
the red curve, the gravitational stress the blue curve, and the acoustic
stress the magenta curve. For the perturbed energies, the kinetic energy is
the blue curve and the enthalpy the red curve.
For the first column, the ratios of the unnormalized maximum values for $|\delta\rho|$/${|\cal W|}$ are
634, 2082 and 3266,
respectively, from top-to-bottom.
}
\label{Iplusmode_plots}
\end{figure*}

I$^+$ modes are illustrated in Figure \ref{Iplusmode_plots} and Table 
3. Equilibrium 
density contours are nearly concentric for small $M_*/M_d$ disks, 
becoming somewhat flatter for high $M_*/M_d$ disks. Corotation falls
outside $r_+$ or, for model I5, just at $r_+$. $|{\cal W}|$ does not show
extrema when $r_{co} > r_+$. ${\cal W}$ is in phase with 
$\delta\rho/\rho_{\circ}$ near the inner edge of the disk,
becoming out-of-phase for $\varpi > r_{\circ}$. There is 
roughly a $\pi/m$ trailing phase shift in $\delta\rho/\rho_{\circ}$ 
that lies close
to $r_{\circ}$. $q$ = 1.5 models have bars near the inner and outer disk 
edges, becoming less perpendicular for higher $M_*/M_d$.  
The work integral
of these models all show two peaks in $E_h$ with the inner peak 
higher, and a region in the middle of the disk that is 
dominated by $E_k$. Unlike the I$^-$ mode, where the peak in $E_k$ lies at 
the zero between the two $E_h$ peaks, the peak in $E_k$ for the I$^+$ mode 
lies within the region of the inner $E_h$ peak. Notably, the 
$q$ = 1.5, $r_-/r_{\circ}$ = 0.50, $M_*/M_d$ = 7 model is dominated more 
strongly by $E_k$ for much of the disk inside $r_{\circ}$. The stress plots 
show domination in the inner and outer regions by ${\partial_t\sigma}_{h}$
while ${\partial_t\sigma}_{G}$ 
dominates in the center of the disk. The Reynolds 
stress is positive with relatively low amplitude. The 
notable exception here is the $q$ 
= 2, $r_-/r_{\circ}$ = 0.60, $M_*/M_d$ = 7 model, 
which has a region inside $r_{\circ}$ that is dominated 
by the Reynolds stress.

There is a region of parameter space that lies between I$^+$ modes and 
edge modes where the characteristics of the models resemble neither. Unlike 
I modes, corotation lies near r$_{\circ}$ and unlike edge modes, the 
${\cal W}$ phase lies significantly away from that of 
$\delta\rho/\rho_{\circ}$, indicating that self-gravity is 
important. There is a large variance in appearance of the models, 
with no strong characteristic identifying modes in this region. 


\subsubsection{ $m$ = 1 Modes } \label{sec_in_m1}

Several kinds of $m$ = 1 modes have been found: 
(i) the eccentric instability discovered by Adams, Ruden, \& 
Shu (1989 and named A modes by Woodward, Tohline,
\& Hachisu 1994) where the central 
star moves off the center of mass of the equilibrium disk in 
response to the nonaxisymmetric gravitational forcing arising 
from the perturbed disk. This leads to the indirect 
potential that couples the star to the disk at the outer 
Lindblad resonance; 
(ii) an instability where the central star moves in response 
to forcing from the perturbed nonaxisymmetric gravitational potential of 
the disk, but the effects of the indirect potential are small and 
the mode is driven by super-reflection of waves at corotation
(Noh, Vishniac, \& Cochran 1992); 
(iii) P and edge-like modes which were discovered in NSG disks 
but can also develop in self-gravitating disks (Kojima 
1986, 1989, Noh, Vishniac, \& Cochran 1992, Woodward, Tohline, \& 
Hachisu 1994); 
(iv) modes in which the disk perturbation arranges 
itself so that its center-of-mass remains fixed at the origin and 
the star does not move (Hadley \& Imamura 2011); and 
(v) elliptical instabilities which arise from noncircular streamlines in 
the equilibrium disk (Ryu \& Goodman 1994). We modelled 
axisymmetric equilibrium disks and so did not consider 
elliptical instabilities.  We do find examples of the 
other one-armed modes in our simulations.

\begin{table*}[t]
\centering
Table 4: Representative $m$ = 1 Modes
\vskip 0.1in
\begin{tabular}{*{11}{c}}
\hline\hline
 &
$M_*/M_d$ &
$q$ &
$r_-/r_+$ &
$r_+/r_{\circ}$ &
$r_{\circ}$ &
$\tau_{\circ}$ &
$J$ &
$y_1,y_2$ &
$\frac{r_{ilr}}{r_{\circ}},\frac{r_{olr}}{r_{\circ}}$ &
$r_{co}/r_{\circ}$ \\
\hline
O1 &1 &2 &0.052 &9.52 &0.126 &0.287 & 0.357 & -0.418,0.0815
& \nodata &1.31 \\
O2&1 &2 &0.101 &4.89 &0.327 &1.18 &0.576 & -0.502,0.248
& \nodata & 1.42 \\
O3 &1 &2  &0.201 &2.60 &1.02 &6.39 &1.04 &-0.428,0.519
& \nodata & 1.32 \\
O4&1 &2 &0.301  &1.91 &2.21 &19.5 &1.57 & -0.363,0.571
& \nodata &1.25 \\
O5 &1 &2 &0.402 &1.59 &4.14 &48.3 &2.23 &-0.358,0.478
& \nodata &1.25 \\
O6 &1 &2 &0.600 &1.28 &12.5 &239 &4.13 &-0.694,0.465
& \nodata & $>$ $r_+$ \\
O7&0 &1.5 &0.100 &2.70 &1.40 &31.8 &0.447 &-0.859,0.0108
& \nodata,\nodata &$>r_+$ \\
O8&0.01&1.5 &0.100 &2.68 &1.42 &30.9 &0.470 &-0.935,1.22
& \nodata,\nodata & $>r_+$ \\
O9&0.1 &1.5 &0.100 &2.60 &1.54 &26.7 &0.647 &-0.607,1.031
& \nodata,\nodata & 1.86\\
O10&1 &1.5 &0.100 &2.35 &2.12 &18.5 &1.72 &-0.311,0.341
& \nodata,2.03&1.28 \\
O11&5 &1.5 &0.100 &2.12 &3.06 &14.9 &4.36 &-1.036,0.0451
& \nodata,\nodata &\nodata \\
O12&0.01 &1.8 &0.100 &3.03 &1.07 &23.3 &0.329 &-0.925,1.13
& 2.41,\nodata & $>r_+$ \\
O13&0.1 &1.8 &0.100 &3.20 &0.974 &15.1 &0.428 &-0.600,0.931
& 0.954,1.16 & 1.66 \\
O14&0.5 &1.8 &0.100  &3.85 &0.669 &4.77 &0.663 & -0.487,0.425
& 0.831,1.90 &1.45 \\
O15&1 &1.8 &0.100 &4.31 &0.486 &2.12 &0.803 &-0.534,0.199
& 0.876,2.01&1.53 \\
O16&100 &2 &0.700 &1.21 &3.37 &389 &18.4 &-0.0420,0.115
& \nodata &1.02 \\
\hline
\end{tabular}
\label{m1_table}
\end{table*}

\begin{figure*}
\begin{center}
\includegraphics[width=3.0in]{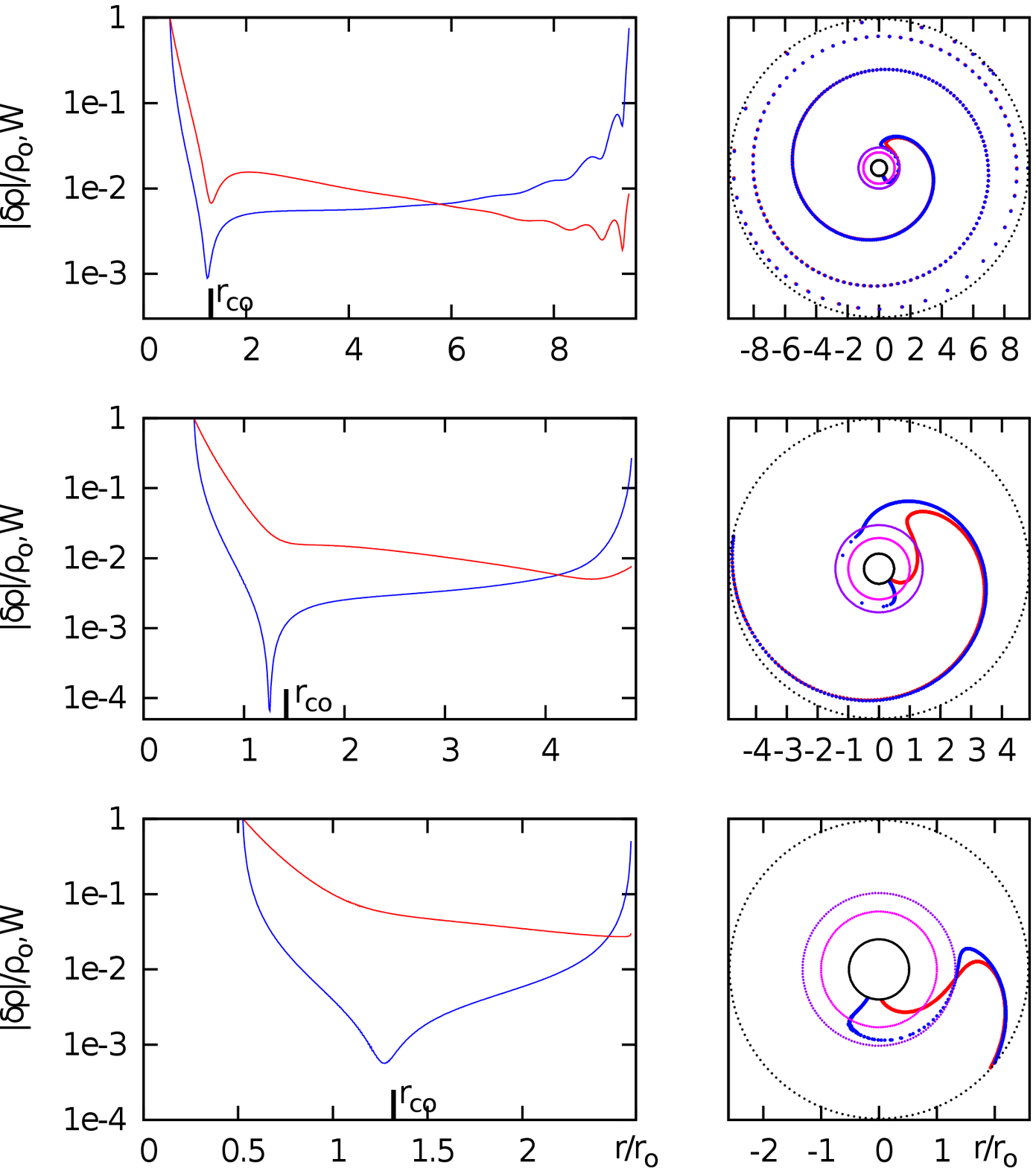}
\includegraphics[width=3.0in]{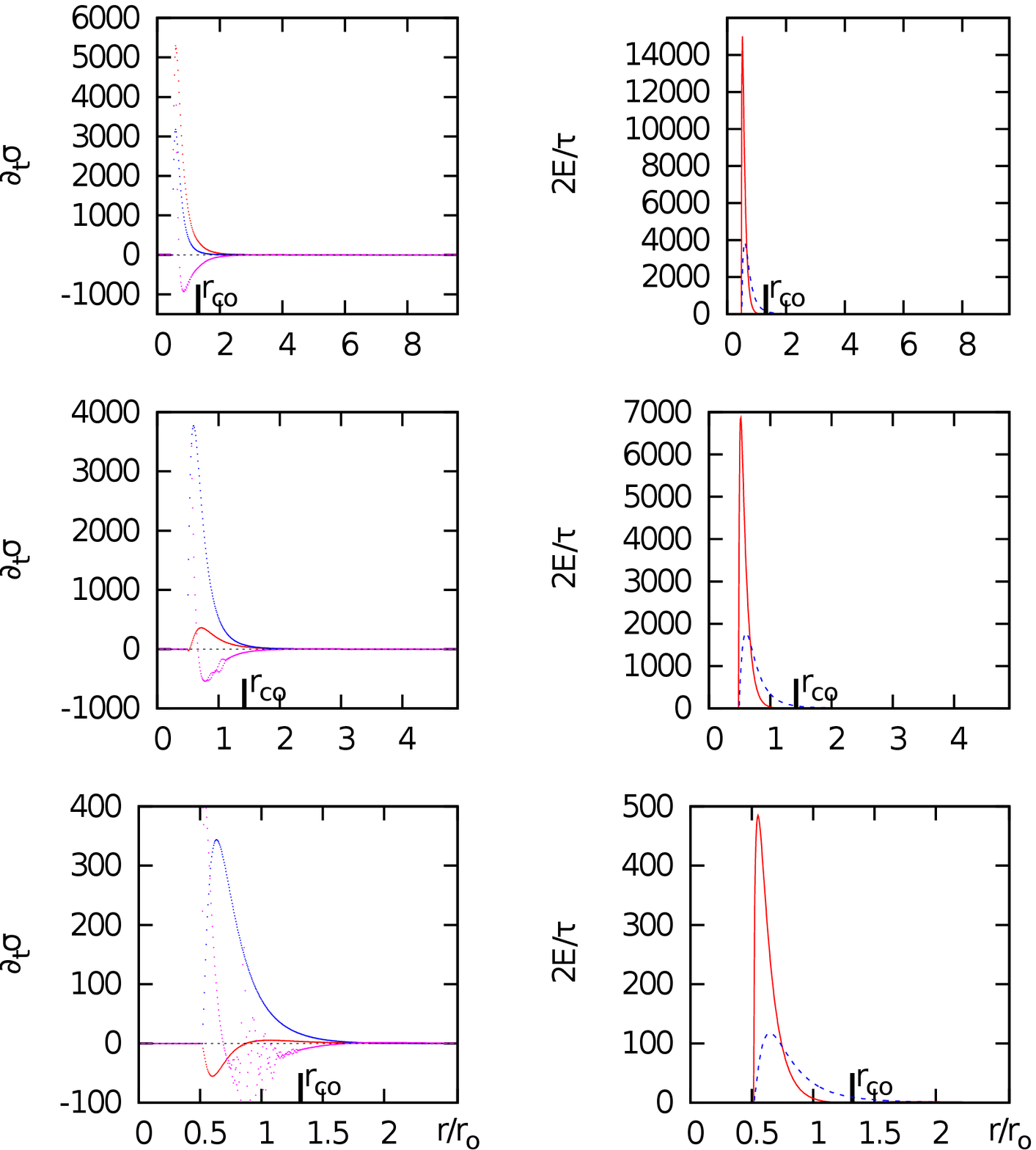}\\
\includegraphics[width=3.0in]{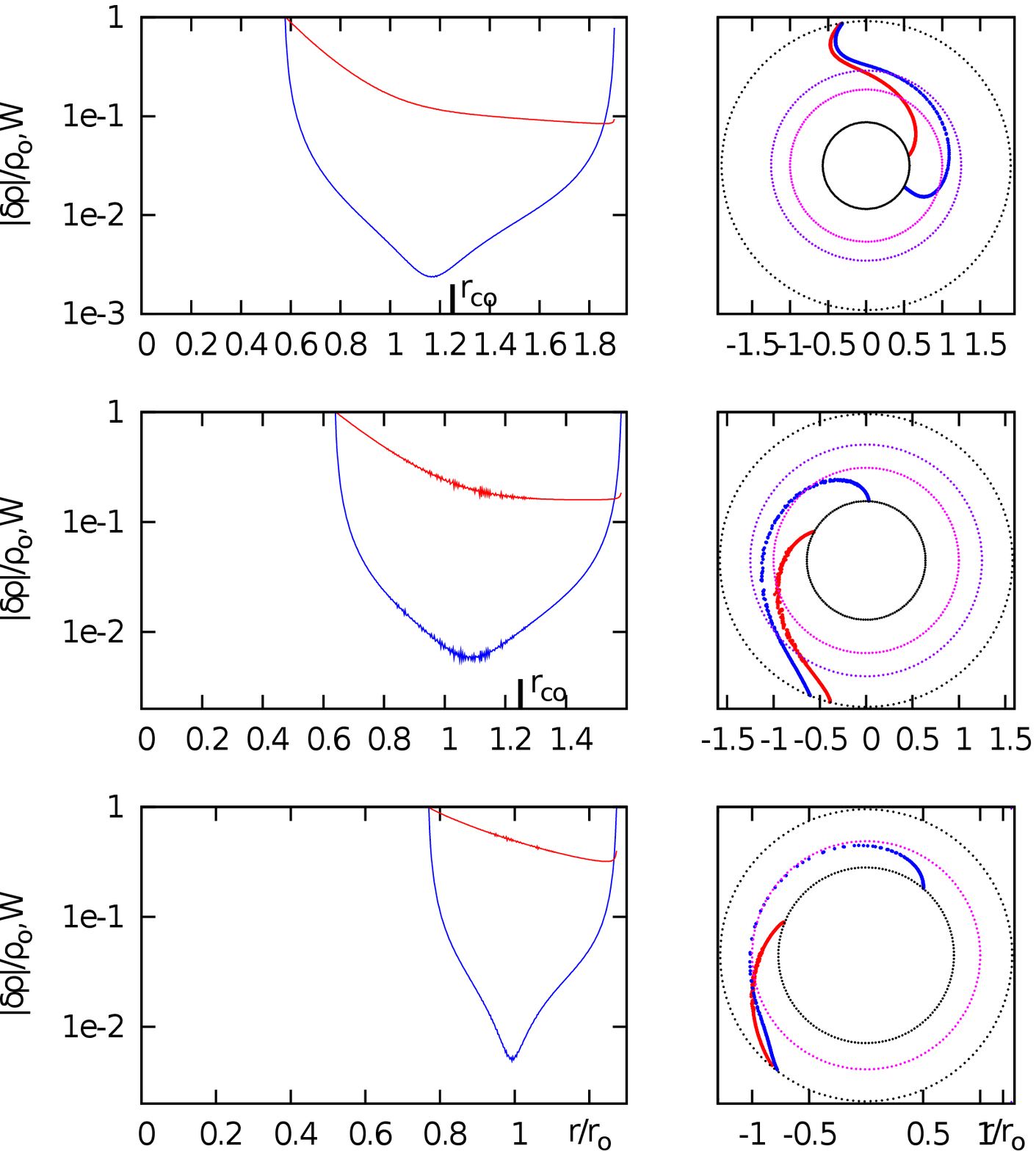}
\includegraphics[width=3.0in]{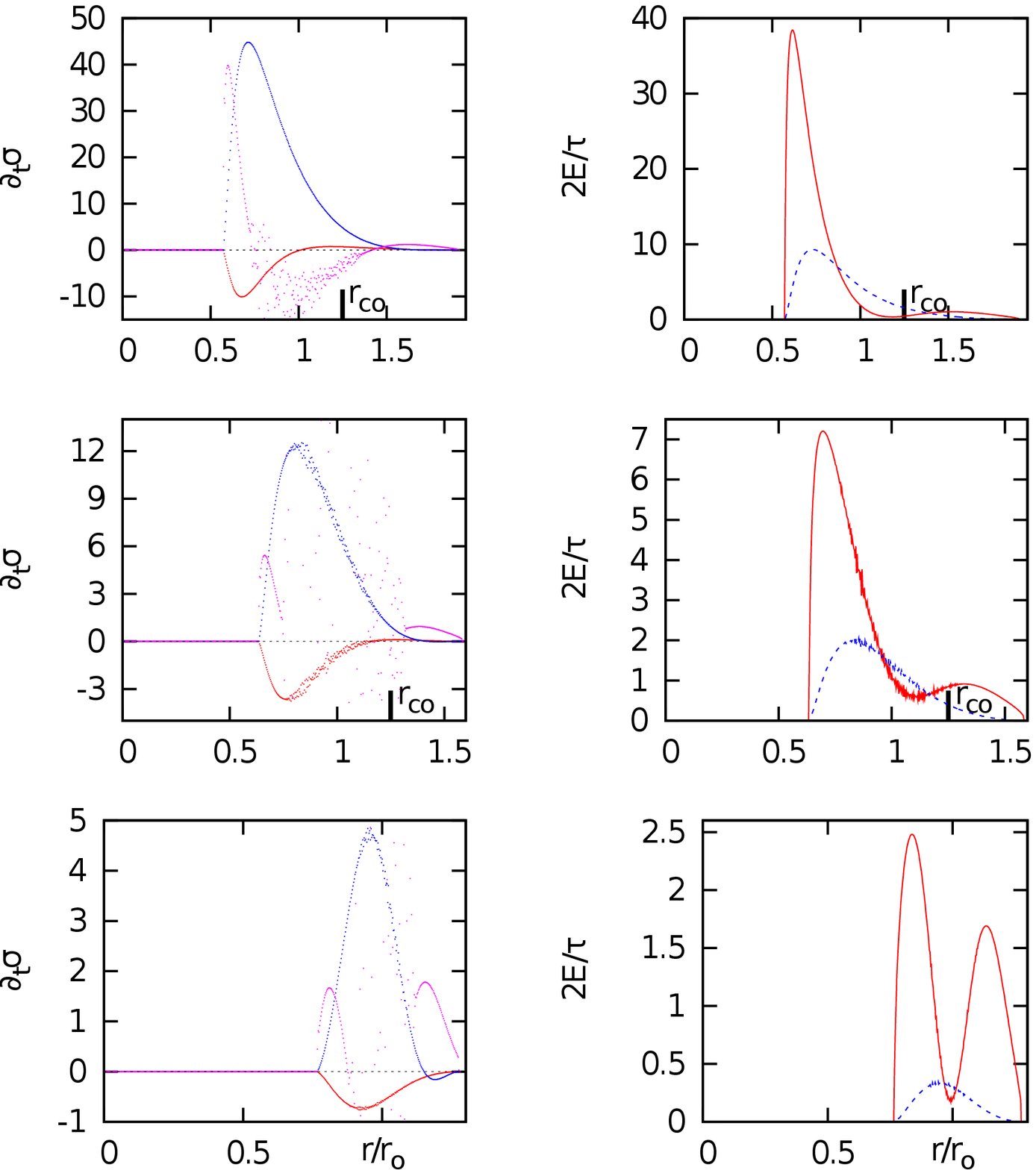}
\end{center}
\caption{
$m$  = 1 modes for models O1 to O6 from top-to-bottom.
These are representative $m$ = 1 modes for $q$ = 2 disks with 
$M_*/M_d$ = 1,
and increasing $r_-/r_+$. 
We show $\delta\rho$ and ${\cal W}$ amplitudes and phases,
$\partial_t\sigma$, and $\delta$J. For the 
eigenfunctions, the blue curve is for $\delta\rho/\rho_{\circ}$ and the
red curve for ${\cal W}$. For the 
$\partial_t\sigma$, the Reynolds stress is
the red curve, the gravitational stress the blue curve, and the acoustic
stress the magenta curve. For the perturbed energies, the kinetic energy is
the blue curve and the enthalpy the red curve.
For the first column, the ratios of the unnormalized maximum values for $|\delta\rho|$/${|\cal W|}$ are
1.54, 7.05, 46.68, 167.48, 431.38 and 3468,
respectively, from top-to-bottom.
}
\label{am1mode_q2_M1}
\end{figure*}

Properties of $m$ = 1 modes are illustrated by 
cuts through $(r_-/r_+,M_*/M_d)$ space.
We first present results for $q$ = 2 disks 
with $M_*/M_d$ = 1, and $r_-/r_+$ 
ranging from 0.05 to 0.60, see Table 4 and Figure \ref{am1mode_q2_M1}. 
$m$ = 1 modes are important in this study, since they dominate higher
order $m$ for small $r_-/r_+$ in many cases. 
For growth rate dependence on azimuthal mode number $m$,
 refer to Section \ref{sec_in_regime}. 
We find that $q$ = 2 disks do not show A modes. They do show unstable $m$ = 1
modes, however, when the indirect potential is suppressed, $m$ = 1 modes
persist. They are similar to those found in the 
full simulations but have 
faster growth rates. This suggests that the
$m$ = 1 modes for $q$ = 2 are not those studied by Adams, Ruden,
\& Shu (1989, see also 
Noh, Vishniac, \& Cochran 1992, Woodward, Tohline, \& Hachisu 1994).  
We see inner coherent bars which make rapid $\pi$-phase changes 
near $r_{co}$ outside of which loosely wound trailing spiral arms form.  
These modes are likely associated with edge-like 
modes. The $r_-/r_+$  = 0.60 sequence, disks with narrow nearly 
circular cross-sections, usually show loose spiral structure, 
with an arm that winds on the order of $\pi$ or less sometimes with 
a small region near the outer edge of the disk where there is a 
phase change and a small trailing arm forms. 
The $m$ = 1 modes of $q$ = 2 sequences approach structures 
found in NSG disks.

\begin{figure*}
\begin{center}
\includegraphics[width=3.0in]{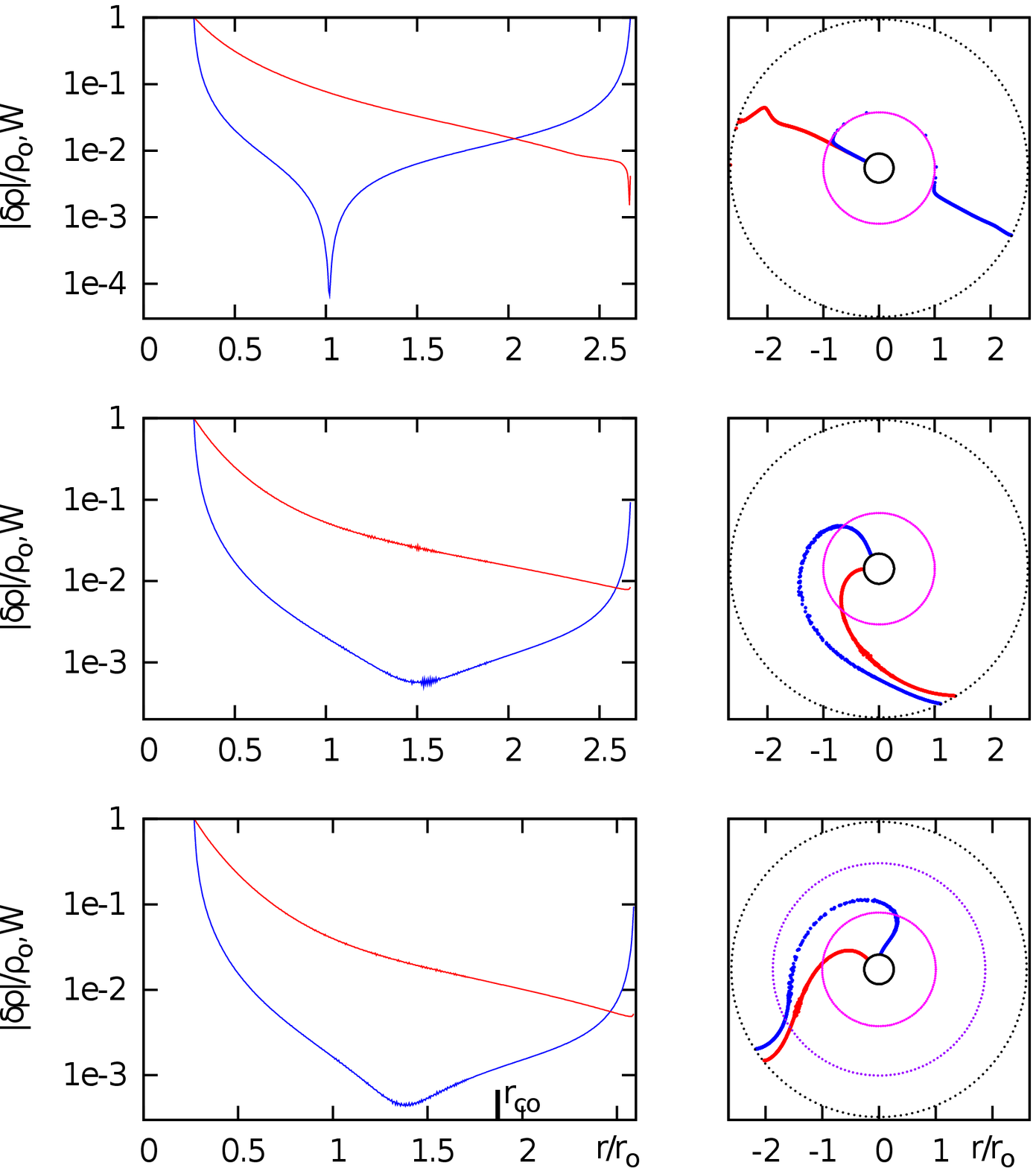}
\includegraphics[width=3.0in]{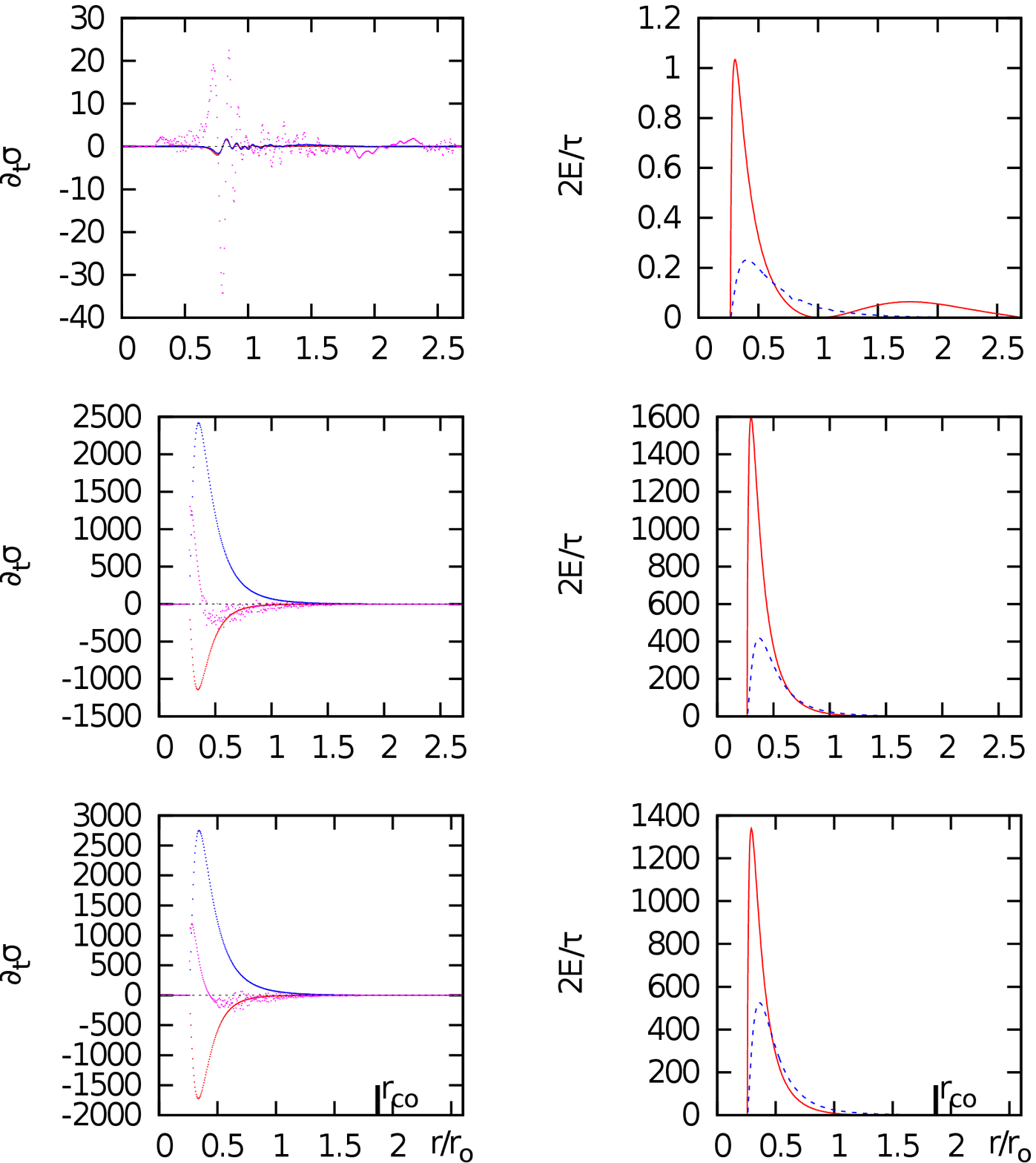}\\
\includegraphics[width=3.0in]{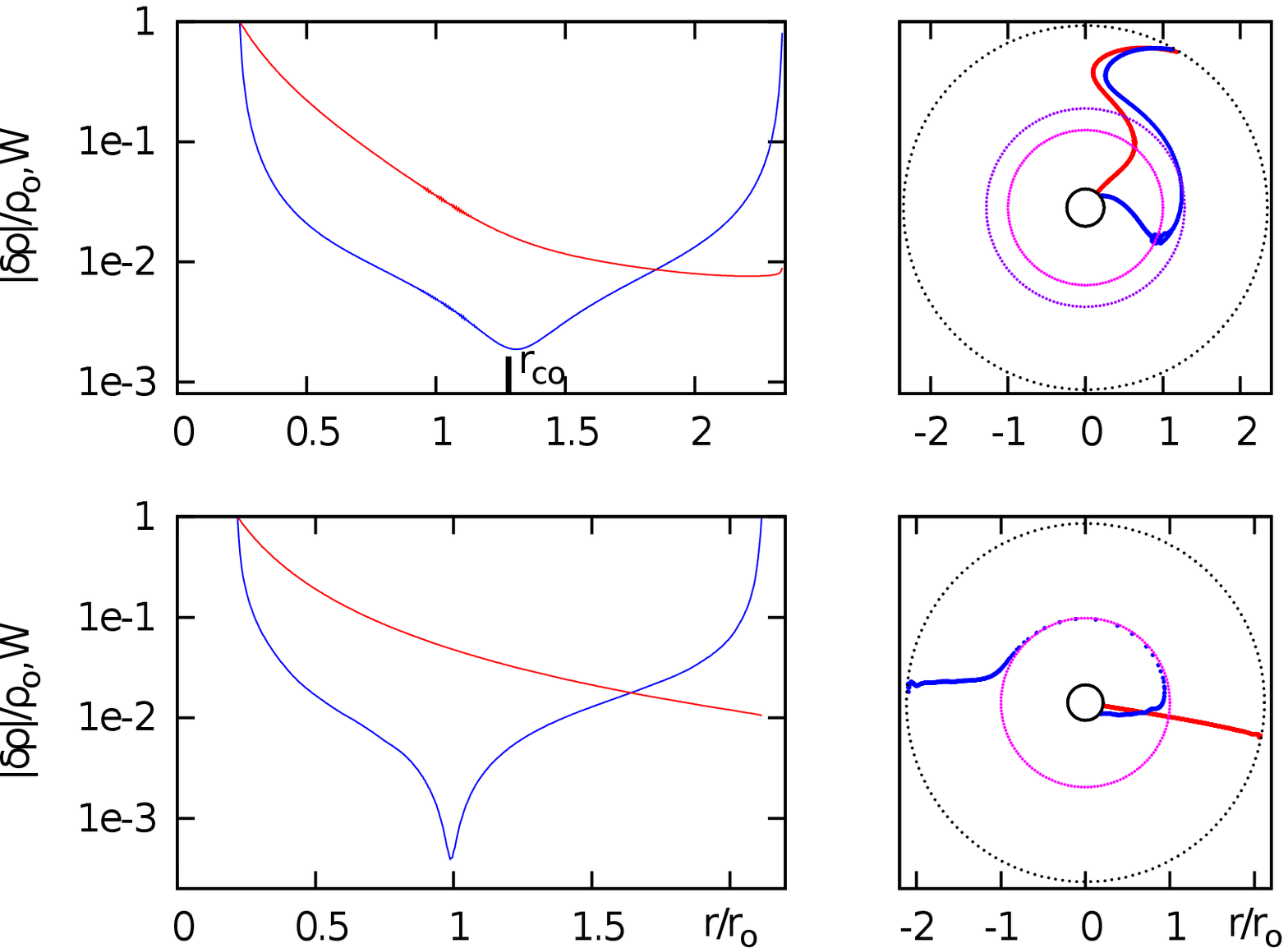}
\includegraphics[width=3.0in]{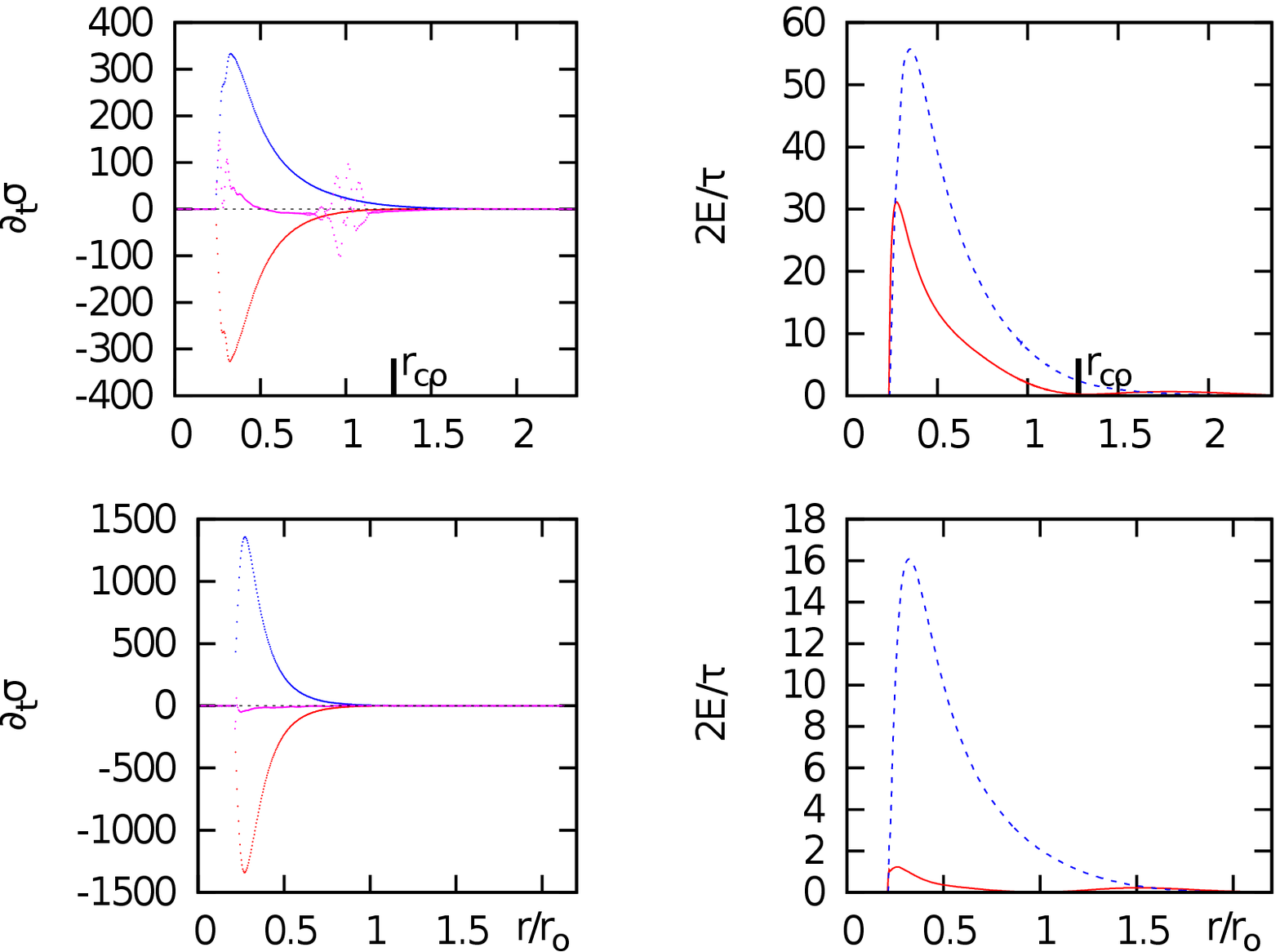}
\end{center}
\caption{
$m$ = 1 modes
in Kepler-like disks with $M_*/M_d$  = 0 to 5
and $r_-/r_+$ = 0.101,
models O7 to O11 from top-to-bottom. The bottom panel
shows a slow retrograde mode nearly 
stationary in the laboratory frame. It strongly 
resembles the $m$ = 1 mode found in toroids (top panel). 
We show $\delta\rho$ and ${\cal W}$ amplitudes and phases,
$\partial_t\sigma$,
and $\delta$J. For the 
eigenfunctions, the blue curve is for $\delta\rho/\rho_{\circ}$ and the
red curve for ${\cal W}$. For the 
$\partial_t\sigma$, the Reynolds stress is
the red curve, the gravitational stress the blue curve, and the acoustic
stress the magenta curve. For the perturbed energies, the kinetic energy is
the blue curve and the enthalpy the red curve.
For the first column, the ratios of the unnormalized maximum values for $|\delta\rho|$/${|\cal W|}$ are
214.83, 172.63, 121.05, 28.12 and 6.61,
respectively, from top-to-bottom.
}
\label{M1plots_q15_rin10}
\end{figure*}

Next consider $q$ = 1.5 systems. Plots in Figure \ref{M1plots_q15_rin10} show 
$m$ = 1 eigenfunctions for $r_-/r_+$ = 0.101
and star-to-disk mass ratios, 0.0 $<$ $M_*/M_d$ $\le$ 5.0.
The low star mass limit, the toroid limit, 
shows a split 
bar structure (Hadley \& Imamura 2011).
With no central star, the disk must conserve linear momentum 
itself. The addition of even a small star changes the nature of 
instability as shown by the $M_*/M_d$ = 0.01 model. The 
star's motion results in the formation of a barlike structure 
outside of which a leading spiral arm appears, 
similar in appearance to a P mode, but with $y_1$ = -0.935.
The mode is slow and corotation falls outside the disk. 
These properties are consistent with an $I^+$ mode although
the appearance of the phase plot does not resemble an 
I mode. For larger $M_*/M_d$, the oscillation frequency 
is higher and corotation moves into the disk. The disks also show outer 
Lindblad resonances. The phase plots also change in this region 
switching to structures composed of
a trailing central bar-like region that abruptly turns to a leading 
spiral arm outside $r_{\circ}$ later switching to an 
outer trailing arm. The trailing arm 
is initially short but grows in size as $M_*/M_d$ is made larger, 
but it never winds more than $\approx$ $\pi$/4 in phase, even for the largest
$M_*/M_d$ disks. These modes bear strong resemblance to the P modes seen 
in $q$ = 2 NSG disks. For $M_*/M_d$ = 5, the mode
again changes character and forms a segmented bar which undergoes 
a $\pi$ phase shift at $r_{\circ}$. 
The disturbance is slowly rotating in the retrograde sense
with $y_1$= -1.04 and $y_2$ = 0.0451. A similar 
structure is found for the $M_*/M_d$ = 25 disk where
($y_1$, $y_2$) = (-0.976, 0.121), but here the mode is prograde. 
The $m$ = 1 mode approaches a neutral point as $M_*/M_d$
increases, consistent with the stability of $q$ $<$ $\sqrt{3}$ NSG
star/disk systems. 
These results depend on $q$. For 
$r_-/r_+$ = 0.05 and $M_*/M_d$ = 10, $m$ = 1 modes
are stable for disks with $q$ = 1.6 and 1.7.

The existence of low-frequency, retrograde $m$  = 1 
modes was first discovered by Kato (1983). 
Kato showed that thin, nearly Keplerian disks, although not 
unstable to nonaxisymmetric modes, supported neutral 
low frequency retrograde $m$ = 1 modes. Later works showed that
prograde low-frequency $m$ = 1 modes also existed, although, similarly
to Kato (1983), excitation mechanisms for the modes were not 
identified (Okazaki 1991, 
Ogilvie 2008, Papaloizou, Savonije, \& Henrichs 1992). 
The $m$ = 1 modes in Keplerian disks 
have oscillation 
frequency given by $\omega_1$ $\approx$ $(c_s/\Omega r)^2\Omega$
(Kato 1983). Using this result, we estimate the frequencies
of our $m$ = 1 modes.  Evaluating 
$\omega_1$ at density
maximum, we find 
$\omega_1$ $\approx$ $0.072\Omega_{\circ}$ where 
$\rho_{\circ}$ = 
$3.67\times10^{-3}$, 
$r_{\circ}$ = 6.56,
and $\Omega_{\circ}$ = 
0.136. For 
the retrograde and prograde $m$ = 1 modes given above,
our simulations yield oscillation frequencies 
-0.024 $\Omega_{\circ}$ and 0.039 $\Omega_{\circ}$,
results smaller than, but similar to the estimated $\omega_1$.

\begin{figure*}
\begin{center}
\includegraphics[width=3.0in]{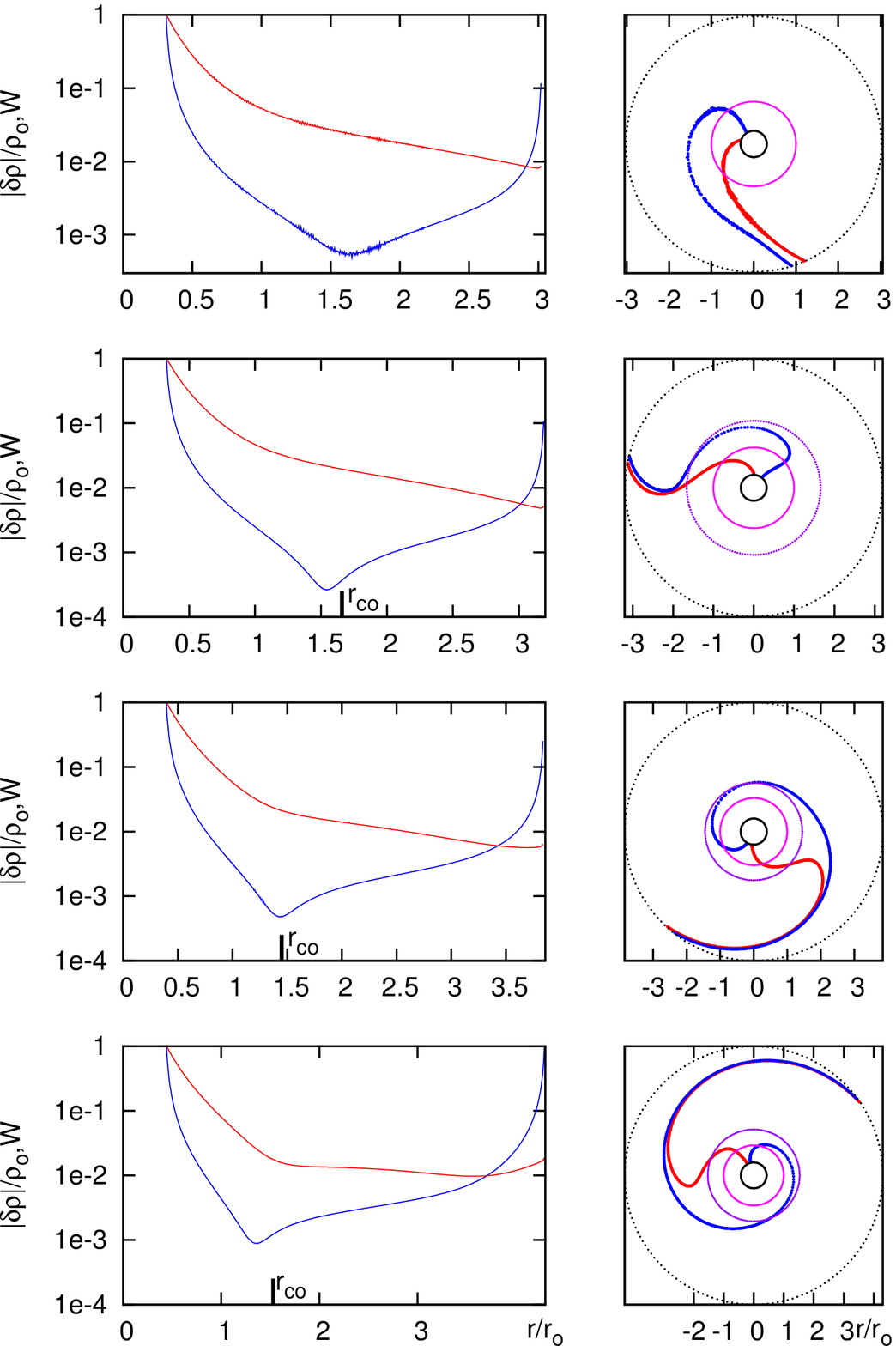}
\includegraphics[width=3.0in]{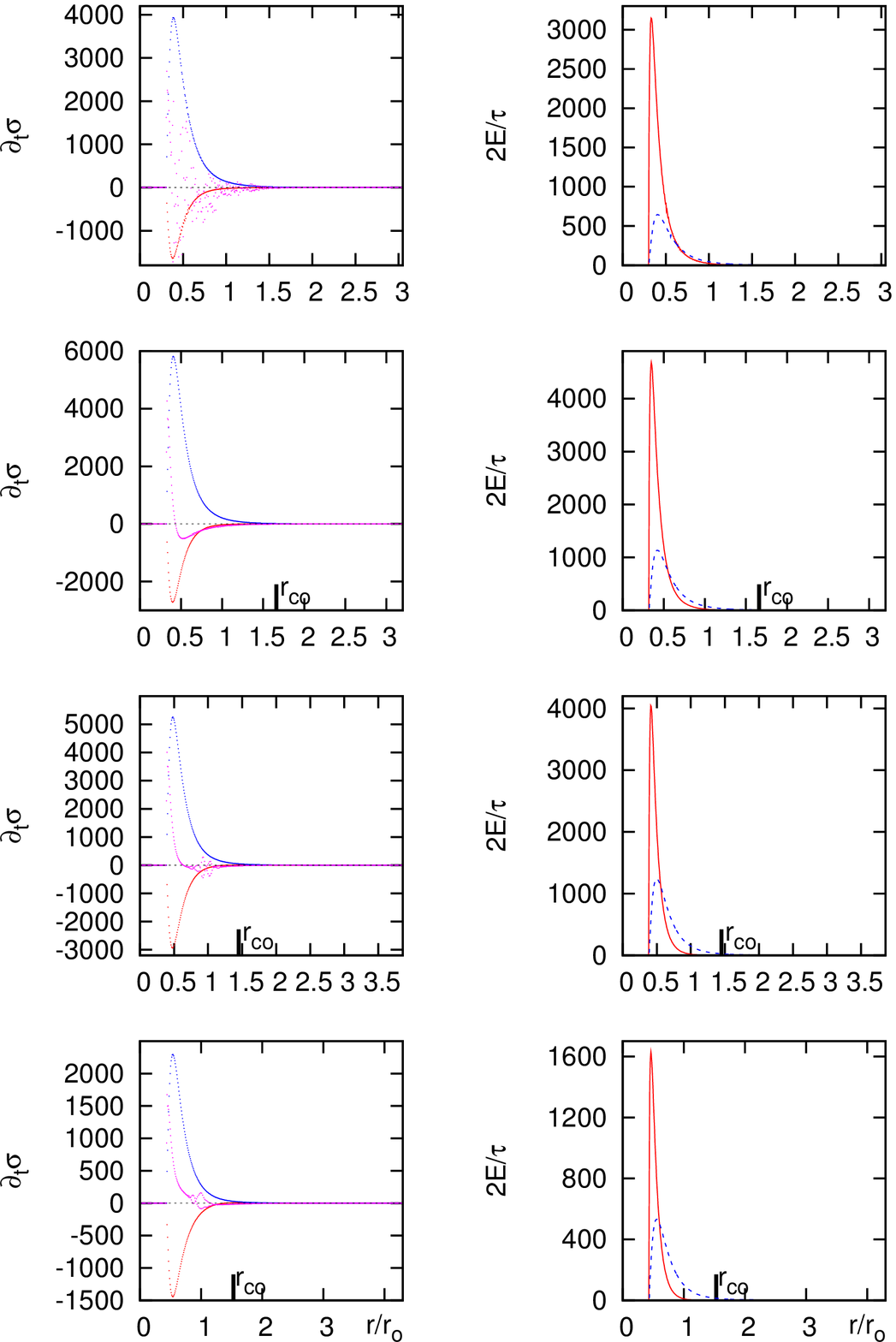}
\end{center}
\caption{
$m$ = 1 modes for disk models O12 to O15. The systems have $q$ = 1.8
and $r_-/r_+$ = 0.1 for
$M_*/M_d$ = 0.01, 0.1, 0.5, and 1,
from bottom-to-top.
The bottom two panels
show $m$ =1 A modes. 
We show $\delta\rho$ and ${\cal W}$ amplitudes and phases,
$\partial_t\sigma$, and $\delta$J. For the 
eigenfunctions, the blue curve is for $\delta\rho/\rho_{\circ}$ and the
red curve for ${\cal W}$. For the 
$\partial_t\sigma$, the Reynolds stress is
the red curve, the gravitational stress the blue curve, and the acoustic
stress the magenta curve. For the perturbed energies, the kinetic energy is
the blue curve and the enthalpy the red curve.
For the first column, the ratios of the unnormalized maximum values for $|\delta\rho|$/${|\cal W|}$ are
115.97, 71.61, 23.95 and 11.51,
respectively, from top-to-bottom.
}
\label{M1plots_q18_rin10}
\end{figure*}

We find A modes for $q$ = 1.8 disks.
A sequence of $q$ = 1.8 disks for $r_-/r_+$ = 0.102
is shown in Figure \ref{M1plots_q18_rin10}.
The $r_-/r_+$ place the models
near the peak of the P mode hump and in the edge mode region. 
We see A mode 
behavior for the $M_*/M_d$ = 0.5 and 1 models when $r_-/r_+$ = 0.102,
a smoothly winding trailing one-armed spiral. The development of the 
arm depends on the indirect potential. When the indirect potential 
is artificially suppressed, the $m$ = 1 mode is stable. 
Outside of the phase plot 
appearance, there is no strong marker for A modes. 

\subsection{ Instability Regimes } \label{sec_in_regime}

Instability regimes for $m$ = 1 to 4 modes
are laid out in ($r_-/r_+,M_*/M_d$) space.
The results naturally break down into three 
mass realms, $M_* \ll M_d$, $M_* \approx M_d$, 
and $M_* \gg  M_d$. For our discussion, we 
present $y_1$ and $y_2$ for 
star/disk systems with $M_*/M_d$ = 0 to 100 and
$q$ = 1.5, 1.75, and 2 in Figures \ref{y1y2_small_plot} to \ref{y1_plots}
and for $q$ = 2 disks in the $M_*$ $\gg$ $M_d$ regime in 
Figure \ref{kojima_plots}.

We did not fully examine $m$ $>$ 4 modes in our current study.
Of $m=1$ to $4$, the fastest growing J mode is always $m=4$ where
the J mode dominates; Presumably higher $m$ J modes will be even more
unstable until they should be cut off as the azimuthal wavelength approaches the
azimuthal Jeans length. Outside of the Jeans dominated region,
low modes are generally observed to have higher growth rates over
larger spans of parameter space.

\subsubsection{ $M_*/M_d$ $\ll$ 1 } \label{sec_in_loM}

\begin{figure}
\begin{center}
\centerline{\includegraphics[width=3.25in]{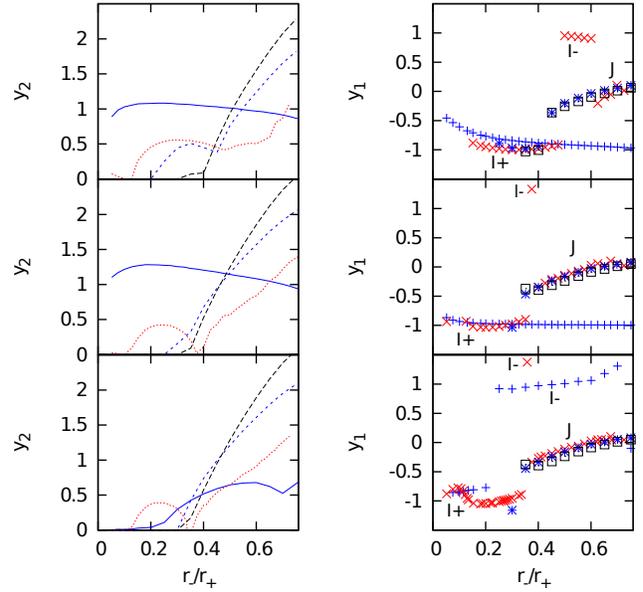}}
\end{center}
\caption{
Oscillation frequencies, $y_1$ and growth rates, $y_2$, 
for $m$  = 1, 2, 3, and 4 modes
for $q$ = 1.5, high disk mass systems, $M_*/M_d$ = 0.0, 0.01, and 0.1
from bottom-to-top.
The $m$ = 1 modes are shown by blue solid lines and crosses, $m$ = 2 modes by
red dotted lines and Xs, $m$ = 3 modes by blue dashed lines and asterisks, and 
$m$ = 4 modes by black long-dashed lines and boxes.
Mode types are demarcated where clear identification could be made; $m=1$ modes generally follow their own scheme (see \ref{sec_in_m1}).
}
\label{y1y2_small_plot}
\end{figure}

In the high disk mass regime, the $M_*/M_d$ $\ll$ 1 regime, there is
only weak dependence on $q$. To illustrate the disk mode properties,
we show results for $q$ = 1.5 disks. Curves are given in Figure \ref{y1y2_small_plot};
Such modes are displayed by modes O7-O9 in Table 4 and form
the left edge in the plots of Figure \ref{eigenvalues_m12a}.
Toroids illustrate nicely the properties of nonaxisymmetric
modes in the high disk mass regime, $M_*/M_d \ll 1$.\footnote
{Toroid results were discussed previously in 
Hadley \& Imamura (2011). Here, we review 
results pertinent to the current discussion 
for the convenience of the reader.}
Their growth rates, $y_2$, and oscillation
frequencies, $y_1$, indicate the presence of 
I modes, $m$ = 1 I$^-$ and I$^+$ modes,
$m$ $\ge$ 2 I$^+$ modes, and $m$ $\ge$ 2 J modes. 
Consider the $m$ $\ge$ 2 modes first. For
$m$ $\ge$ 2, toroids are stable for $r_-/r_+ \lesssim$ 0.05. 
For $r_-/r_+ \gtrsim$ 0.05, 
weak growth in $m$ = 2 is seen which peaks 
in strength around $r_-/r_+$ $\sim$ 0.23
where $y_2$ $\sim$ 0.5, and then falls to zero near $r_-/r_+$
$\sim$ 0.4. The mode has slow oscillation frequency, $y_1$ $\sim$ -1
so that corotation
sits outside the location of density maximum, $r_{\circ}$. 
Hadley \& Imamura (2011) classified these $m$ = 2 modes as I$^+$ modes. 
Around $r_-/r_+$ $=$ 0.40, a second set of $m$ $\ge$ 2 modes appears,
modes with faster growth rates and faster oscillation frequencies.
Near threshold, these $m$ = 2 modes have
$y_1$ $\sim$ -0.4 which then slowly increases reaching $y_1$ $\sim$ 0
near $r_-/r_+$ $\sim$ 0.7. This  
second set of modes has $r_{co}$ nearly at $r_{\circ}$.
Hadley \& Imamura (2011) classified these fast modes as J modes. 
Toroid $m$ = 1 modes exhibit similar behavior but without showing J mode
characteristics. For low $r_-/r_+$, $m$ = 1 modes are weakly unstable
with $y_2$ $<$ 0.05 near $r_-/r_+$ $\sim$ 0.05. These $m$ = 1 
modes are nearly stationary in the laboratory frame, $y_1$ $\sim$ -1. 
Hadley \& Imamura (2011) classified these slow $m$ = 1 modes 
as I$^+$ modes. Around
$r_-/r_+$ $\sim$ 0.3, a faster growing $m$ = 1 mode with faster
oscillation frequency, $y_1$ $\sim$ 1, appears. 
Corotation for these fast $m$ = 1 modes
sits inside the inner radius of the disk leading to their
classification as I$^-$ modes.

The extent to which toroids serve as the limiting case for 
high disk mass systems is shown by the $M_*/M_d$ = 0.01 system. 
The only difference seen for the
$m$ $\ge$ 2 modes is that the transition from the I$^+$ to J
modes moves to slightly larger $r_-/r_+$ when $M_*/M_d$ = 0.01. 
Toroids are the limiting case for $m$ $\ge$ 2 modes.
For $m$ = 1 modes, however, qualitative differences arise.
For the $M_*/M_d$ = 0.01 system, $m$ = 1 modes show  
faster growth rates than those shown by toroids and are 
the dominant mode for $r_-/r_+$ up to 0.50-0.60. They
are nearly stationary in the laboratory frame with $y_1$ 
similar to, but smaller than those for the toroid 
$m$ = 1 modes. The $y_1$ do not, however, show a jump 
as $r_-/r_+$ increases. The $y_1$ values remain close to 
-1 for all $r_-/r_+$. The presence of the star
qualitatively alters the properties of $m$ = 
1 modes even for this low $M_*/M_d$ system. Similar results and
conclusions are reached when the toroid results are compared to 
less massive disk systems, {\it e.g.,} the $M_*/M_d$ = 0.1 system.

\subsubsection{ $M_*/M_d$ $\approx$ 1 } \label{sec_in_midM}

For systems where $M_d$ and $M_*$ are comparable, the effects on
the $m$ = 1 mode are larger because the gravitational potential of 
the star and disk have comparable strengths. These are exhibited by
modes O1-O6 of Table 4 and Figure \ref{am1mode_q2_M1} as
well as the middle region of figures \ref{eigenvalues_m12a} and \ref{eigenvalues_m12b}.
 Although $m$ = 1 modes dominate low $r_-/r_+$ disks,
as they did for the $M_*/M_d$ $<$ 0.1 disk systems, they
show different character. 
The $M_*/M_d$ $\approx$ 1 systems show  two types of
$m$ = 1 modes: (i) for $q$ = 1.5, the oscillation frequency
$y_1$ shows a {\it humped} structure between
$r_-/r_+$ = 0.05 and 0.3-0.4 where it first rises from 
-0.5 to -0.25 and then returns
to -0.5. The growth rate $y_2$ rises from 0.25 to 0.4 
and then falls to 0.3 over this range. Similar behavior
is seen for the $q$ = 1.75 and 2 systems except 
that the upper end of the hump moves to larger
$r_-/r_+$ as $q$ increases. (ii) For $r_-/r_+$
$>$ 0.3-0.4, the character 
shown by the $m$ = 1 mode $y_1$ and $y_2$
is similar to that shown by the $m$ = 1 modes of the 
$M_*/M_d$ = 0.01 and 0.1 systems;
$y_2$ starts small and then slowly increases to
$\sim$ 0.4, and 
$y_1$ smoothly falls from -0.5 to -1 as 
$r_-/r_+$ increases. Both types of $m$ = 1 
modes show $r_{co}$ outside $r_{\circ}$ and 
are classified as $I^+$ modes.

The $m$ $\ge$ 2 J modes also show
strong changes as $M_*/M_d$ increases but here it is because
self-gravity weakens. Effects of
weaker self-gravity are apparent 
by the $M_*/M_d$ = 1 systems. We find that: 
(i) the $r_-/r_+$ where $y_1$ makes the transition from $\sim$ -1 
to 0, the change that marks the transition from I$^+$ modes 
to J modes, moves to larger $r_-/r_+$ as $M_*/M_d$
increases for the $m$ = 2 modes. By the 
$M_*/M_d$ = 1 sequence, the transition has disappeared
altogether.
J modes are never dominant for systems where the star and disk
masses are comparable.
(ii) The slow $m$ $\ge$ 2
I$^+$ modes become dominant 
starting around $r_-/r_+$ $\sim$ 0.3-0.4 giving way to $m$ = 3 
and $m$ = 4 I$^+$ modes as $r_-/r_+$ increases. 
Even though there is always only
one dominant mode, all $m$ $\ge$ 2 modes 
grow quickly and several modes may play 
roles in nonlinear simulations.
(iii) At large $r_-/r_+$, $r_-/r_+ \gtrsim$ 0.6, a fast 
$m$ = 2 I$^-$ mode appears which dominates the disks.
(iv) For $q$ = 2 disks, a fast $m$ = 2 mode also
appears at small $r_-/r_+$. The mode is only weakly unstable 
and not expected to play a significant role
in nonlinear simulations. These fast $m$ = 2 modes do not 
arise in $q$ = 1.5 and 1.75 disk systems.
Overall, we find dependence on $q$ but that the dependence
is weak 
in systems where $M_*$ $\approx$ $M_d$.

\begin{figure*}
\begin{center}
\centerline{\includegraphics[width=4.5in]{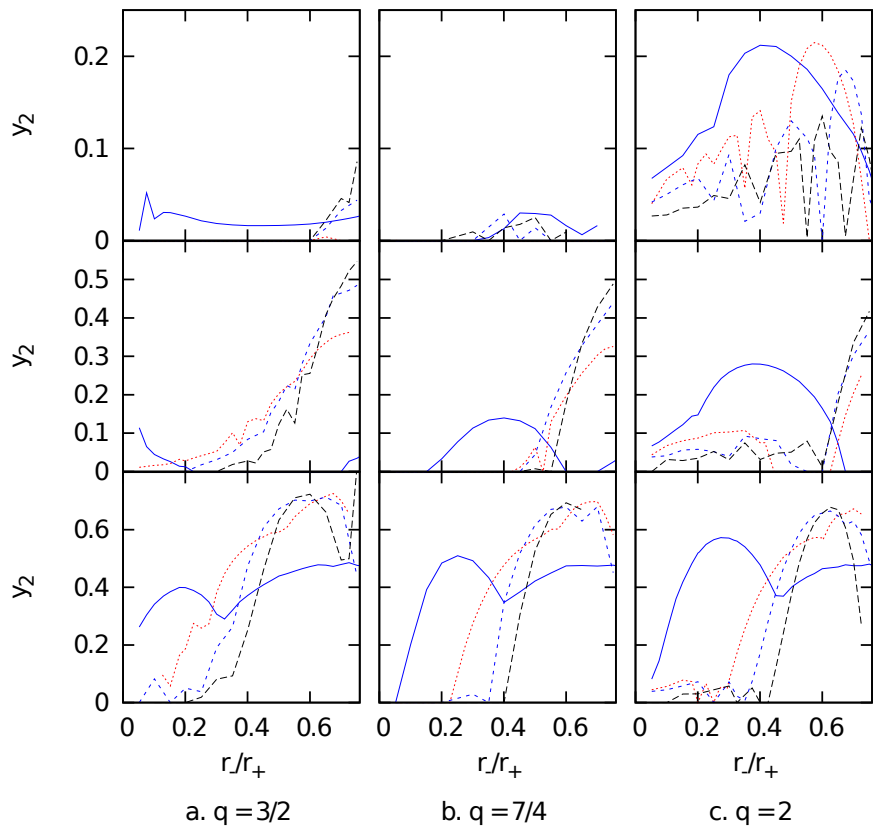}}
\end{center}
\caption{
Growth rates, $y_2$, for the low $m$ modes, $m$ = 1, 2, 3, and 4 
for systems with $M_*/M_d$ = 1, 10, and 100
and $q$ = 1.5, 1.75, and 2 disks.
The columns are for $q$ = 1.5, 1.75, and 2, respectively. The rows are 
for $M_*/M_d$ = 1, 10, and 100 from bottom-to-top.
The $m$ = 1 modes are shown by blue solid lines, $m$ = 2 modes by
red dotted lines, $m$ = 3 modes by blue dashed lines, and 
$m$ = 4 modes by black long-dashed lines.
}
\label{y2_plots}
\end{figure*}

\begin{figure*}
\begin{center}
\centerline{\includegraphics[width=4.5in]{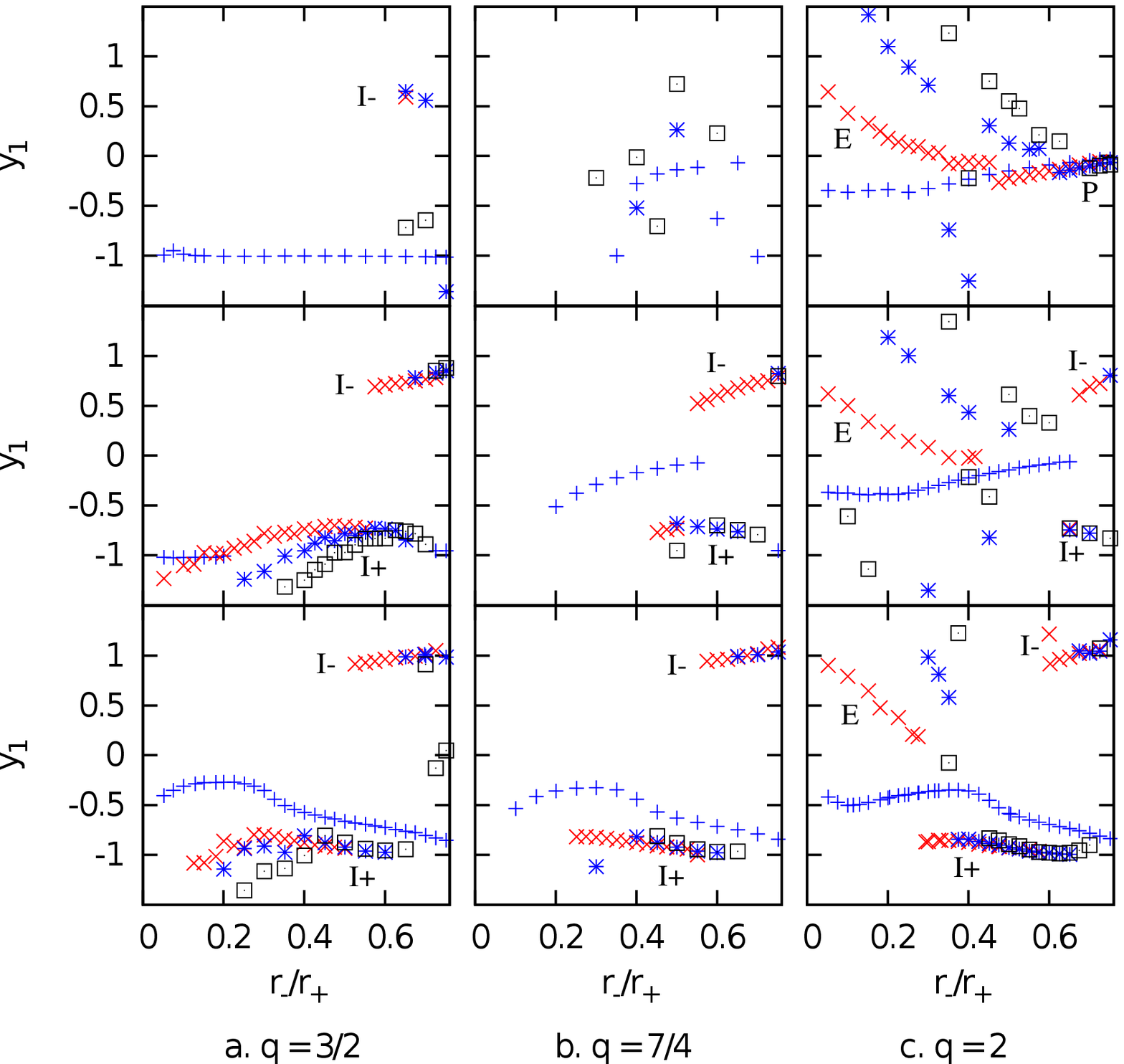}}
\end{center}
\caption{
Oscillation frequencies, $y_1$, for the low $m$ modes, $m$ = 1, 2, 3, and 4 
for systems with $M_*/M_d$ = 1, 10, and 100
and $q$ = 1.5, 1.75, and 2 disks.
The columns are for $q$ = 1.5, 1.75, and 2, respectively. The rows are 
for $M_*/M_d$ = 1, 10, and 100 from bottom-to-top.
The $m$ = 1 modes are shown by blue crosses, $m$ = 2 modes by
red Xs, $m$ = 3 modes by blue asterisks, and 
$m$ = 4 modes by black boxes.
Mode types are demarcated where clear identification could be made; $m=1$ modes generally follow their own scheme (see \ref{sec_in_m1}).
}
\label{y1_plots}
\end{figure*}

\subsubsection{ $M_*/M_d$ $\gg$ 1 } \label{sec_in_hiM}

The character of instability changes and its dependence on $q$ 
becomes stronger in the $M_*/M_d \gg 1$ regime, the NSG disk 
regime. This is explored in Figures \ref{y1_plots} and \ref{y2_plots},
with detailed plots for elements O11 and O16 of Table 4.
These form the righthand side of Figures 
\ref{eigenvalues_m12a} and \ref{eigenvalues_m12b}.

\begin{figure}
\begin{center}
\includegraphics[width=3.0in]{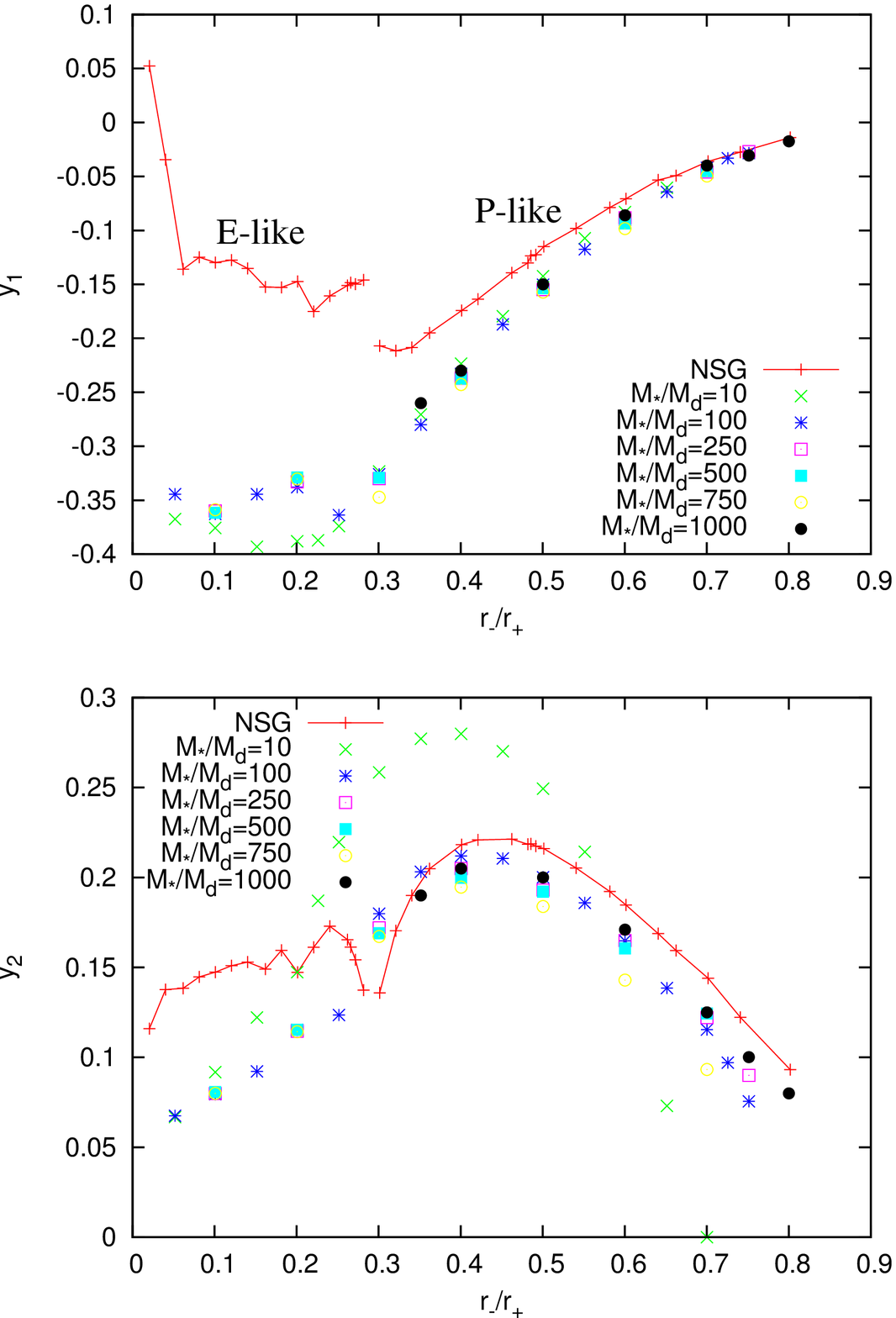}
\includegraphics[width=3.0in]{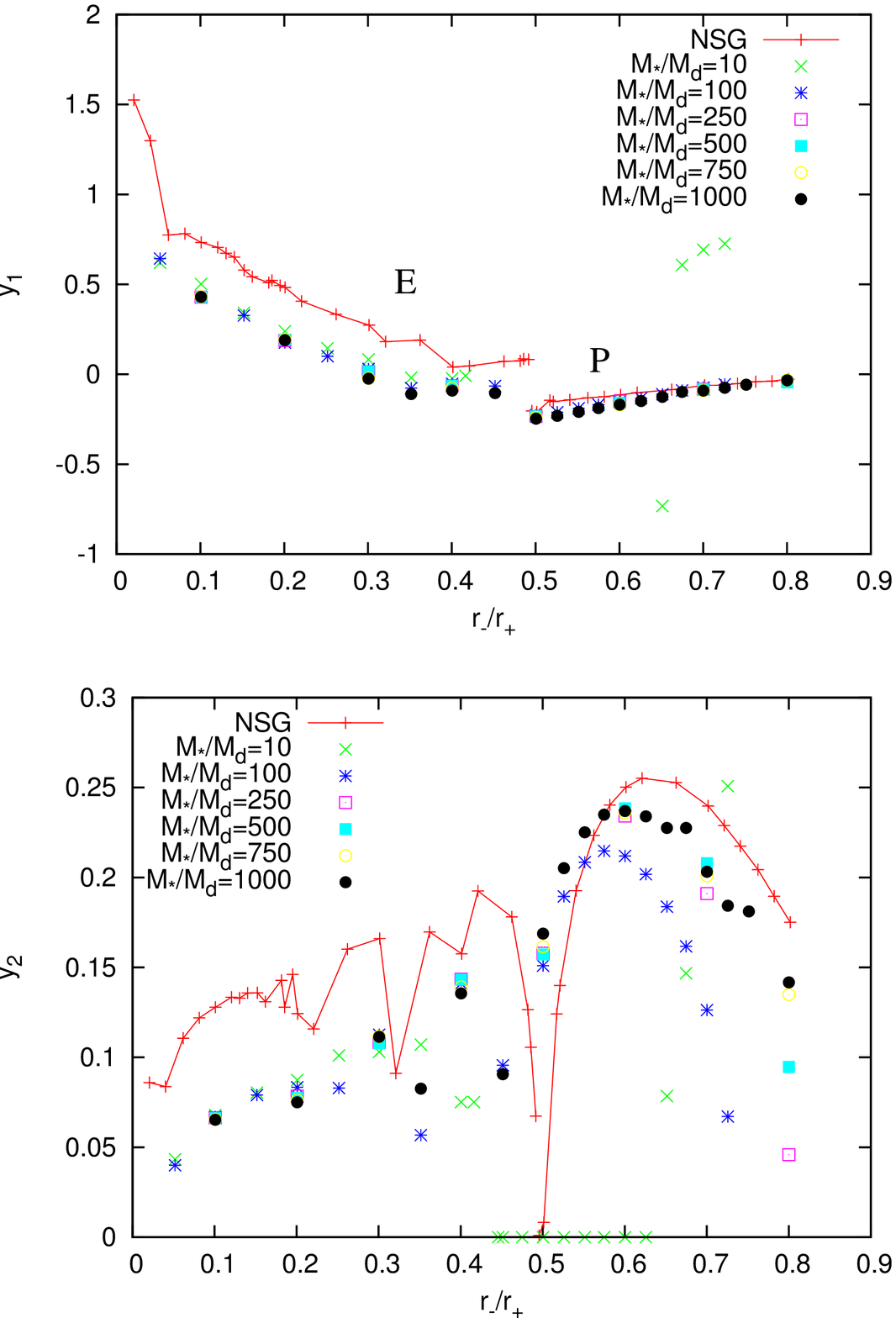}
\end{center}
\caption{
Comparison of the eigenvalues $y_1$ and $y_2$ for the 
(a) $m$ = 1 mode (left column), and (b) $m$  = 2 mode 
(right column) for $q$ = 2 
NSG disks and $M_*/M_d$ = 10, 100, 250, 500, 750, 
and 10$^3$ disk systems. 
}
\label{kojima_plots}
\end{figure}

 To explore these effects, we begin by 
showing $y_1$ and $y_2$ values for the $m$ = 1 and 2 
modes of $q$ = 2 disks for disks with $M_*/M_d$ = 10 to 
10$^3$ and for NSG disks in Figure \ref{kojima_plots}. 
We first describe the NSG disk results. In the NSG limit, 
$m$ = 1 and 2 modes show two distinct behaviors. For small
$r_-/r_+$, the modes are weakly unstable, with  
$y_2 < 0.05$, but have fast oscillation frequencies, 
$y_1 \approx$ 1 to 2 for both $m$ = 1 and 2. 
At higher $r_-/r_+$, a second type of mode
appears, one with smaller 
oscillation frequency, $y_1 \sim 0$,  
and faster growth rate, $y_2 >$ 0.14-0.15. 
The transition falls at 
$r_-/r_+$ $\approx$ 0.28 and 0.5 for the $m$ = 1 and 2 modes, 
respectively. 
The growth rate $y_2$ grows, peaks at higher $r_-/r_+$, and then
smoothly falls off approaching zero at large $r_-/r_+$. The peaks
fall at $r_-/r_+$ = 0.55 and 0.6 for $m$ = 1 and 2 modes, respectively. 
This behavior is shown by the other low-m modes differing only 
in the $r_-/r_+$ where the transition lies, $r_-/r_+$ = 
0.62 and 0.7 for $m$ = 3 and 4, respectively. 
Below the transitions, the instabilities
are classified as edge modes. Above the transitions, they 
are classified as P modes (see $\S\ref{sec_BE}$).

The $y_1$ and $y_2$ values for the $M_*/M_d$ $>$ 100 disks
converge as $M_*/M_d$ increases. Their dependence on $r_-/r_+$ 
qualitatively follows that shown by NSG disk models
for $r_-/r_+$ $>$ 0.3; NSG disks are not self-consistent below 
$r_-/r_+$ $\sim$ 0.3. For $m$ = 1, the first quantitative 
differences for systems with decreasing $M_*/M_d$ appear 
in the $M_*/M_d$ = 100 sequence where the $y_2$ 
fall below those of larger $M_*/M_d$ sequences at 
$r_-/r_+$ $>$ 0.7. Differences are noticeable
for all $r_-/r_+$ in the $M_*/M_d$ = 10 sequence 
where the peak growth rate,
$y_2$ = 0.28, falls at $r_-/r_+$ $\sim$ 0.40 
close to where peak falls 
for the higher $M_*/M_d$ sequences but the growth rate
is roughly twice as large. In line with this, the $M_*/M_d$ = 1
growth rate is about twice as large at for the 
$M_*/M_d$ = 10 model. 
For $m$ = 2 similar behavior is seen in that the sequences
converge toward the NSG results as $M_*/M_d$ increases 
but significant differences appear as early as the 
$M_*/M_d$ = 500 sequence for $r_-/r_+$ $>$ 0.7.

For $q$ = 1.75 disks, the $m$ = 1 mode shows 
the high growth rate region, but the peak growth rate is smaller 
than for $q$ = 2 disks. The weak growth rate regime seen for 
$q$ = 2 disks does not 
appear in $q$ = 1.75 disks. For $q$ = 1.5 disks, 
the $m$ = 1 mode does not show the high growth
rate regime, rather it shows only a weak growth rate region 
at $r_-/r_+$ $<$ 0.15, where it has very small oscillation 
frequency, $y_1$ $\sim$ -1. These slow $m$ = 1 modes are unstable 
for $M_*/M_d \gtrsim$ 5 to the NSG regime, with slow
oscillation frequency, $|\omega_1|$ $\ll$ $\Omega_{\circ}$ (Kato 1983)
and may be retrograde or prograde depending on $r_-/r_+$. This behavior 
is not shown by $q$ = 2 systems. 
For higher $m$ modes, $y_1$ values decline 
as $r_-/r_+$ increases with a transition to higher values for 
$m$ = 2 at $r_-/r_+$ = 0.70.
The $m$ = 2, 3, and 4 modes are unstable for 
$r_-/r_+$ $>$ 0.5 for the $q$ = 1.75 models. 
The $m$ = 2 mode dominates $q$
= 1.5 models for $r_-/r_+$ $\sim$ 0.10 to 0.60 and the $m$ = 3 and 4 
modes dominate for $r_-/r_+$ $\ge$ 0.65. 

\subsubsection{ Summary of Eigenvalue Results } \label{sec_Msummary}

\begin{figure*}
\begin{center}
\includegraphics[width=2.75in]{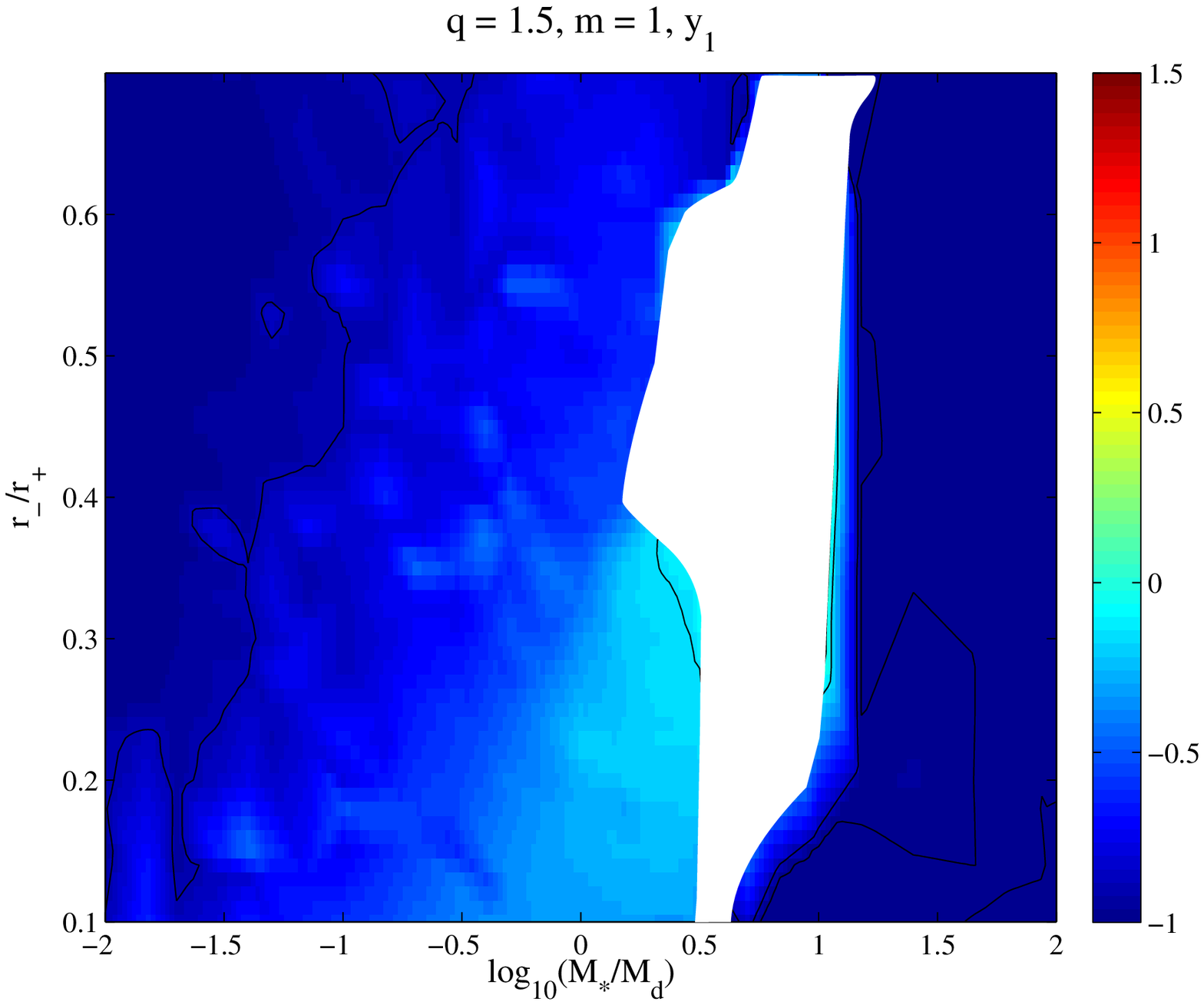}
\includegraphics[width=2.75in]{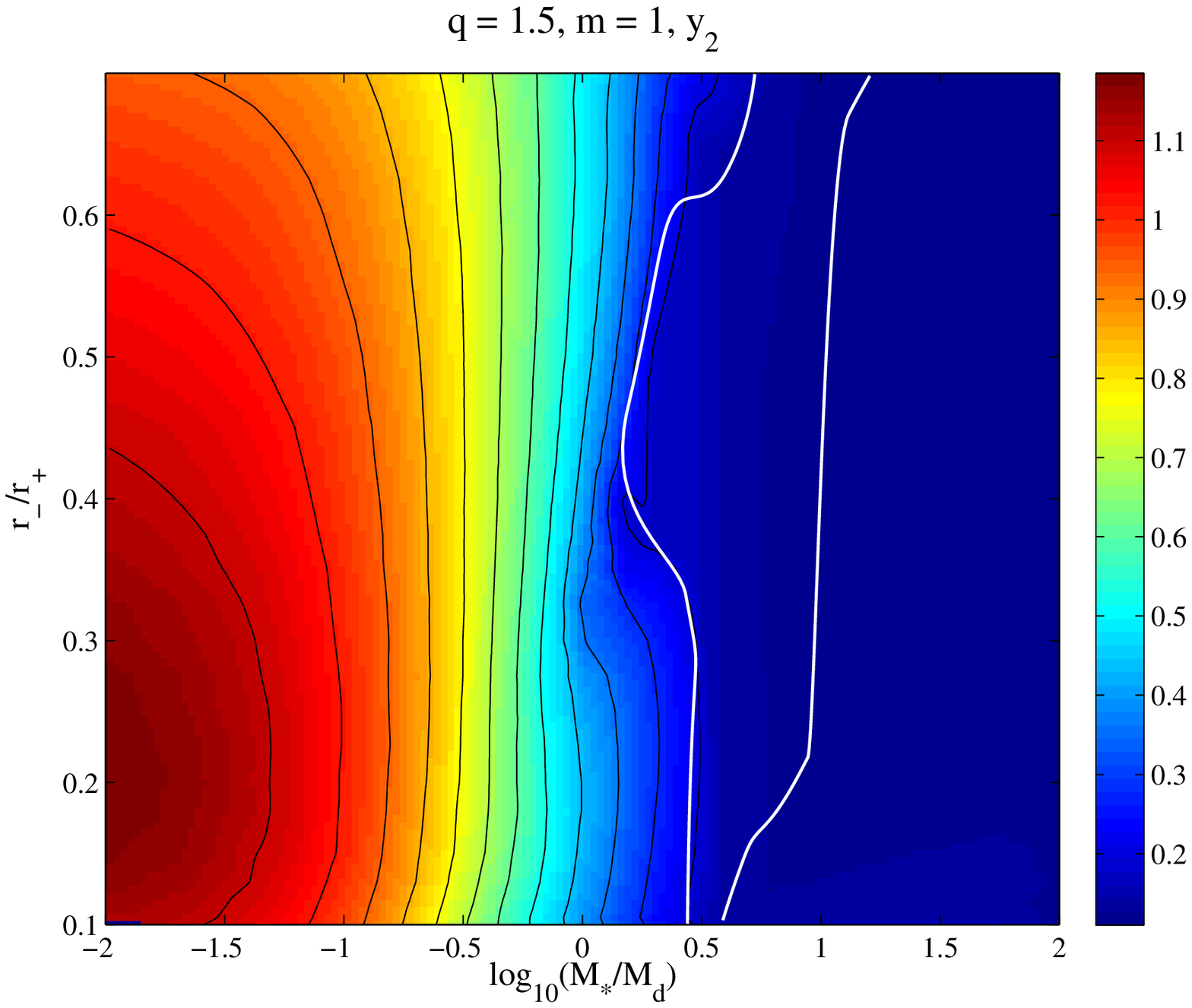}\\
\includegraphics[width=2.75in]{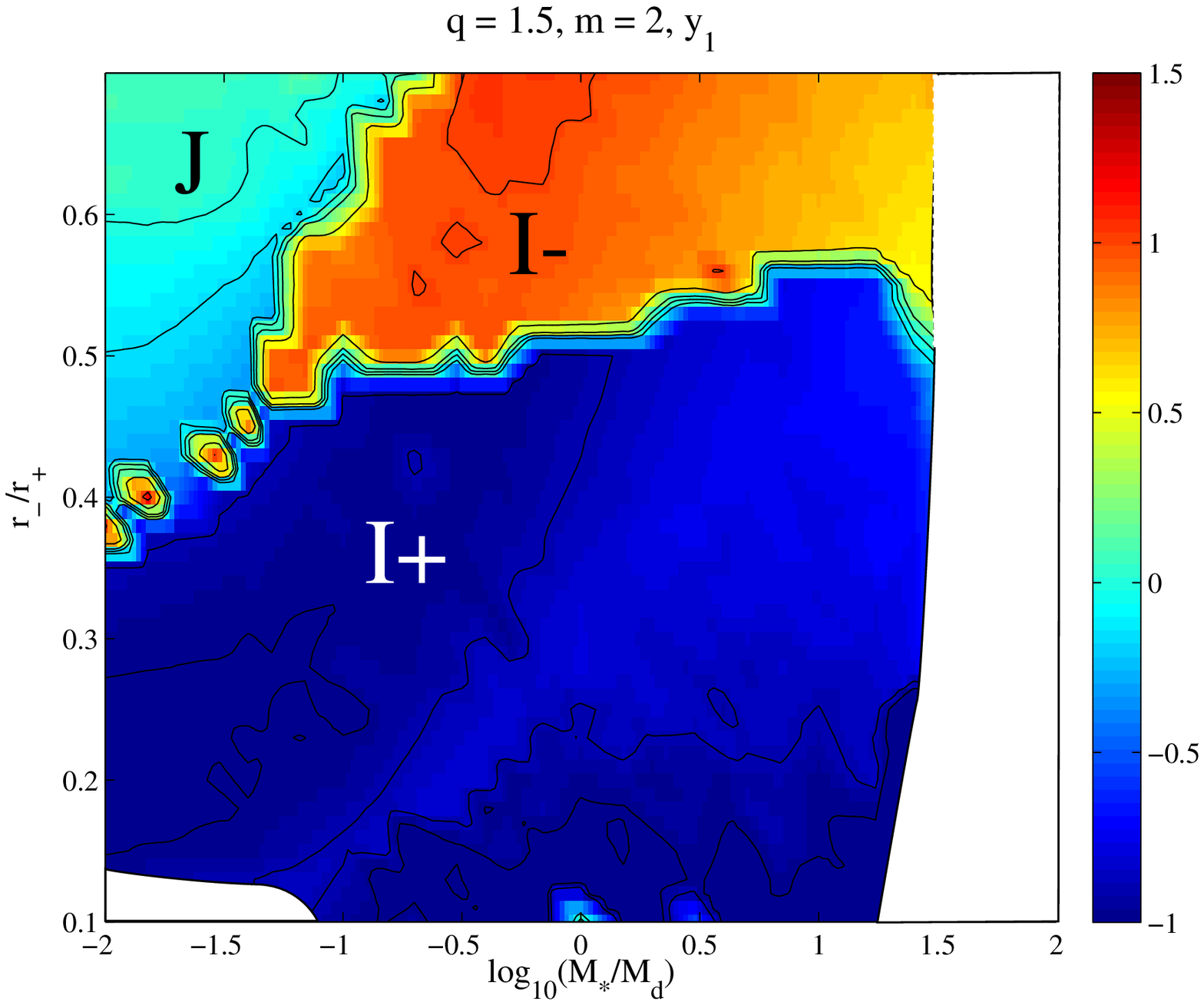}
\includegraphics[width=2.75in]{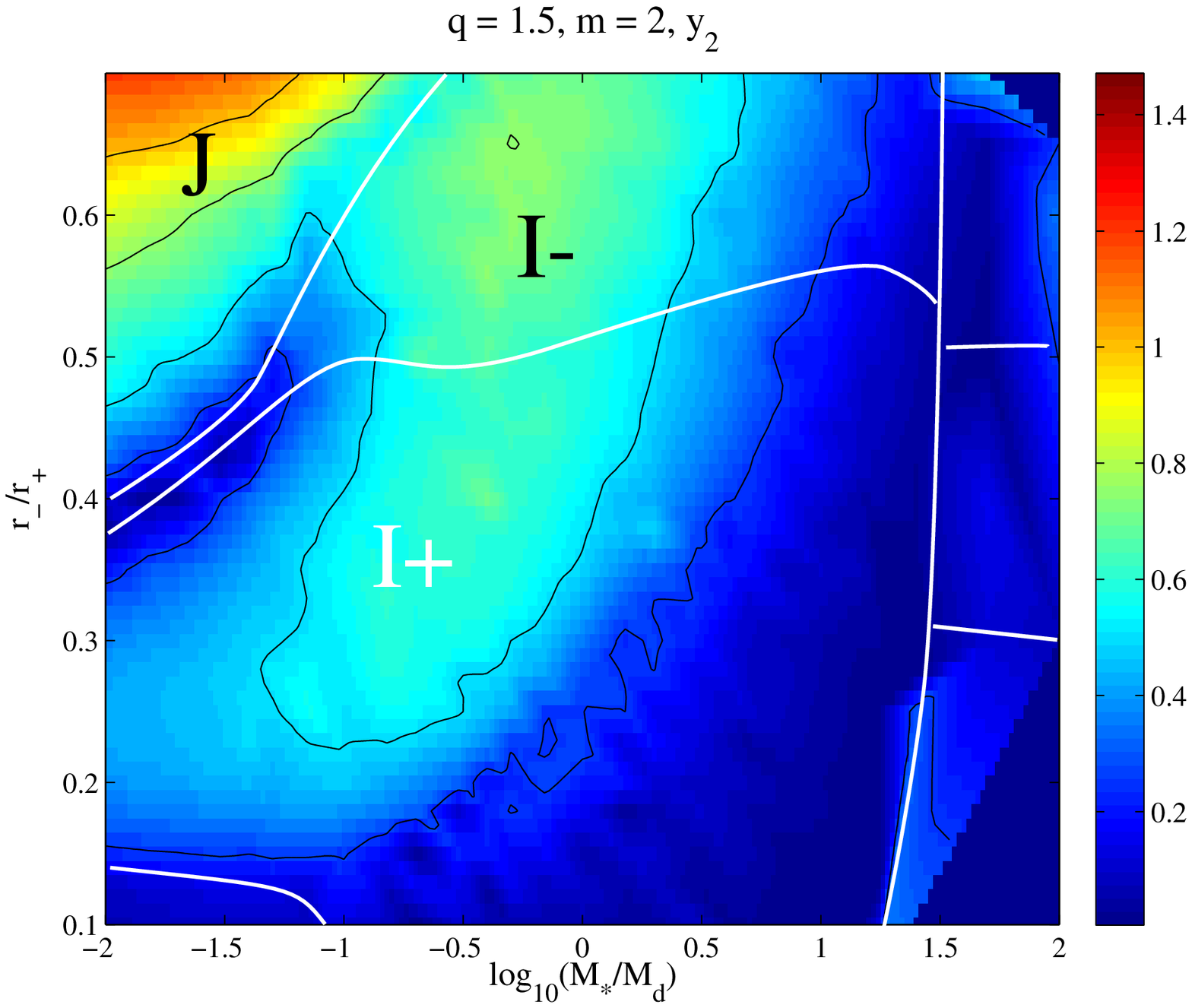}\\
\includegraphics[width=2.75in]{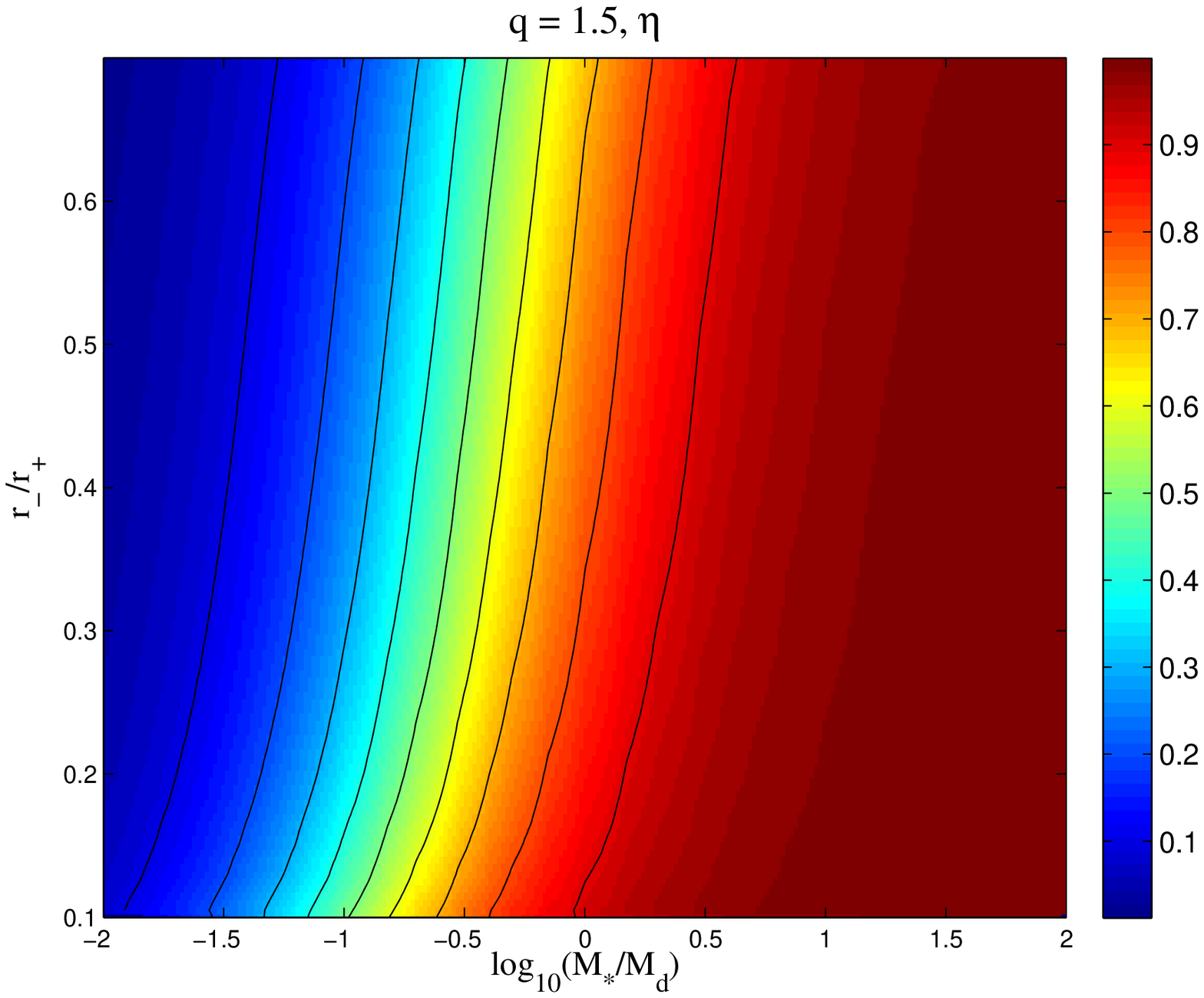}
\includegraphics[width=2.75in]{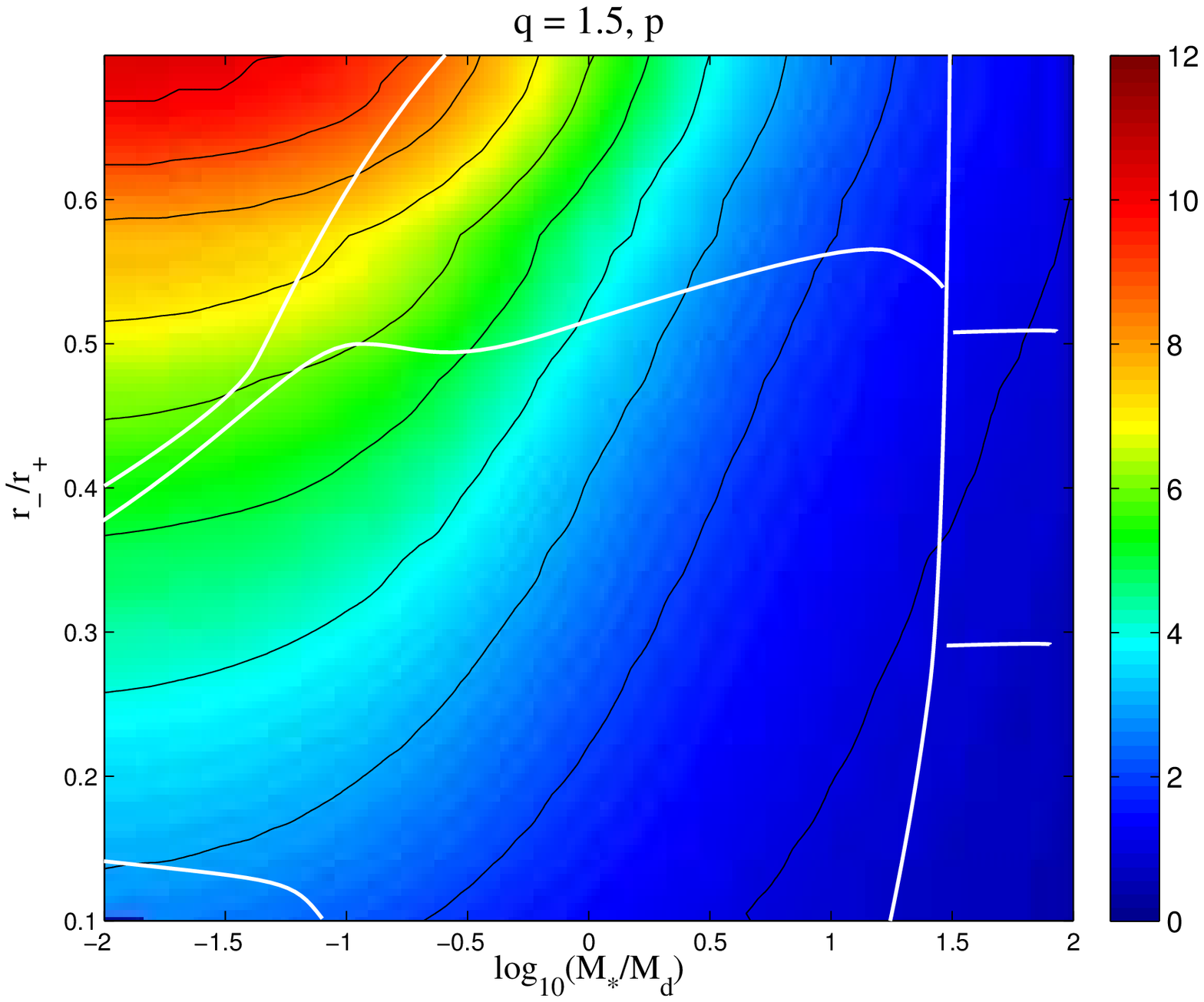}\\
\end{center}
\caption{
Oscillation ($y_1$) and growth rate ($y_2$) eigenvalues, $\eta$ and p for 
$q = 1.5$ disks. $\eta$ contours step by .1, p contours step by 1 (smallest $p \approx .5$).
Regions without a resolved pattern frequency are whited out. For $m=2$ at $M*/M_d > 30$, the 
boxed regions extrapolate stable, I$^+$ and I$^-$ modes, bottom to top, 
based on 0, 3 and 3 unstable models respectively.
}
\label{eigenvalues_m12a}

\end{figure*}

\begin{figure*}
\begin{center}
\includegraphics[width=2.75in]{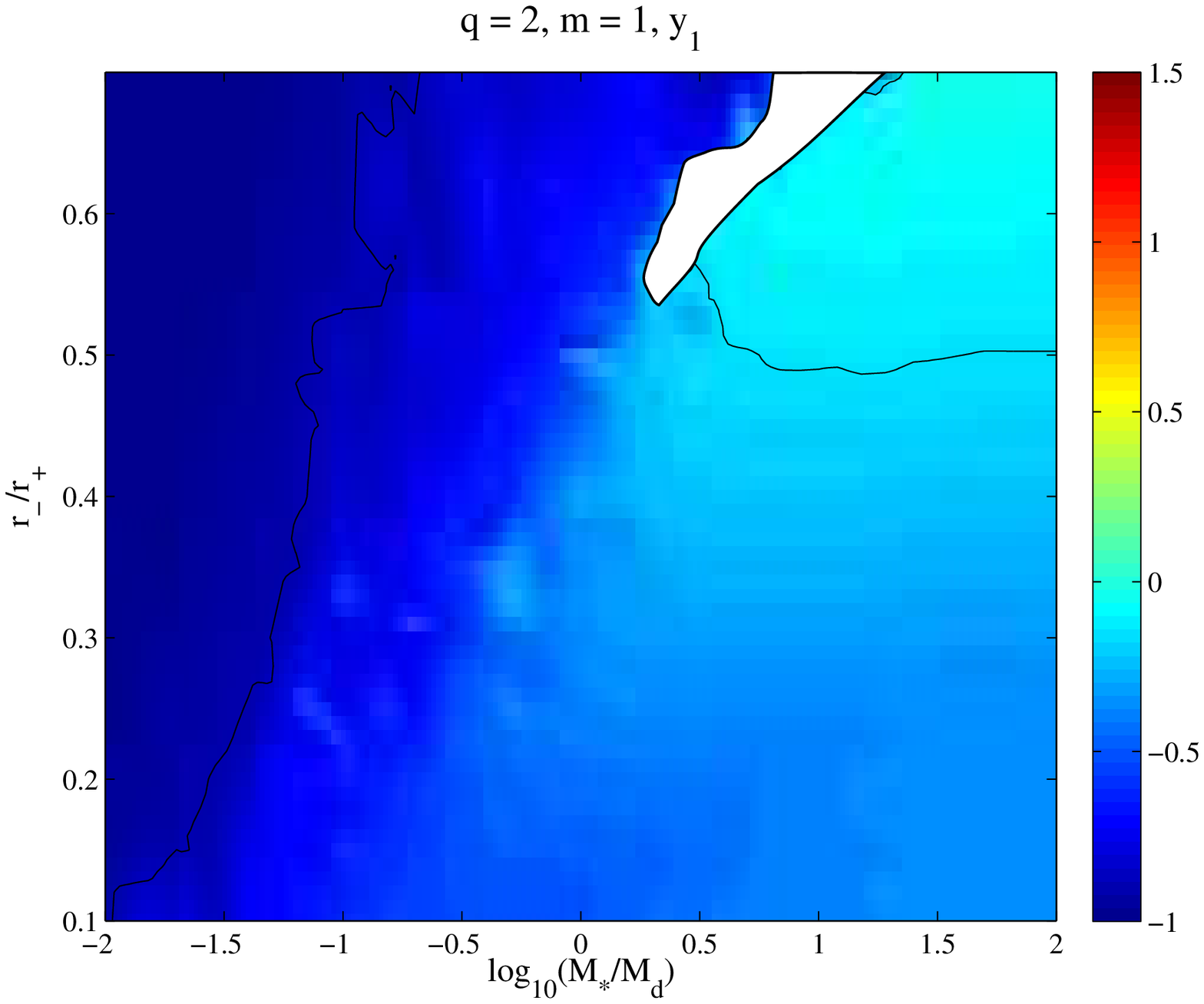}
\includegraphics[width=2.75in]{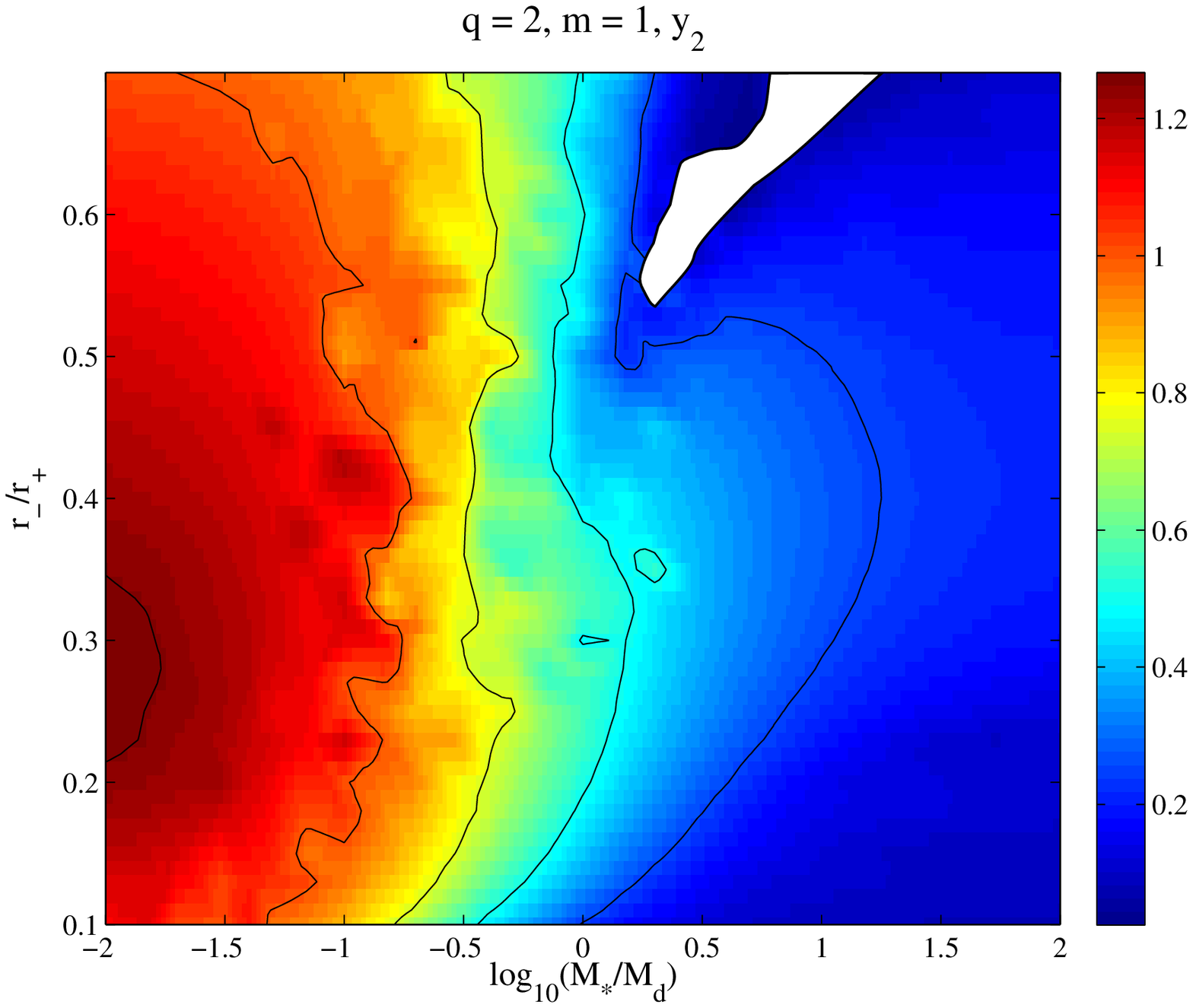}\\
\includegraphics[width=2.75in]{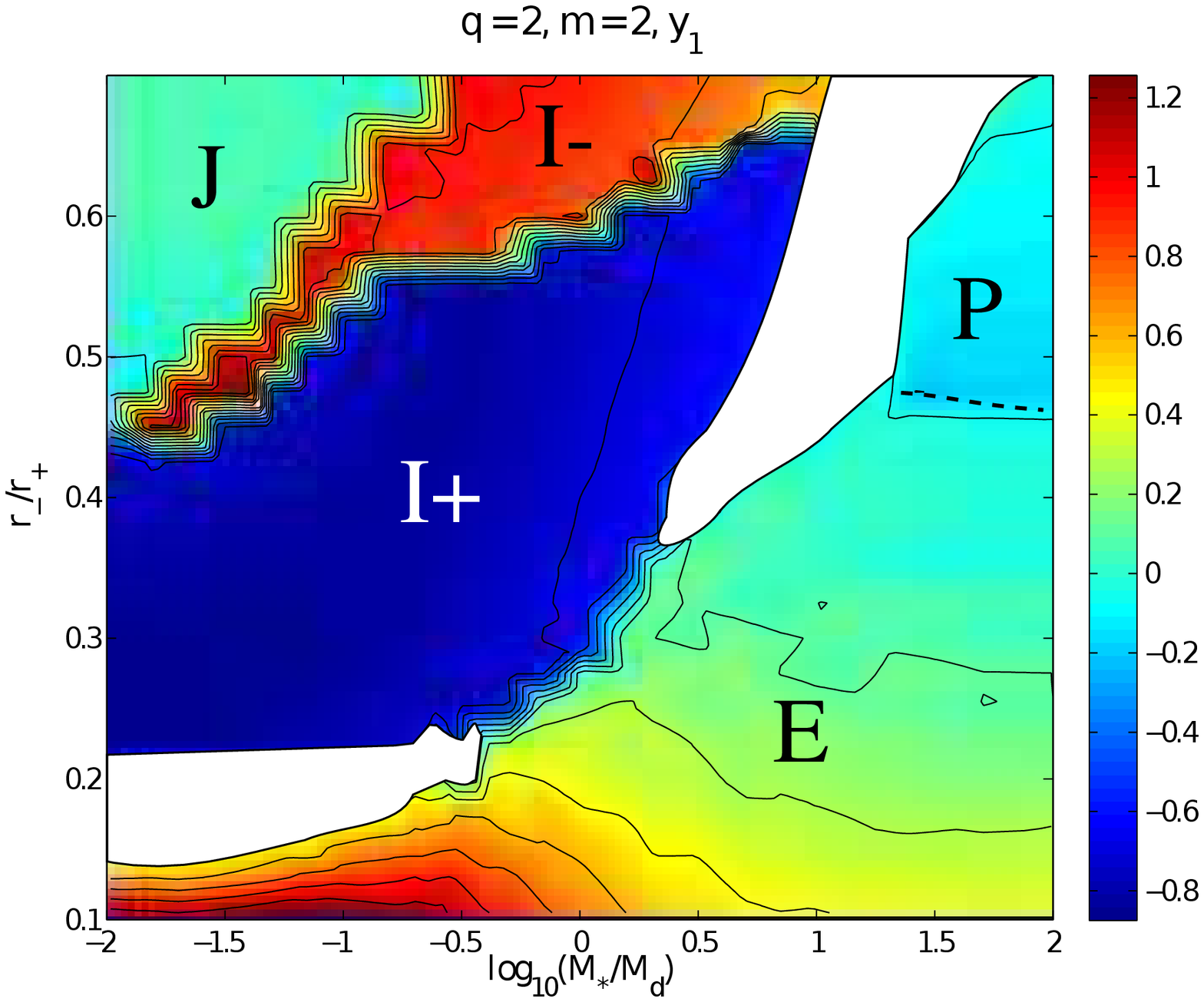}
\includegraphics[width=2.75in]{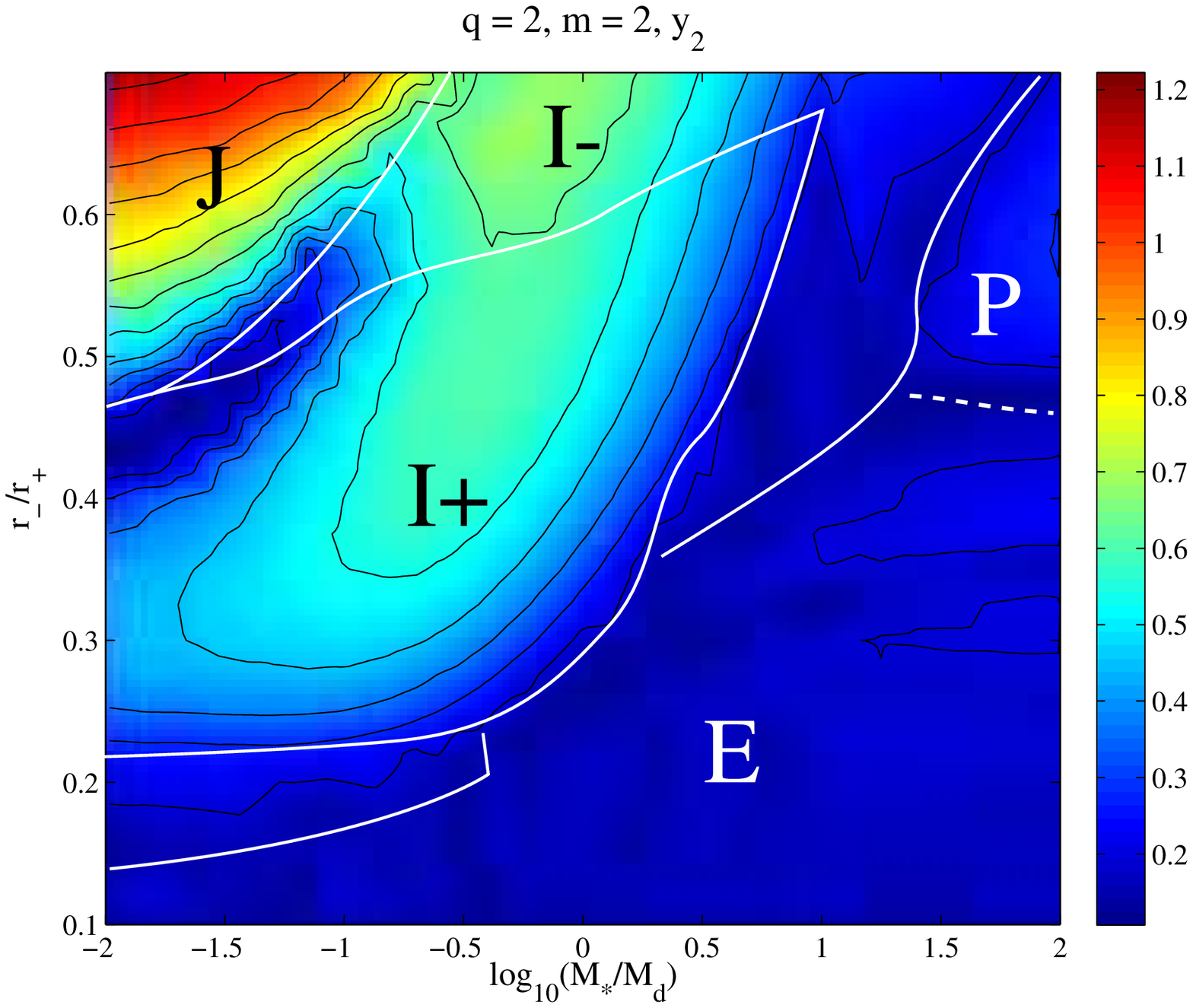}\\
\includegraphics[width=2.75in]{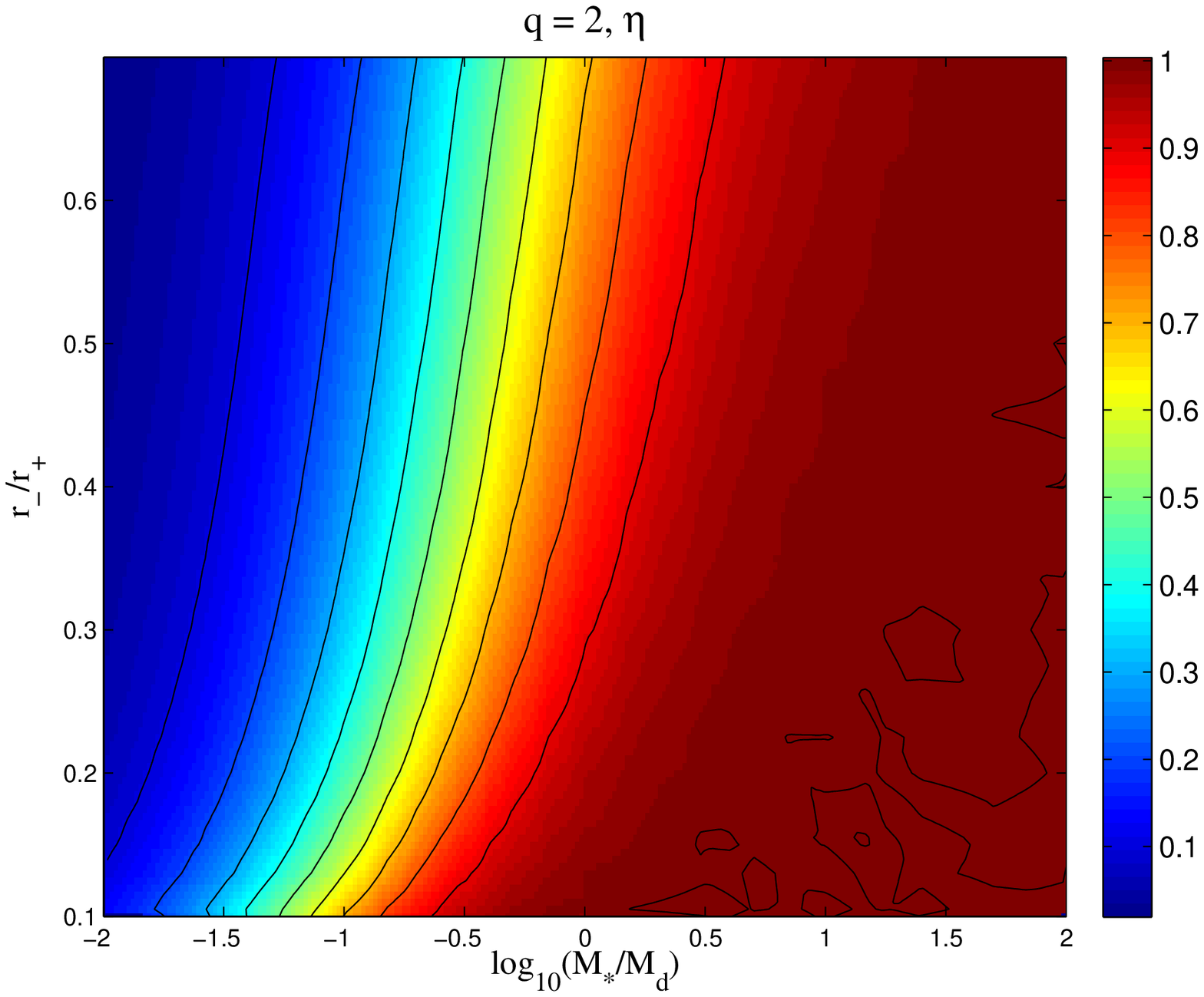}
\includegraphics[width=2.75in]{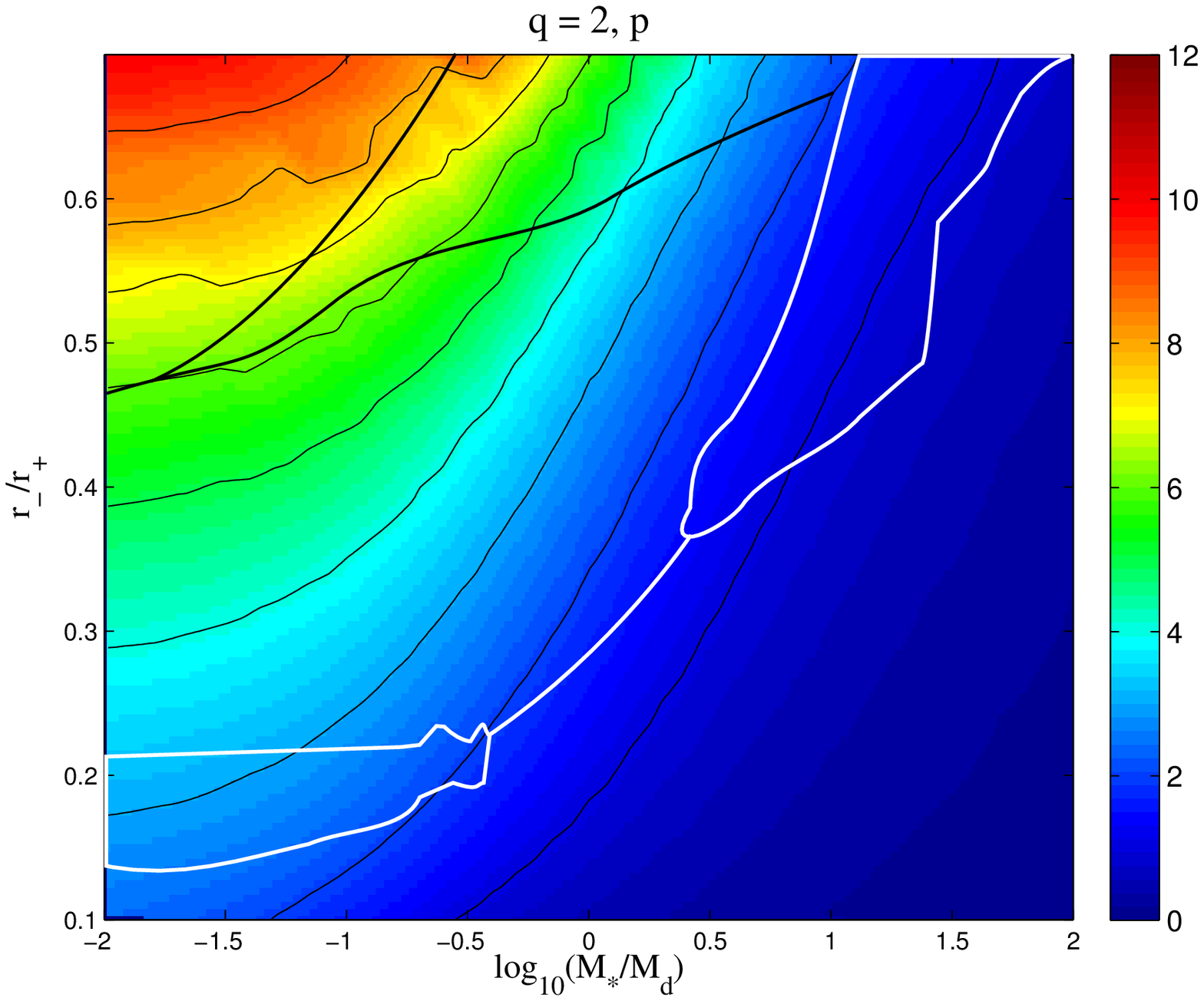}\\
\end{center}
\caption{
Oscillation ($y_1$) and growth rate ($y_2$) eigenvalues, $\eta$ and p for 
$q = 2$ disks. $\eta$ contours step by .1, p contours step by 1 (smallest $p \approx .05$).
Regions without a resolved pattern frequency are whited out.
}
\label{eigenvalues_m12b}
\end{figure*}

\begin{figure}
\begin{center}
\includegraphics[width=3.0in]{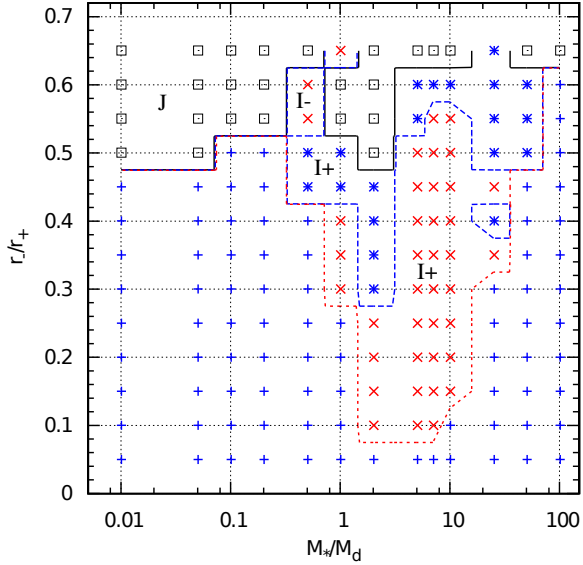}
\end{center}
\caption{
Dominant modes in $(r_-/r_+,M_*/M_d)$ space for $q$ = 1.5 disks.
The modes are denoted as follows: $m$ = 1 blue crosses,
$m$ = 2 red Xs, $m$ = 3 blue asterisks, and $m$ = 4 black squares. Stable 
models are filled black squares. Note that the large $m=2$ I$^+$ region
showed numerical difficulties with $m=1$ at $q=1.5$, and $m=1$ may in
fact dominate here.
}
\label{dominant_modes_q15}
\end{figure}

\begin{figure}
\begin{center}
\includegraphics[width=3.0in]{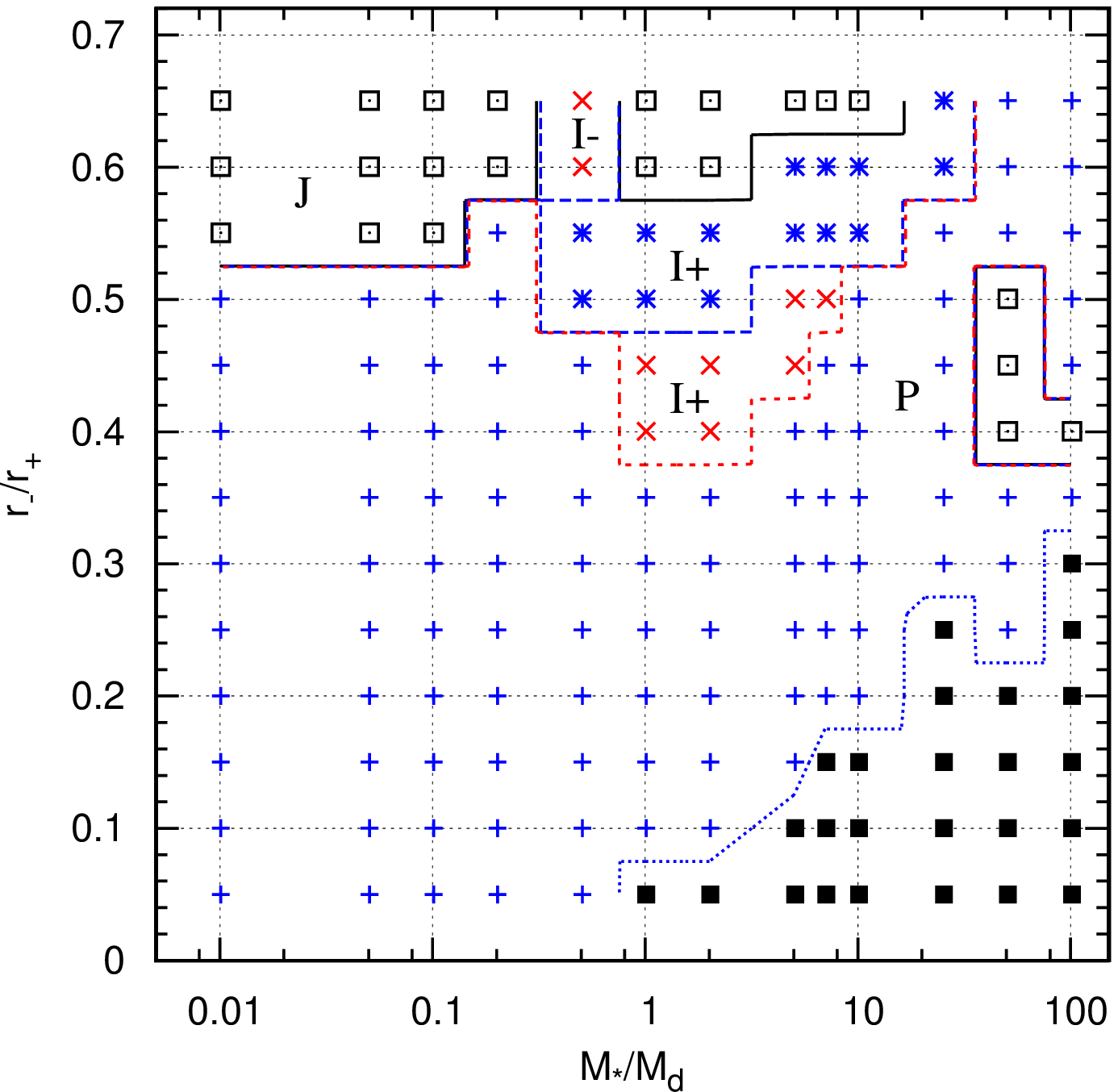}
\end{center}
\caption{
Dominant modes in $(r_-/r_+,M_*/M_d)$ space for $q$ = 1.75 disks.
The modes are denoted as follows: $m$ = 1 blue crosses,
$m$ = 2 red Xs, $m$ = 3 blue asterisks, and $m$ = 4 black squares. Stable models are filled black squares.
}
\label{dominant_modes_q175}
\end{figure}

\begin{figure}
\begin{center}
\includegraphics[width=3.0in]{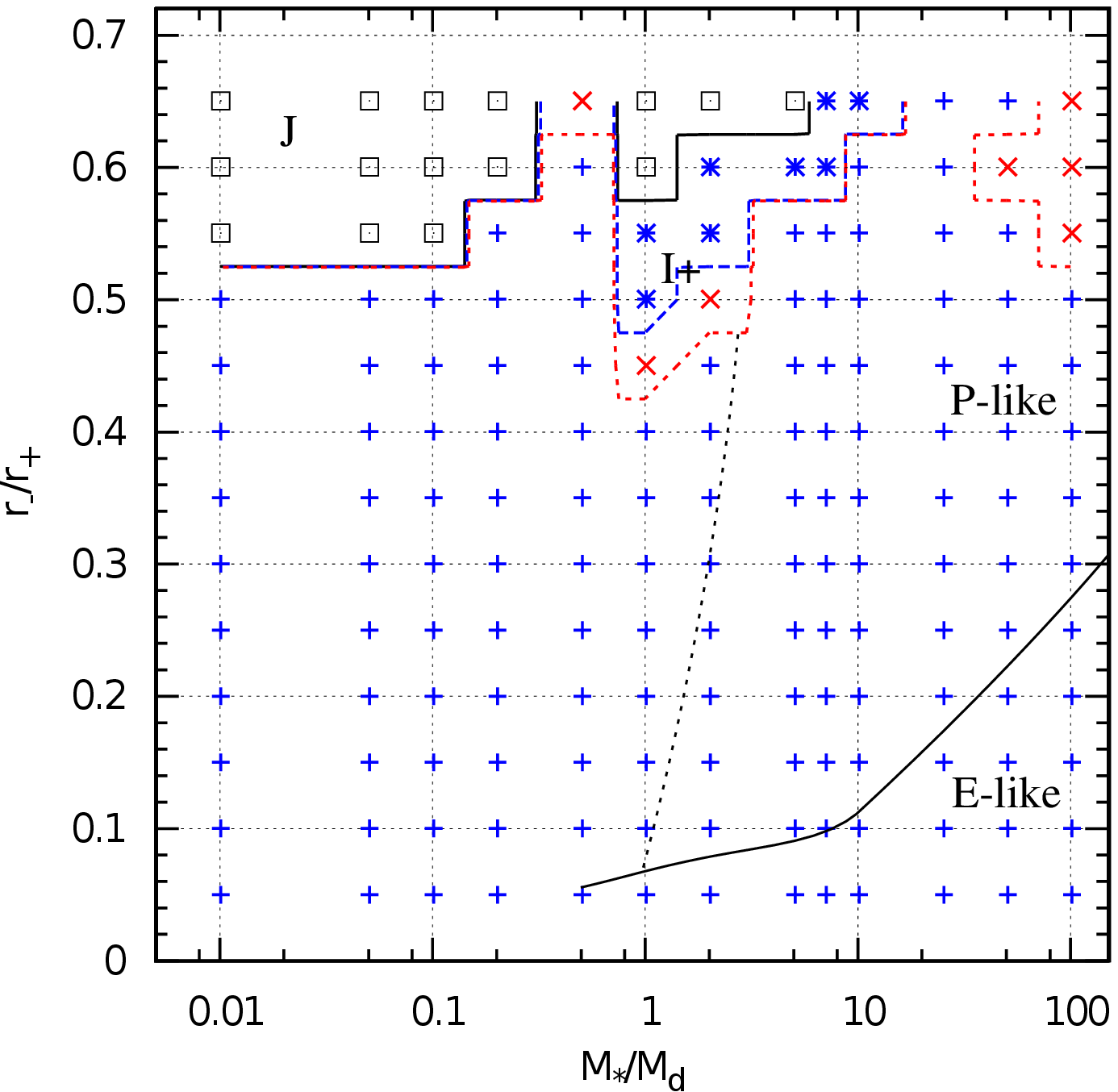}
\end{center}
\caption{
Dominant modes in $(r_-/r_+,M_*/M_d)$ space for $q$ = 2 disks.
The modes are denoted as follows: $m$ = 1 blue crosses,
$m$ = 2 red Xs, $m$ = 3 blue asterisks, and $m$ = 4 black squares. Stable models are filled black squares.
}
\label{dominant_modes_q2}
\end{figure}

Our results are highlighted in Figure \ref{eigenvalues_m12a}
where we show $y_1$ and $y_2$ values as well as equilibrium 
parameters $\eta$ and $p$ for $q = 1.5$ disks, 
and in Figure \ref{eigenvalues_m12b} where we do the same for
$q=2$ disks. 
Our results are summarized in Figures \ref{dominant_modes_q15} 
to \ref{dominant_modes_q2} where we
show the dominant modes for $q$ = 1.5, 1.75, and 2 disks.
Star/disk systems are generally unstable. For $q$ = 2 disks, which
have a Toomre Q parameter of zero, $m$ = 1 modes usually dominate.
Multi-armed modes, $m$ $\ge$ 2, are only dominant
at large $r_-/r_+$, with some exceptions. Similar 
behavior is seen in the $q$ = 1.75 and 1.5 disks 
except that the range over which multi-armed modes 
dominate covers more of parameter space. For $q$ = 1.5 disks, 
the region stretches from high $r_-/r_+$ to $r_-/r_+$$\sim$ 0.1 for systems with
moderate to low disk mass, $M_*/M_d$ $\approx$ 1 to 20. 
Protostellar and protoplanetary disks are expected to show
nearly ${\it Keplerian}$ rotation with $M_*/M_d$ $\approx$ 1 to 20
and are thus expected to be dominated by $m$ = 2, 3, and 4 modes. 
If they, however, are closer to constant specific angular momentum disks, $q$ = 
2 disks, they are are then expected to be dominated by $m$ = 1 modes.

The instability regimes for multi-armed modes in 
($r_-/r_+,M_*/M_d$) space track the strength of 
self-gravitational effects as measured by the parameter $p$
(independent of $q$). In Figures \ref{eigenvalues_m12a} and
\ref{eigenvalues_m12b}, mode type boundaries have been overlaid on
$p$ which clearly show this. The most unstable are the J modes
in the upper left hand corner, where $p$ is largest.

Moving away from this corner we encounter I modes. At $m=2$ they
are present for $r_-/r_+ \ge ~0.30$ and $M_*/M_d \le 10$, excluding
the J mode region itself. At lower $r_-$ we observe I$^+$ modes
with corotation outside density max. At higher $r_-/r_+$ we 
observe I$^-$ modes with corotation well inside density max.

A stable region with $y_2$ $\sim$ 0 exists between I$^-$ and the J modes, 
in a short arc sweeping from $r_-/r_+$ $\approx$ 0.425, $M_*/M_d$ = 
0.01 to $r_-/r_+$ $\approx$ 0.475, $M_*/M_d$ = 0.05. A second long 
arc of stability sweeps through parameter space from 0.1 $\le$
$r_-/r_+$ $\le$ 0.20, 0.01 $\le$ $M_*/M_d$ $\le$ 0.1 to 
$r_-/r_+$ $\approx$ 0.70, 17.5 
$\le$ $M_*/M_d$ = 50. The two arcs roughly follow $p$ $\approx$ 
7.5 and 3. 
In the $y_1$ plots in Figures \ref{eigenvalues_m12a} and
\ref{eigenvalues_m12b} we see that these stable regions are where the dominant
mode type changes. Sometimes, as in the transition from J to I$^+$,
there is no ``forbidden region'' and the transition is marked by more
subtle changes in the growth rates.

For one-armed spirals, $m$ = 1 modes, the instability regimes more
closely follow the constant $\eta$ curves than constant $p$
curves. One-armed spirals are more unstable at large $\eta$
than small $\eta$. For $q$ = 2 disks, 
a tongue of weak instablity extends from intermediate 
$M_*/M_d$ to high $M_*.M_d$ at intermediate $r_-/r_+$. This feature
is not prominent in $q$ = 1.5 disks and is somewhat present in $q=1.75$ disks.
 Outside of this feature, the
instability regimes for the $m$ = 1 modes are similar for $q$ 1.5, 1.75, and 2 disks.
This does not mean the same types of $m$ = 1 modes populate
the regions, however. For $q$ = 1.5 disks, $m$ = 1 modes are 
slow for small $M_*/M_d$ and nearly stationary for large 
$M_*/M_d$. For $q$ = 2 disks,
$m$ = 1 modes are also slow for small $M_*/M_d$, but show faster modes,
modes with corotation near
$r_{\circ}$ and just outside $r_{\circ}$ at large $M_*/M_d$. For 
Keplerian disks, $q$ = 1.5 disks, $m$ = 1 modes
show $r_{co}$ just outside $r_{\circ}$ for systems with small $r_-/r_+$ and
$M_*/M_d$ $\approx$ 0.5-3.


\begin{figure}
\begin{center}
\includegraphics[trim=0cm 5cm 0cm 0cm, clip=true, width=3.3in]
{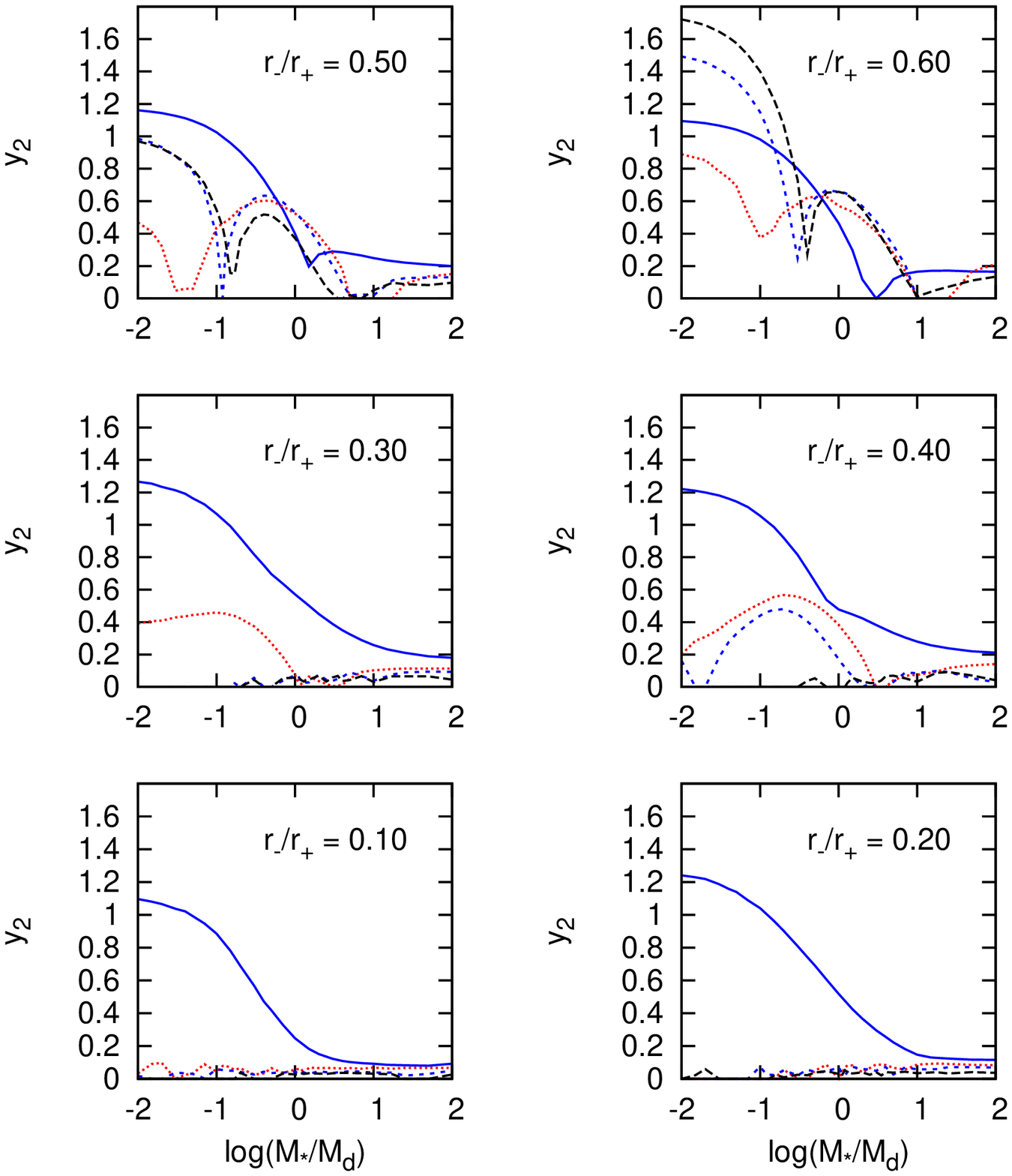}
\end{center}
\caption{Growth rates $y_2$ of $q=2.0$ disks as 
functions of log($M_*/M_d$) for selected values of $r_-/r_+$.
The $m$ = 1 modes are shown by blue solid lines, $m$ = 2 modes by
red dotted lines, $m$ = 3 modes by blue dashed lines, and
$m$ = 4 modes by black long-dashed lines.
\label{q20y_2_r-+}
}
\end{figure}

We show the growth rates, $y_2$, as functions of $M_*/M_d$ for $r_-/r_+$ = 0.1, 0.2,
0.3, 0.4, 0.5 and 0.6 in Figure \ref{q20y_2_r-+}. For $r_-/r_+$ = 0.1, 
we see that m = 1 dominates models with $M_*/M_d < 10$ and shows comparable growth 
rates for high $M_*/M_d$ models. As $r_-/r_+$ increases, higher order m modes
increase their growth rates, overtaking m = 1 modes at $r_-/r_+$ = 0.5 and 0.6. 
At high $r_-/r_+$, we see the growth rates cascading upward in m
for small $M_*/M_d$.

\subsection{ Angular Momentum Transport } \label{sec_amom}

The spiral arms in nonaxisymmetric disk modes 
exhibit both leading and trailing
nature. This suggests that gravitational torques resulting 
from disk instabilities can manifest themselves through 
transport of angular momentum inward and outward
through disks. Disks dominated by 
I and P modes show Reynolds stress torques
which can dominate gravitational 
torques. For these cases, angular momentum transport is
not determined by whether 
the gravitational perturbation leads or lags the
density perturbations. Locally
defined transport is expected. 
Here, we analyze torques for selected disk models. Transport is
studied in the quasi-linear (QL) approximation using a model developed 
from the equation of conservation of angular momentum in its
conservative form ({\it e.g.}, Shariff 2009).

We form the torque density,
\begin{equation}
{\bf N} = {\bf r} \times {\bf f}
\end{equation}
where {\bf f}, the force density is given by
\begin{equation}
{\rm {\bf f}} = 
\partial_t(\rho{\bf v}) = -{\bf \nabla}\cdot{\cal {\bf S} }.
\end{equation}
In {\bf f}, ${\cal {\bf S}}$ is the stress tensor which has the form 
\begin{equation}
{\cal {\bf S}} = \rho{\bf vv} + {\bf {\cal P}} + \frac{1}{4\pi G}\left(
{\bf \nabla}\Phi_g{\bf \nabla}\Phi_g-\frac{1}{2}{\cal I}{\bf \nabla}\Phi_g\cdot
{\bf \nabla}\Phi_g
\right),
\end{equation}
where ${\bf {\cal P}}$ is the pressure tensor and ${\cal I}$ is the unit 
tensor.
The $z$-component of ${\bf N}$ is responsible for radial
angular momentum transport and is given by
\begin{equation}
N_z = \partial_t(\rho v_{\phi}\varpi)
= -\nabla\cdot(\rho v_{\phi}\varpi{\bf v})
- \rho\partial_{\phi}\Phi_g.  \label{torque_z}
\end{equation}
The Reynolds stress is in the first term on the right hand
side of Equation \ref{torque_z}. The gravitational torque, composed of the
disk self-interaction torque and the star-disk coupling torque, is
the last term in the right hand side of the equation. The azimuthal
pressure gradient term integrates to zero in our approximation.

Quasi-linear forms for the torque are found from substitution of
\begin{equation}
\rho = \rho_{\circ} + \delta\rho e^{im\phi}, 
{\bf v} = {\bf v}_{\circ} e^{im\phi},\;{\rm and}\; \Phi_g = \Phi_{g,\circ} 
+ \delta\Phi_ge^{im\phi}
\end{equation}
into Equation \ref{torque_z}.
The gravitational self-interaction torque over the annulus between $\varpi$
and $\varpi+\Delta\varpi$ is
\begin{equation}
\Gamma_g = m\pi\int_{-Z_*}^{Z_*}{\rm dz}\int_{\varpi}^{\varpi+\Delta\varpi}
|\delta\rho||\delta\Phi|{\rm sin}(\phi_{\rho}-\phi_{\Phi})
\varpi{\rm d}\varpi
\end{equation}
where $\phi_{\rho}$ and $\phi_{\Phi}$ are the phases of the
density and gravitational perturbations (Imamura, Durisen, \&
Pickett 2000). The Reynolds stress arises from coupling
of the azimuthal and radial velocity perturbations. The radial
flux of angular momentum carried by waves is the Reynolds
stress integrated over the cylindrical surface with radius $\varpi$,
\begin{equation}
{\cal F}_R = \pi\int\rho_{\circ}|\delta v_{\varpi}||\delta v_{\phi}|
{\rm cos}(\phi_{v_{\phi}}-\phi_{v_{\varpi}})
\varpi^2{\rm d}z
\end{equation}
({\it e.g.,} see Shariff 2009). The Reynolds stress torque on the
volume between $\varpi$ and $\varpi+\Delta\varpi$ is 
$\Gamma_R$ = -${\cal F}_R(\varpi+\Delta\varpi)$+${\cal F}_R(\varpi)$. 
The $\Gamma_g$ and $\Gamma_R$ are calculated using 
eigenfunctions normalized such that
\begin{equation}
{\cal M}_m = \int|\delta\rho|{\rm d}^3{\rm x} = 1.
\end{equation}
The torques scale as ${\cal M}_m^2$ for arbitrary $\delta\rho$ 
and are given in 
units of $\Gamma_{\circ}$ $=$ $J/\tau_{\circ}$ where
$J$ is the total angular momentum. The  
characteristic angular momentum transport time,
\begin{equation}
\tau_{\Gamma} = \frac{J_z(\varpi)}{\Gamma_m}
\end{equation}
where $J_z(\varpi)$ is the z-component of the angular 
momentum in the annulus
$[\varpi,\varpi+\Delta\varpi]$,
\begin{equation}
J_z(\varpi) = \int_0^{2\pi}{\rm d}\phi\int_{-Z_*(\varpi)}^{Z_*
(\varpi)}{\rm d}z
\int_{\varpi}^{\varpi+\Delta\varpi}{\bf \hat{z}}\cdot({\bf r\times\rho v})
\varpi{\rm d}\varpi,
\end{equation}
$Z_*(\varpi)$ is the vertical height of the disk at given 
$\varpi$, $\Gamma_m$ = $\Gamma_g+\Gamma_R$
and $\tau_{\Gamma}$ is given in units of $\tau_{\circ}$.
Properties of models selected for presentation are 
described in Tables 1 to 5 with 
their $\Gamma_g$, $\Gamma_R$, $\delta$J,
and $\tau_{\Gamma}$ shown in Figures \ref{Jmode_m4_plot} 
to \ref{Pmode_m1_plot}.
We designate the smallest radius where  $\tau_{\Gamma}$ = $\tau_{\circ}$
as $r_{1,-}$ and the next radius where $\tau_{\Gamma}$ = $\tau_{\circ}$
as $r_{1,+}$.
In Table 5, the regions in the disks where
$\tau_{\Gamma}$ $<$ $\tau_{\circ}$, r $<$ $r_{1,-}$ and r $>$ 
$r_{1,+}$, and the mass contained in those
regions, $m_{1,-}$ and $m_{1,+}$,
the radius at which the first sign change in $\Gamma_m$ 
falls, $r_{\Gamma}$, 
and the mass inside that radius, $m_{\Gamma}$ are given. 

\begin{table*}[t]
\centering
Table 5: Mass Transport Properties
\vskip 0.1in
\begin{tabular}{*{5}{c}}
\hline\hline
Model &
Mode &
r$_{(1,-)}$/r$_{\circ}$,m$_{(1,-)}$ &
r$_{(1,+)}$/r$_{\circ}$,m$_{(1,+)}$ &
r$_{\Gamma}$/r$_{\circ}$,m$_{\Gamma}$ \\
\hline
J1&$m$=4 J& 0.85,0.12 & 1.24,0.12 & 1.02,0.44\\
J1&$m$=1 I$^+$& 0.69,0.0053 & 1.11,0.36 & 0.77,0.040 \\
I1&$m$=2 I$^-$& 0.84,0.023 & 1.14,0.090 & 1.07,0.71 \\
I2&$m$=2 I$^-$& 0.80,0.0023 & 1.25,$<$0.001 & 0.96,0.28\\
I4&$m$=2 I$^+$&0.83,0.10& 1.29,0.071 & 1.07,0.73 \\
I5&$m$=2 I$^+$&0.71,0.050 & 1.32,$<$0.001 & 0.76,0.0060;1.08,0.33 \\
O13&$m$=1 A& 1.37,0.16 & 3.90,0.004 & 1.75,0.34 \\
E1&$m$=2 Edge& 0.73,0.0032 & 5.4,$<$0.0001 & 1.12,0.056 \\
P2&$m$=2 P& 0.79,0.0043 & 1.42,0.0050 & 1.09,0.51 \\
O16&$m$=1 P & 0.85,$<$ 0.0001 & 1.21,$<$ 0.0001 & 1.02,0.51 \\
\hline
\end{tabular}
\end{table*}

\subsubsection{ $m$ = 4 J Mode }

\begin{figure}
\begin{center}
\includegraphics[width=3.0in]{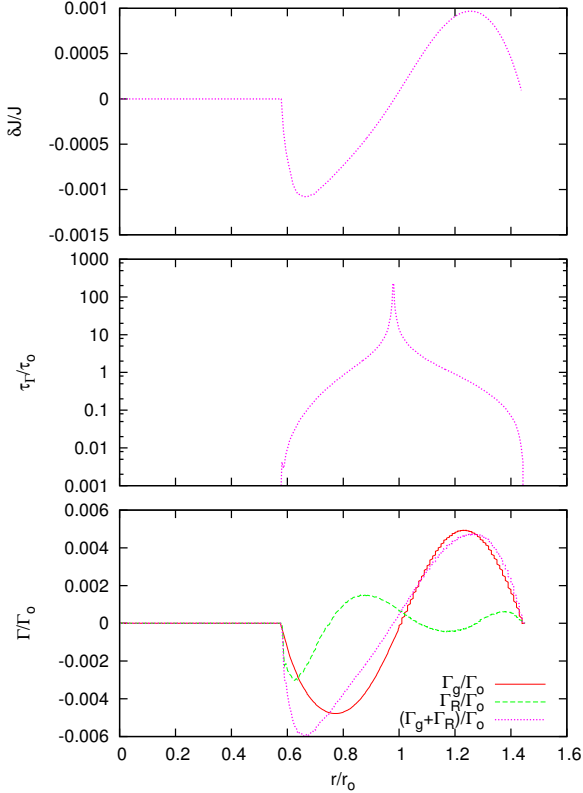}
\end{center}
\caption{
The $\tau_{\Gamma}$, $\delta$J,
$\Gamma_g$, and $\Gamma_R$ for the
$m$ = 4 J mode of the $M_*/M_d$ = 0.01, $q$ = 1.5, 
$r_-/r_+$ = 0.402 disk, model J1.
}
\label{Jmode_m4_plot}
\end{figure}

J modes are dominated by the gravitational stress
$\Gamma_g$. The $\tau_{\Gamma}$, $\delta$J, $\Gamma_g$,
and $\Gamma_R$ for a representative J mode, the
dominant $m$ = 4 J mode for the $M_*/M_d$ = 0.01, $q$ = 1.5, 
$r_-/r_+$ = 0.402 disk, Model J1, are shown in Figure
\ref{Jmode_m4_plot}. The eigenvalues for $m$ = 4 mode are
$(y_1,y_2)$ = (-0.378,0.576). Corotation 
sits at 1.068 r$_{\circ}$. The inner Lindblad radius
r$_{ilr}$ falls near an extremum in $\Gamma_m$ while 
r$_{olr}$ is near r$_+$. The gravitational torque, 
$\Gamma_g$ is negative inside $\sim$ r$_{\circ}$ and positive outside 
$\sim$ r$_{\circ}$. $\delta$J and $\Gamma_m$ also both change sign 
near $r_{\circ}$. Although the total torque $\Gamma_m$  
is dominated by $\Gamma_g$, the Reynolds stress torque $\Gamma_R$
is not negligible in the inner half of the disk. 
The torque is strong in this disk and the disk 
expected to spread on time scales of $\sim$ $\tau_{\circ}$ as 
instability approaches ${\cal M}_4$ $\sim$ 1. 
Material with inflow times $< \tau_{\circ}$
fall at radius $< 0.848 r_{\circ}$, a region which contains 12 \% of the
disk mass and for $r > 1.24 r_{\circ}$, a region which contains 12 \%
of the disk mass. The sharp peak in 
$\tau_{\Gamma}$ indicates where $\Gamma_m$ goes 
to zero. Because this is a discrete
calculation, the exact zero is stepped over in 
our simulation and the peak has finite amplitude. 
The narrow peak in the
transport time indicates that only matter 
close to the zero in $\Gamma_m$
will remain relatively stationary as the disk evolves.
Up to 44 \% of the disk mass has flow time less than  
2 $\tau_{\circ}$ at saturation. The 
perturbed angular momentum $\delta$J and $\Gamma_m$ 
follow each other's behavior
suggesting that how and where instability is 
driven follows from $\Gamma_m$.
Several other modes in this disk model are also unstable and, 
in fact, the dominant mode is not the $m$ = 4 J mode 
but, rather, it is an $m$ = 1 $I^+$ mode (see $\S3.4.5$). 

\subsubsection{ $m$ = 2 I$^-$ Mode }

\begin{figure*}
\begin{center}
\includegraphics[width=3.0in]{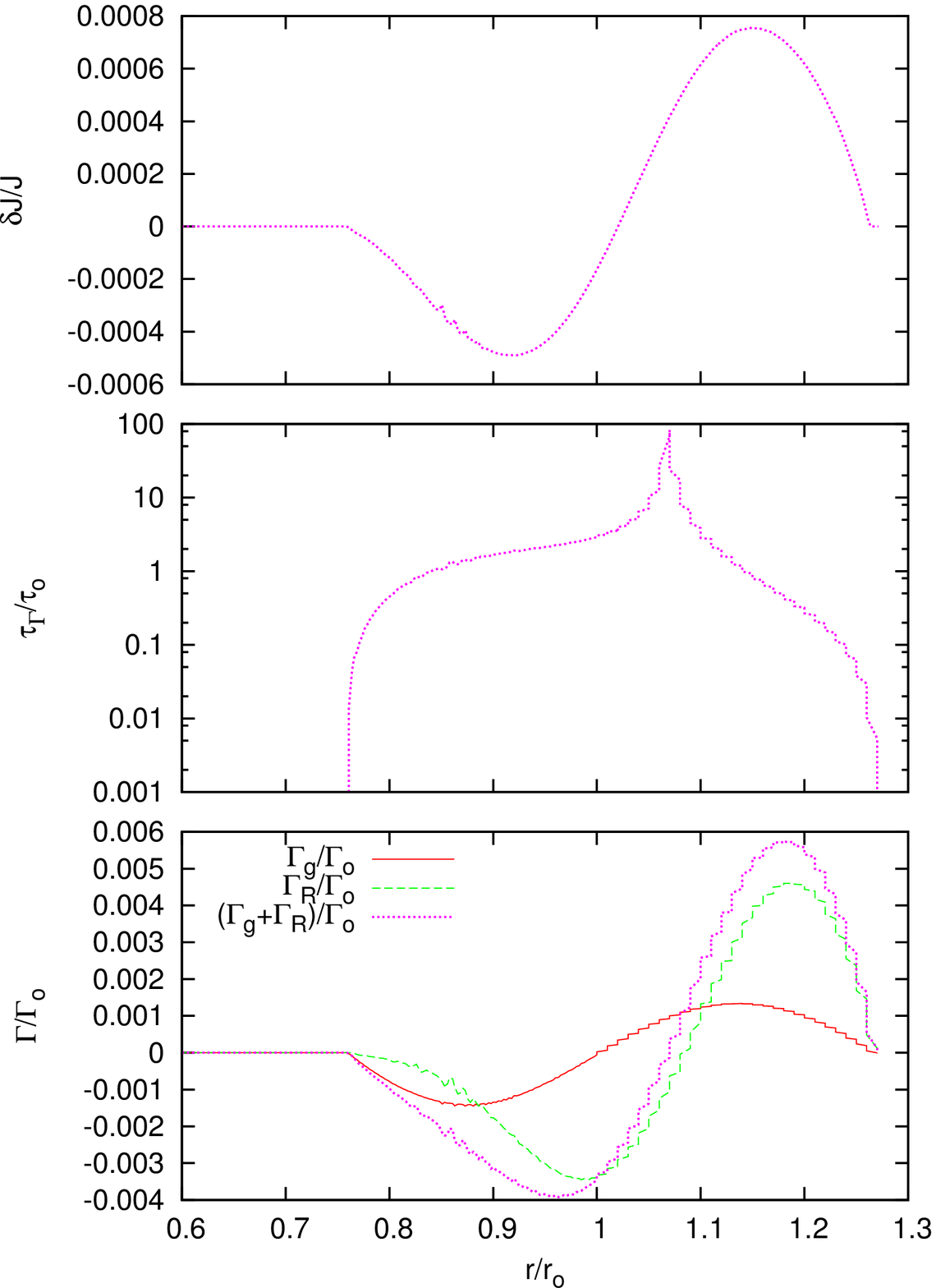}
\includegraphics[width=3.0in]{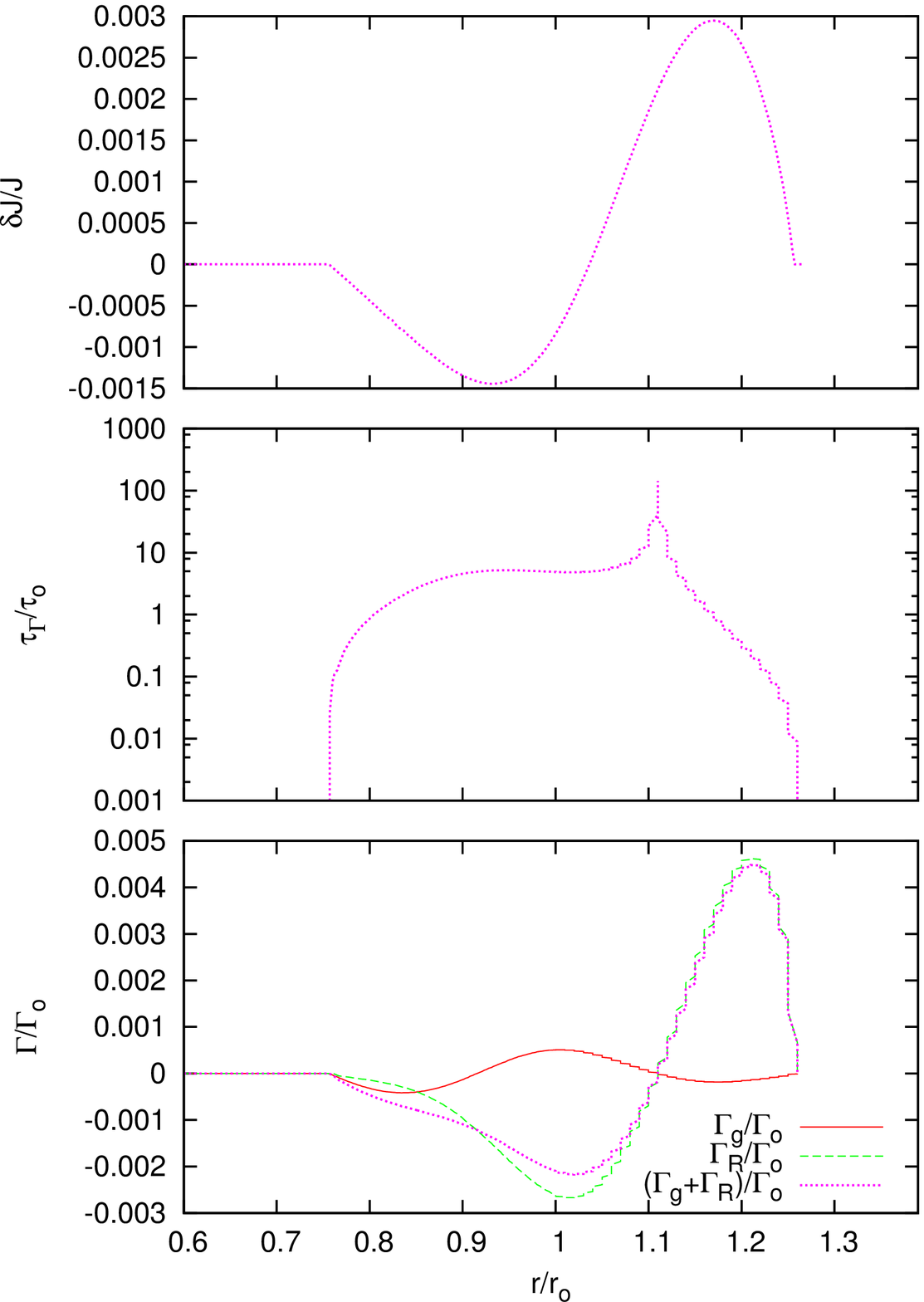}
\end{center}
\caption{
The $\tau_{\Gamma}$, $\delta$J,
$\Gamma_g$, and $\Gamma_R$ for the
$m$ = 2 I$^-$ mode for $q$+ 1.5, $r_-/r_+$ = 0.602 disks 
with $M_*/M_d$ = 0.1 and 5,
Models I1 (left column) and I2
(right column), respectively.
}
\label{Iminus_m2_plot}
\end{figure*}

The $\tau_{\Gamma}$, $\delta$J,
$\Gamma_g$, and $\Gamma_R$ for the $m$ = 2 $I^-$ modes 
for the $q$ = 1.5 disks with  $M_*/M_d$ = 0.1 and 
$r_-/r_+$ = 0.402 and $M_*/M_d$ = 5 and  
$r_-/r_+$ = 0.602, Models I1 and I2, respectively
are shown in Figure \ref{Iminus_m2_plot}. For Model I1,
corotation sits inside r$_-$ for this disk and r$_{olr}$ 
sits near r$_{\circ}$ which is around where $\Gamma_g$ peaks.
$\Gamma_g$ is negative in the inner part of
the disk and positive in the outer part of the disk.
The Reynolds stress dominates overall, but the gravitational 
stress cannot be ignored.
The general shapes of $\Gamma_m$ and $\delta$J are similar, 
strongly negative in the inner half of the disk and positive
in the outer half of the disk. 
The zero of $\Gamma_m$ lies at 1.07 $r_{\circ}$, a
region which contains 71 \% of the disk mass. 
The disk transport times are generally 
long. Material with inflow time $<$ $\tau_{\circ}$ falls
within radius $r$ $<$ 0.84 $r_{\circ}$, 
a region which contains 2.3 \% of the disk mass and
material with outflow time $<$ $\tau_{\circ}$ fall outside
radius 1.14 r$_{\circ}$ a region which contains 9 \% of the disk
mass. The similar forms for $\delta$J and $\Gamma_m$ support
the suggestion that $\Gamma_m$ indicates where and how
instability is driven. For the  I$^-$ mode of Model I2, $\Gamma_g$
changes character. $\Gamma_g$ is negative near the inner and outer
edges of the disk and positive around $r_{\circ}$. $\Gamma_R$ tracks
$\delta$J more closely than does $\Gamma_g$. 
Instability is driven primarily by 
$\Gamma_R$ although $\Gamma_g$ is not negligible. 
For this mode only 28 \% of the disk mass
falls within the sign change in $\Gamma_m$ with only 0.23 \% of the
disk mass having inflow time $<$ $\tau_{\circ}$. This disk model is 
expected to evolve only slowly.

\subsubsection{ $m$ = 2 I$^+$ Mode }

\begin{figure*}
\begin{center}
\includegraphics[width=3.0in]{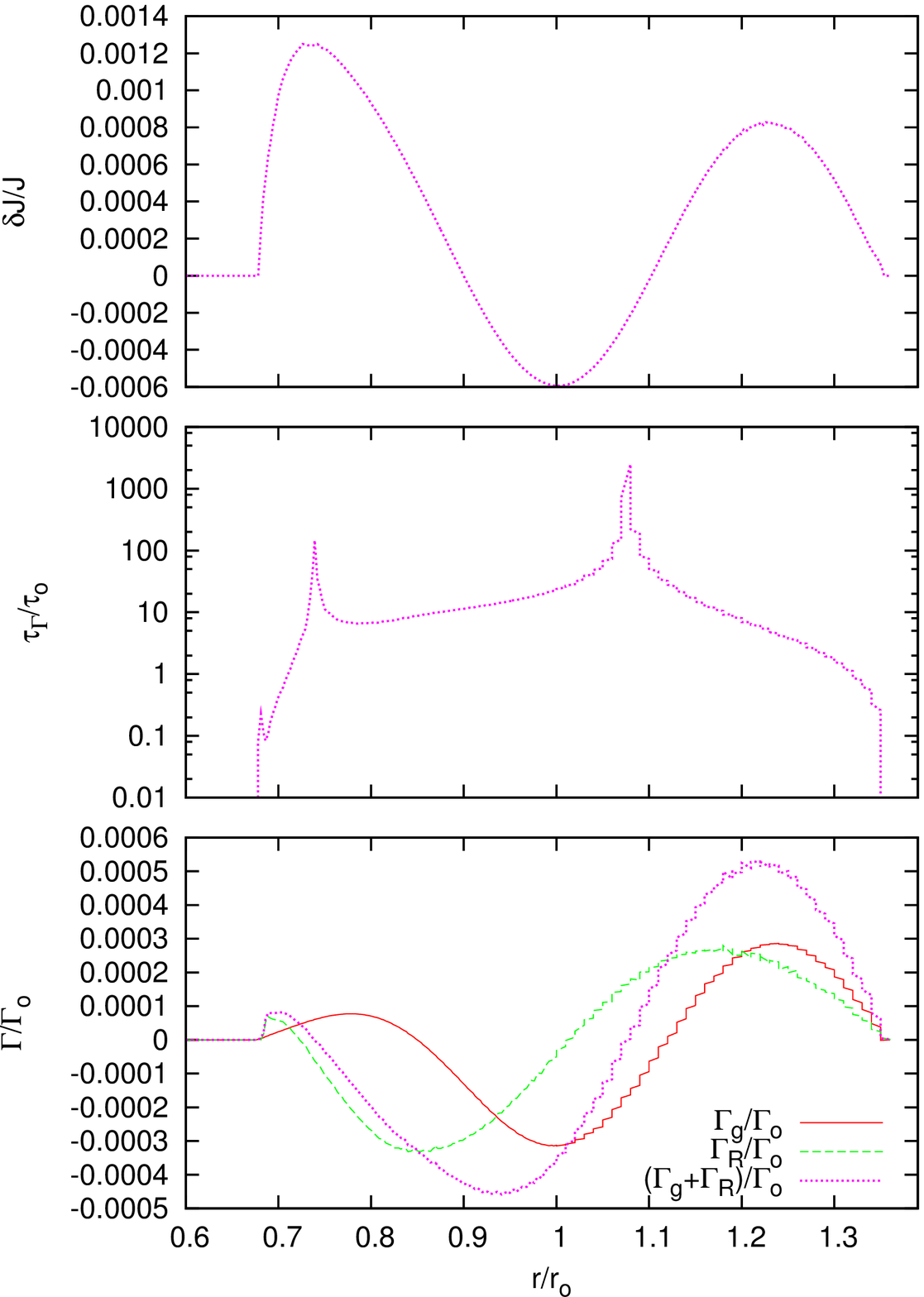}
\includegraphics[width=3.0in]{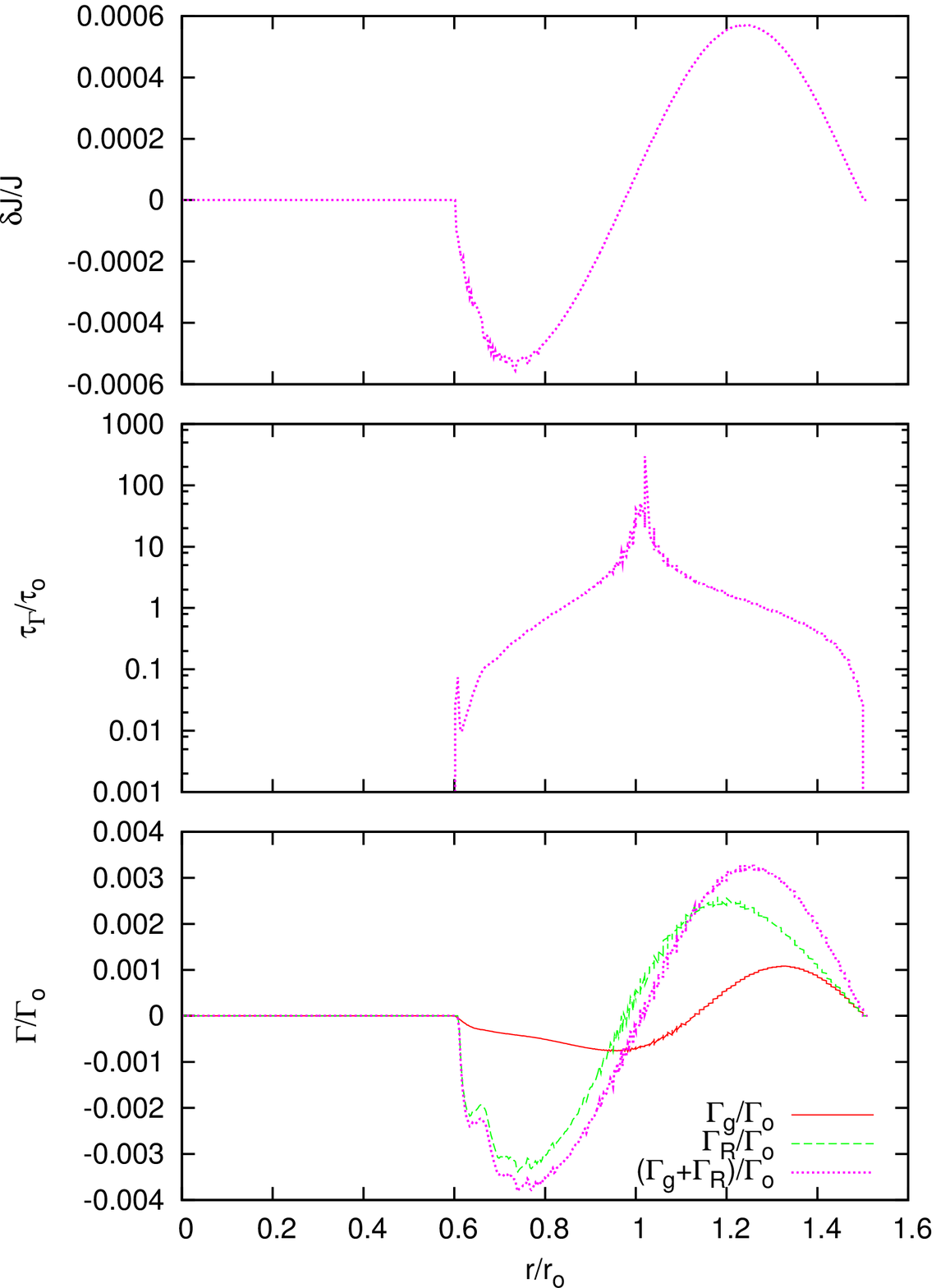}
\end{center}
\caption{
The $\tau_{\Gamma}$, $\delta$J,
$\Gamma_g$, and $\Gamma_R$ for the
$m$ = 2 $I^+$ modes for $q$ = 1.5 disks with 
$M_*/M_d$ = 7, 
and $r_-/r_+$ = 0.5, Model I5 (left column).
$M_*/M_d$ = 0.2 and 
$r_-/r_+$ = 0.402, Model I4 (right column).
}
\label{Iplus_m2_plot}
\end{figure*}

The $\delta$J, $\tau_{\Gamma}$, $\Gamma_g$, and 
$\Gamma_g$ for the $m$ = 2 I$^+$ mode from $q$ = 1.5 disks, and
$M_*/M_d$ = 7 and $r_-/r_+$ = 0.5, and $M_*/M_d$ = 0.2 and 
$r_-/r_+$ = 0.602, Models I5 and I4,
respectively are shown in Figure \ref{Iplus_m2_plot}. For Model I5
corotation sits outside r$_{\circ}$ just inside $r_+$.
The $\delta$J differs qualitatively from that in the I$^-$ mode.
$\delta$J is positive in the inner and outer thirds of the disk and
negative in the central region of the disk.
The Reynolds stress and gravitational stress 
are both important for this mode, however,
$\Gamma_g$ tracks $\delta$J more closely than does 
$\Gamma_R$. $\Gamma_g$ 
is positive in the inner and outer thirds of 
the disk and negative in the middle portion of the disk.
Near the inner edge of the disk, r $<$ 0.71 r$_{\circ}$,
the transport will be outward on times less than
$\tau_{\circ}$. It is possible that 
the inner 0.05 \% of the disk mass inihibits accretion 
onto the star. Between 0.76 and 1.08 $r_{\circ}$, a region which
contains two-thirds of the disk mass, $\Gamma_m$ $<$ 0 and the 
material is expected to flow inward. Outside 1.08 $r_{\circ}$,
$\Gamma_m$ changes sign again and we expect the outer third of 
the disk mass to flow outward. 
The results for the
I$^+$ mode for the small $M_*/M_d$ disk, model I4 are more similar to 
those for the I$^-$ mode. The I$^+$ mode 
results discussed here  and the I$^-$ mode results discussed
in $\S3.4.2$ suggest that the properties of I modes are more strongly
affected by $M_*/M_d$ than by whether they are an I$^-$ or an I$^+$ 
mode.

\subsubsection{ $m$ = 2 P \& Edge Modes }

\begin{figure}
\begin{center}
\includegraphics[width=3.0in]{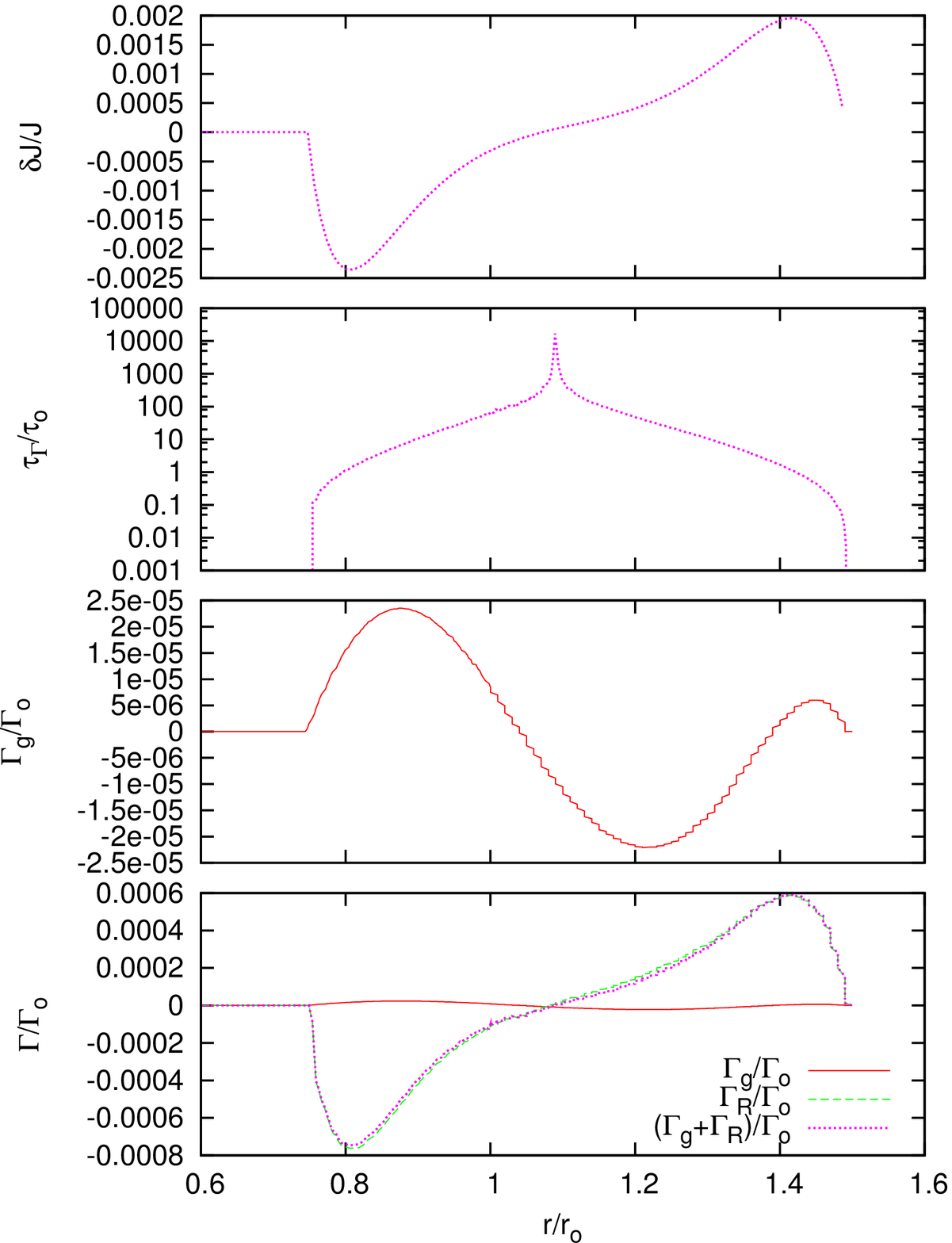}
\end{center}
\caption{
The $\tau_{\Gamma}$, $\delta$J,
$\Gamma_g$, and $\Gamma_R$ for the
$m$ = 2 P mode in the $M_*/M_d$ = 100, $q$ = 2, 
$r_-/r_+$ = 0.5 disk, model P2.
}
\label{Pmode_m2_plot}
\end{figure}

\begin{figure}
\begin{center}
\includegraphics[width=3.0in]{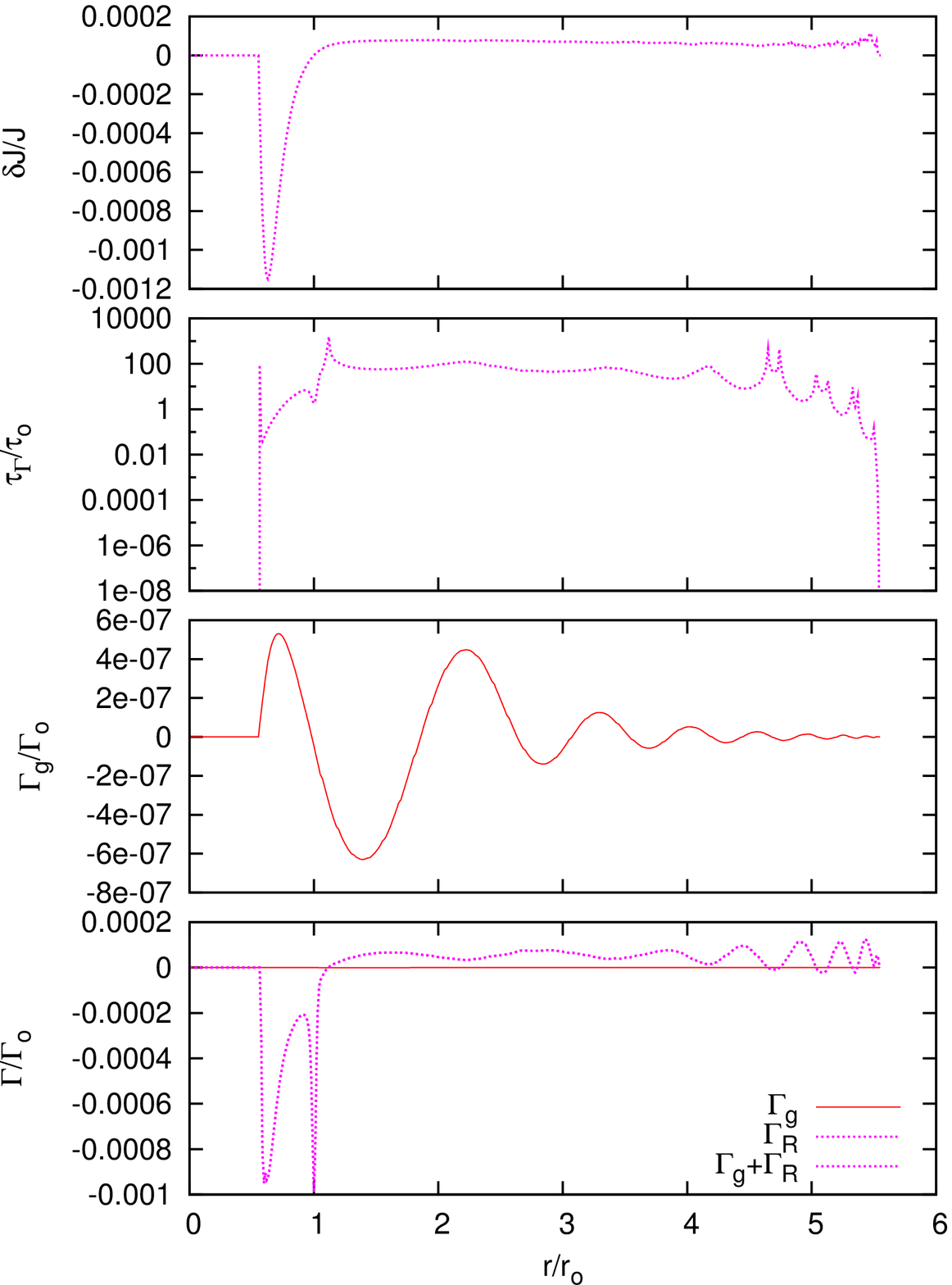}
\end{center}
\caption{
The $\tau_{\Gamma}$, $\delta$J,
$\Gamma_g$, and $\Gamma_R$ for the
$m$ = 2 edge mode in the $M_*/M_d$ = 100, $q$ = 2, 
$r_-/r_+$ = 0.101 disk, model E1.
}
\label{edge_m2_plot}
\end{figure}

The evolution of P modes and edge modes is expected to 
be qualitatively different from that of J modes and to 
show differences in detail from I modes. We show 
$\Gamma_g$, $\Gamma_R$, $\tau_{\Gamma}$, and $\delta$J
for representative $m$ = 2 modes of an $M_*/M_d$ = 10$^2$, $q$ = 2,
system with $r_-/r_+$ = 0.5 and 0.101, Models P2 and E1,
in Figures \ref{Pmode_m2_plot} and \ref{edge_m2_plot}.
The small $r_-/r_+$ disk shows an edge mode and the larger
$r_-/r_+$ disk shows a P mode. First consider the P mode.
The Reynolds stress dominates the disk, the gravitational 
stress is negligible for this mode.  
$\Gamma_m$ is strongly negative in the inner region of the disk, 
decreasing angular momentum and driving the
inner disk inward. Material inside $r$ $=$ 0.80 $r_{\circ}$ (the 
inner 0.5 \% of the disk mass) has inflow time $<$ $\tau_{\circ}$.
The region where $\Gamma_m$ $<$ 0 ends at $r$ = 1.09 $r_{\circ}$
(and contains 51 \% of the disk mass). The stress is weak most of 
the disk has $\tau_{\Gamma}$ $>$ $\tau_{\circ}$. 
Comparison of  
$\delta$J and $\Gamma_m$ suggests that, although
$\delta$J changes sign slightly outside where $\Gamma_m$ changes
sign their shapes are nearly the same, suggesting that $\Gamma_m$ 
once again indicates where and how instability is driven.

Edge modes are dominated by the Reynolds stress with the total
stress $\Gamma_m$ strongly negative near the inner edge of the
disk as can be seen in Figure \ref{edge_m2_plot}
where we show $\tau_{\Gamma}$, $\delta$J,
$\Gamma_g$, and $\Gamma_R$ for an $m$ = 2 edge mode
for an $M_*/M_d$ = 10$^2$, $q$ = 2, 
$r_-/r_+$ = 0.101 disk, Model E1. The bottom panel shows the $\Gamma$
plotted on the same axes. The next panel up shows $\Gamma_g$ on 
rescaled axes to show its form compared to $\Gamma_R$.
The transport time is smallest
near the inner edge of the disk where for material
inside $r$ = 0.73 $r_{\circ}$ (0.32 \% of the disk 
mass) the inflow time $<$ $\tau_{\circ}$. The
region inside the first zero of $\Gamma_m$ contains only 5.6 \%
of the disk mass, it does contain the density 
maximum, however. Based on quasi-linear modeling, we expect rapid 
accretion of 0.3 \% of the disk mass to occur; the bulk of the 
disk mass will not take part. The angular momentum perturbation 
$\delta$J and $\Gamma_m$ vary together for the most part. There
are differences, however, and the case for direct correlation is
less compelling. As also true for P modes, instablility for this
multi-armed mode is driven by the Reynolds stress.

\subsubsection{ $m$ = 1 I$^+$, A, and P Modes }

\begin{figure}
\begin{center}
\includegraphics[width=3.0in]{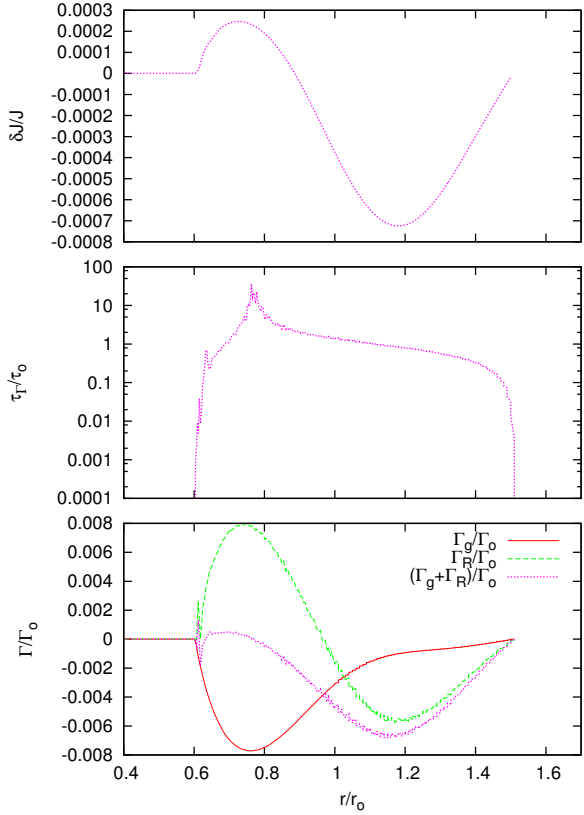}
\end{center}
\caption{
The $\tau_{\Gamma}$, $\delta$J,
$\Gamma_g$, and $\Gamma_R$ for the
$m$ = 1 I$^+$ mode in the $M_*/M_d$ = 0.01, $q$ = 1.5, 
$r_-/r_+$ = 0.402 disk, model J1.
}
\label{iplus_m1_plot}
\end{figure}

The $\tau_{\Gamma}$, $\delta$J,
$\Gamma_g$, and $\Gamma_R$ for the dominant $m$ = 1 I$^+$ 
mode from the $M_*/M_d$ = 0.01, $q$ = 1.5, and $r_-/r_+$ = 0.402
disk, Model J1, are shown in Figure \ref{iplus_m1_plot}. The mode is  
slow with $(y_1,y_2)$ = (-0.986,1.210) so that r$_{co}$ $>$ r$_+$.
$\Gamma_g$ and $\Gamma_R$ both play roles in the torque, 
however, $\Gamma_R$ dominates outside $r_{\circ}$ while 
$\Gamma_g$ and $\Gamma_R$ are comparable in magnitude in the inner
half of the disk. Similarly to the I$^-$ and I$^+$ multi-armed modes, 
$\Gamma_R$ is positive in the inner region of the disk.
The large $\Gamma_g$ in the inner half of the disk 
arises even though this is an I$^+$ mode because of the star/disk
couple. We find that $\Gamma_m$ is negative over most of the disk 
because of the star/disk couple. The inner region of the disk,
r $<$ 0.69 r$_{\circ}$ with mass $5\times10^{-3}$, feels positive torque and 
not expected to accrete onto the star.  For r $>$ 0.76 r$_{\circ}$
which contains 96 \% of the mass of the disk, the torque is 
negative and material is expected to flow inward.
The angular momentum lost by the disk drives orbital motion 
of the star about the center-of-mass of the system. However, the outer 36 \% of the disk mass 
will flow inward on time scales $<$ $\tau_{\circ}$ tending to
narrow the disk. The perturbed angular momentum $\delta$J is positive
in the inner region of the disk similarly to the earlier I$^+$ mode
discussed in contrast to the other cases shown where $\delta$J $<$ 0
near the inner edge of the disk. Comparison of
$\delta$J and $\Gamma_m$ suggests that, although
$\delta$J changes sign slightly outside where $\Gamma_m$ changes
sign, their similarity indicates that $\Gamma_m$ shows where and how
instability is driven.

\begin{figure}
\begin{center}
\includegraphics[width=3.0in]{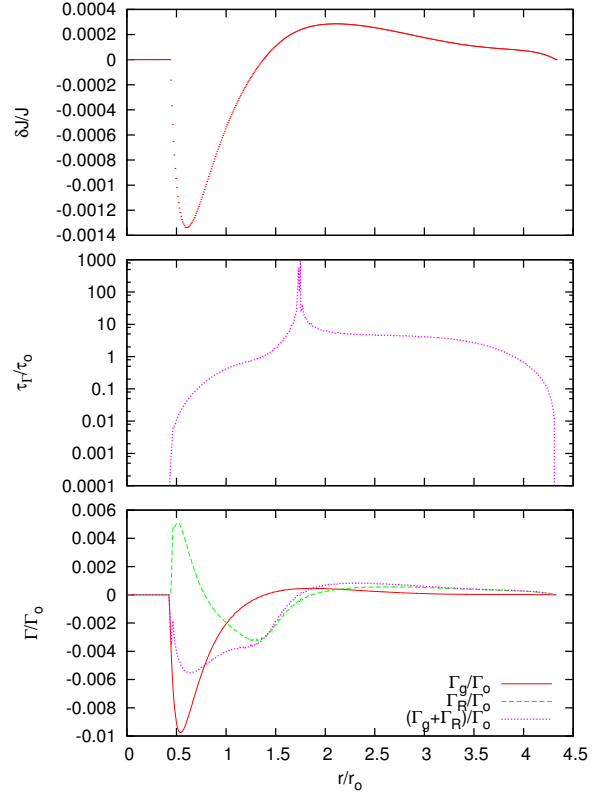}
\end{center}
\caption{Q5
$\tau_{\Gamma}$, $\delta$J,
$\Gamma_g$, and $\Gamma_R$ for the $m$ = 1 A mode in the $M_*/M_d$ = 
1.0, $q$ = 1.8, $r_-/r_+$ = 0.101 disk, model O13.
}
\label{amode_m1_plot}
\end{figure}

A modes arise in $q$ $\sim$ $\sqrt{3}$ systems with $M_*/M_d$ $\sim$ 
0.5 to 5, and
small $r_-/r_+$. We did not find A modes in $q$ = 1.5 and 2 systems.
$\Gamma_g$, $\Gamma_R$, $\tau_{\Gamma}$, and $\delta$J for the
$M_*/M_d$ = 1, $q$ = 1.8 disk with $r_-/r_+$ = 0.101 disk, Model O13,
are shown in 
Figure \ref{amode_m1_plot}
For this mode, corotation sits outside 
r$_{\circ}$ similar to I$^+$ modes and similar to I$^+$ modes, $\Gamma_R$
is positive in the inner region of the disk and 
$\Gamma_g$ and $\Gamma_R$ are competitive
in this disk with $\Gamma_g$ dominating near the inner edge of 
the disk while $\Gamma_R$ plays a more important role in the middle
to outer regions of the disk. The overall torque is generally
negative and the disk loses angular momentum to the central
star. As noted earlier, the mode is slow with $r_{co}/r_{\circ}$ = 1.53 and 
Lindblad outer resonance at $r_{olr}/r_{co}$ = 2.01.
The transport time for the A mode is short;
inside $r$ $=$ 1.37 $r_{\circ}$ the inflow time is $<$ $\tau_{\circ}$
a region which contains 16 \% of the disk mass. The region which 
has $\Gamma_m$ $<$ 0 is within $r$ $<$ 1.75 $r_{\circ}$ with 
contains 34 \% of the disk mass.
The angular momentum transport time, $\tau_{\Gamma}$ $<$ 2-3
$\tau_{\circ}$ over much of the disk and disk spreading is
expected as the nonlinear regime is approached. Comparison of  
$\delta$J and $\Gamma_m$ suggests that, although $\delta$J
changes sign slightly outside where $\Gamma_m$ changes
sign, their general similarity indicates that $\Gamma_m$ 
shows where and how instability is driven.

\begin{figure}
\begin{center}
\includegraphics[width=3.0in]{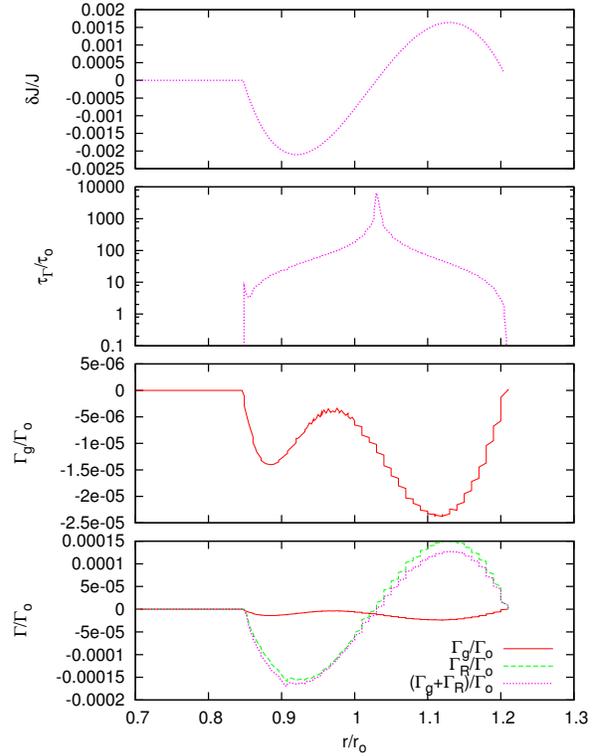}
\end{center}
\caption{Q8
$\tau_{\Gamma}$, $\delta$J,
$\Gamma_g$, and $\Gamma_R$ for
an $m$ = 1 P mode in an $M_*/M_d$ = 100, $q$ = 2, 
$r_-/r_+$ = 0.7 disk, model O16.
}
\label{Pmode_m1_plot}
\end{figure}

We show $\Gamma_g$, $\Gamma_R$, $\tau_{\Gamma}$, and $\delta$J
for a disk with large $M_*/M_d$, Model O16, in  Figure 
\ref{Pmode_m1_plot}. The
mode has $(y_1,y_2)$ = (-0.0222,0.134) and
corotation sits near r$_{\circ}$.
We classify the mode as a P mode. 
It is different from other 
P modes because even though the disk is dominated by 
$\Gamma_R$, $\Gamma_g$ is not negligible.
Self-gravity is not negligible likely because the 
disk is narrow, $r_-/r_+$ = 0.70;
the inner edge of the disk sits far from the star and
so is less affected by the star's tidal field.
$\Gamma_m$ is strongly negative inside $r_{\circ}$ 
and positive outside $r_{\circ}$ suggesting that the disk will 
have a tendency to spread as instability grows. However, $\Gamma_m$ is 
weak and  $<$ 0.01 \% of the mass is expected to flow inward and
outward on timescales of $\tau_{\circ}$ as ${\cal M}_1$ approaches 1.
The disk does not tend to redistribute angular momentum efficiently
as nonlinearity is approached. 
Comparison of $\delta$J and $\Gamma_m$ suggests $\Gamma_m$
again offers support for the suggestion that $\Gamma_m$  indicates
where and how instability is driven.

\section{ DISCUSSION }  \label{sec_discuss}

\subsection{ Comparison to Previous Work }  \label{sec_prev}

\begin{figure*}
\begin{center}
\includegraphics[width=5.0in,angle=0]{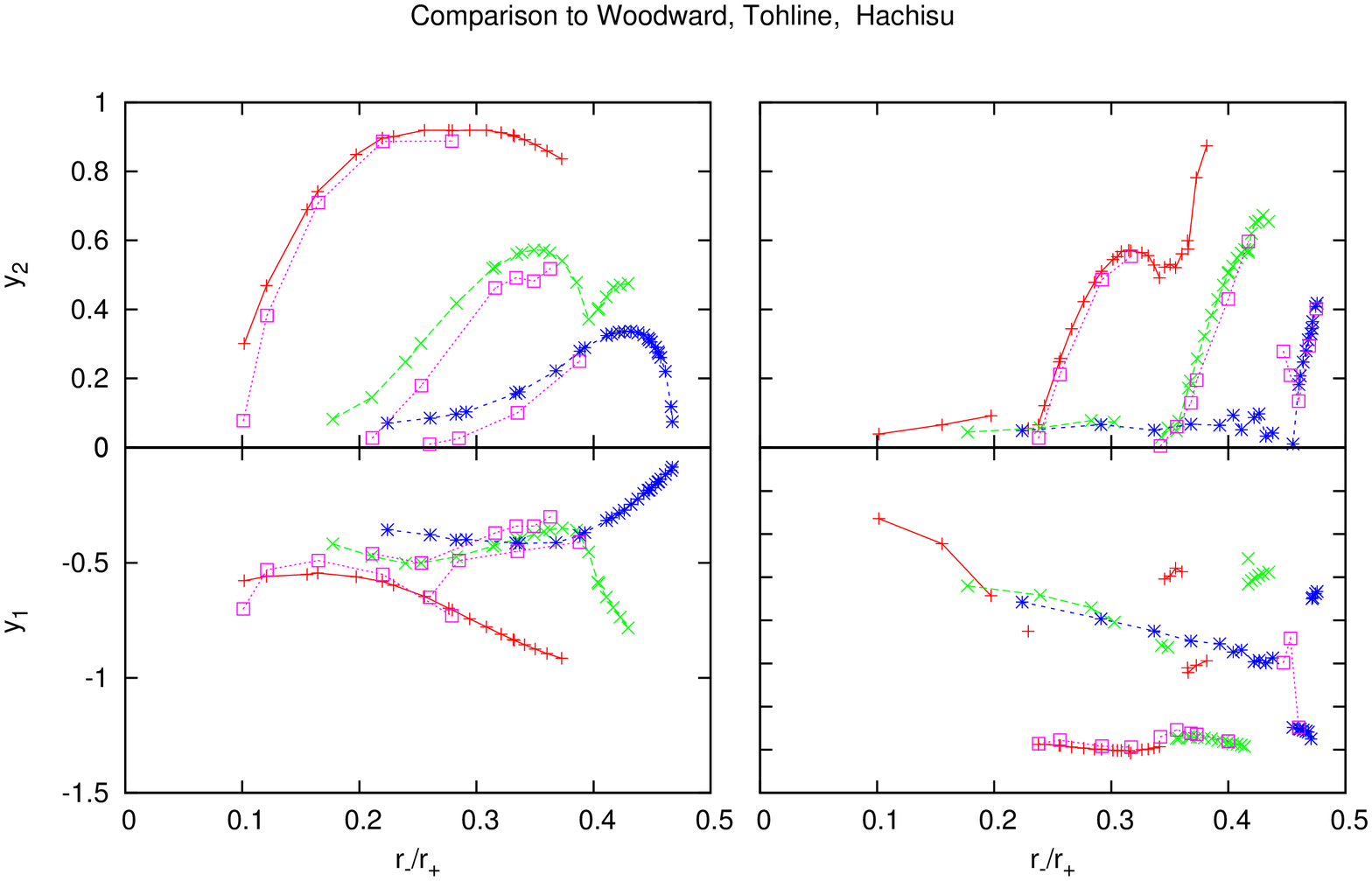}
\end{center}
\caption{
The $m$ = 1 and 2 eigenvalue comparison for disk sequences with $n$ = 1.5, 
$q$ = 2 and $M_*/M_d$ = 0.2, 1, and 5. Our eigenvalues are the green line 
symbols and those from WTH are marked by the red dashed-line symbols.
}
\label{wth_comparison_plot}
\end{figure*}

The work most directly comparable to ours is that of Woodward, 
Tohline, \& Hachisu (1994, WTH). WTH performed
nonlinear simulations of $(n,q)$ = (1.5,2) disks; 15 simulations
with full 2$\pi$ coverage in azimuth to study $m$ = 1 modes and
19 simulations with $\pi$-symmetry to study even $m$ modes. WTH 
modeled star/disk systems with $M_*/M_d$ = 0.2, 1, and 5 and 
repeated seven toroid ($M_*/M_d$ = 0.0) simulations
reported by Tohline \& Hachisu (1990) 
(see also Hadley \& Imamura 2011). WTH usually used numerical
grids of size $64\times32\times64$, radial$\times$vertical$\times$azimuthal
zones. For some narrow disks, WTH used $128\times32\times64$ resolution. 
For our simulations, we typically used grids of dimension, radial $\times$ vertical zones,
given by $512\times512$. For some models we used grids as large as 
$1,024\times1,024$. Because of the difference 
in zoning, we opted for $r_-/r_+$ values which produced disks
that matched the $T/|W|$-values given in WTH rather than 
matching $r_-/r_+$ values.
Figure \ref{wth_comparison_plot} shows comparisons for the $y_1$ 
and $y_2$ values. Overall, the agreement is good. 
Differences, when they arise, are quantitative in nature.

We begin our comparison with the $m$ = 1 results. For the lowest star 
mass case, $M_*/M_d$ = 0.2, the eigenvalues agree to within 10 \% and 
the eigenfunction phase plots are similar exhibiting
edge-like mode character through $T/|W|$ $\sim$ 0.22 after which
the phase plots take on characteristics associated with I
and P modes. However, the existence
of P modes for $M_*/M_d$ = 0.2 disks would be
surprising because even weak self-gravity has been shown
to suppress P modes (Goodman \& Narayan 1988). The 
{\it low} oscillation frequency measured for the modes,
$y_1 \sim -0.7$, places corotation radius well
outside $r_{\circ}$, in contrast to massive star systems and
non-self-gravitating disks where we find that corotation sits near
$r_{\circ}$. For NSG disks where P modes are unambiguously identified, 
even for the system with a very wide disk, $r_-/r_+ \sim 0.4$, the
oscillation frequency $y_1$ = -0.16. NSG narrow disks where
$r_-/r_+ > 0.6$, show $y_1 < -0.1$. 
That is, even for cases where 
corotation does not sit precisely at $r_{\circ}$, corotation
is still close to $r_{\circ}$ for small $T/|W|$ 
and approaches $r_{\circ}$ as $T/|W|$ increases in NSG 
systems. These results suggest that the modes in SG systems are
I$^+$ modes, not P modes. Our results and those of WTH are in
good agreement.
For the higher star mass sequence, $M_*/M_d$ = 1 sequence, we find
similarly good agreement. We again see 
edge-like mode behavior at low $T/|W|$, persisting to $T/|W|$ = 
0.253. Beyond  $T/|W|$ = 0.253, the disturbance again
takes on the appearance of an I and/or P mode. By $r_-/r_+ \sim 0.6$, 
where $T/|W| \sim 0.42$, the phase diagram strongly resembles
that for modes in NSG disks. The $y_1$ are, however, still 
small, -0.36. Although the modes have appearances of 
P and/or I modes, because 
corotation falls well outside $r_{\circ}$ we identify the modes
as I$^+$ modes.
Our results for $M_*/M_d$ = 1 
match closely those of WTH. The results for the
$M_*/M_d$ = 5 sequence are not in as good quantitative agreement. However,
direct comparison of our work to that of WTH for 
$M_*/M_d$ = 5 models is difficult 
because the nonlinear simulations of WTH were only in the $linear$
regime for, at most, a couple of rotation periods.

Consider next the results for $m$ $\ge$ 2 modes. 
Start with the $M_*/M_d$ = 0.2 sequence. Our results
and those from  WTH agree well. Both
indicate a stable model at $T/|W|$ = 0.22, an unstable 
leading phase shift I mode at $T/|W|$ = 0.26,
changing to a trailing arm
I mode at $T/|W|$ = 0.29. 
The plot of the comparison of $y_1$ and $y_2$ values for 
$m$ = 2, $M_*/M_d$ = 1 sequences are similar with our
$y_2$ values approximately 10 \% larger in magnitude than those
in WTH. Also, our $y_1$ values are close to those in WTH and 
indicate that corotation is near 
the inner edge of the disk. 
The WTH phase plot for the $T/|W|$ = 0.422 is not well resolved 
and is difficult to use for comparison
(Woodward 1994). Our linear model took 10 $\tau_{\circ}$ 
to settle into a mode and the nonlinear calculation had 
organized and saturated before that much time had elapsed.
There were discrepancies in the 
last sequence of models, the $M_*/M_d$ = 5 sequence. To clarify
the situation, we added results for disk systems over a wider
range in $T/|W|$ than considered by WTH to the comparison plot. We see 
that the discrepancies arise in the region where a mode change occurs. 
In fact, the eigenvalues for the $T/|W|$ = 0.42 and 0.43 
models have been indicated to be uncertain by 
WTH, with the modes identified as L modes.\footnote
{
WTH suggested that some star/disk
systems exhibited {\it supercritical stability} (Landau \& Lifshitz 
1987, Drazin \& Reid 2004) in that some linearly unstable modes would 
saturate at low amplitudes and not disrupt the systems. WTH referred to 
these modes as L modes.
}
The phase plots of these models (Woodward 1994) are hard to 
compare with ours, but we agree that corotation lies 
nearly at $r_{\circ}$. Our eigenfunction plots in this region indicate 
a small second dip just beginning to emerge. Such a feature
would not be visible in the phase plots of WTH which are too 
noisy (see Woodward 1994).
The next two higher $T/|W|$ models considered by WTH were also classed 
by them as L modes with no uncertainty indicated in their 
growth rates. However, growth in these models saturated at low amplitude. 
Our calculations for these two models 
disagree. We find that $m$ = 2 modes in these models are stable showing 
no hint of growth after 40 $\tau_{\circ}$. The last three models qualitatively 
agree between the studies as far as the I mode nature of the 
plots and the values of the growth rates. There is agreement in 
the $y_1$ values for the $T/|W|$ = 0.46 model while the other $y_1$
values were not reported by WTH. We investigate
this regime in a later paper where in a larger study we
apply our linear and quasi-linear theories and nonlinear 
modeling to study the early nonlinear behavior
in a wide range of disk models unstable to I, J, 
P, edge, and $m$ = 1 modes in order to determine how 
they manifest themselves  in the nonlinear regime in 
terms of their angular momentum transport properties 
(Hadley {\it et al.} 2013).

\subsection{ Be Stars } \label{sec_BE}

Be stars are emission line systems composed of rapidly rotating Main 
Sequence B type stars with circumstellar disks ({\it e.g.,} see 
Okazaki 1997).  In general, they exhibit double-peaked hydrogen 
emission lines indicating
the existence of circumstellar disks.  Further, several Be star 
systems show long-period quasi-periodicity in the 
relative intensity of the violet (V) and red (R) components of their 
double-peaked emission lines.  The V/R quasi-variability occurs 
on periods ranging from years to decades with a statistical mean on the 
order of 7 y, a period much longer than the rotation periods
of the stars and much longer than the orbital periods of material in
typical circumstellar disks unless the disks extend to great distances
from the Be star.
A model proposed for the V/R variability was the 
excitation of $m$ = 1 modes in nearly Keplerian disks ({\it e.g.,} 
Kato 1983). Typically, the excitation mechanism for such $m$ = 1 modes
has not been identified except for binary star 
systems where tidal forcing due to the companion could drive $m$ = 1
modes and for massive disk systems where coupling between the disk and 
central star could drive oscillations ({\it e.g.,} Adams, Ruden, \& Shu 1989).
Optical interferometric observations have indicated the existence
of prograde $m$ = 1 modes in the binary Be star system $\zeta$ Tau (Vakili
{\it et al.} 1998), and the early-type Be star system 
$\gamma$ Cas (Berio {\it et al.} 1999) while 
the disks are in Keplerian motion ({\it e.g.,} see Vakili, Mourard, \&
Stee 1994).

Early theoretical work on the $m$ = 1 modes of
nearly Keplerian disks showed that pressure forces naturally 
led to low-frequency retrograde modes (Kato 1983, Okazaki 1991).
Later works showed that a gravitational quadrupole moment (as expected for 
the rapidly rotating stars in Be systems) overcame pressure effects and led  
to disks with prograde $m$ = 1 oscillations. Okazaki (1997) argued
that such a solution was viable only for late-type Be star systems where disks 
were likely to be cold. The early-type $\gamma$ Cas system required another
explanation which Okazaki (1997) suggested was radiative line forcing. 
Papaloizou \& Savonije (2006) then showed that for a free inner boundary, 
the $m$ = 1 modes were prograde. The aforementioned works
all considered thin, two-dimensional disks. Ogilvie (2008) showed that in 
fat disks (three-dimensional disks), prograde $m$ = 1 modes were possible 
for point mass stars independent of the inner boundary condition. In our 
work, self-excited, long-period, $m$ = 1 modes appear in self-gravitating, 
thick Kepler-like disks. Kepler-like disks for
$M_*/M_d$ $>$ 5, and $r_-/r_+$ =  0.05 to 0.2 show low-frequency, 
$|\omega_1|$ $\ll$ $\Omega_{\circ}$ $m$ = 1 modes which can be either 
prograde or retrograde depending upon the system. At the lower end of the 
parameter range, $m$ = 1 modes are retrograde. For the higher $M_*/M_d$ 
end of the range, $m$ = 1 modes switch to prograde for small $r_-/r_+$,
that is, when self-gravity and pressure forces in the disk are weaker, 
$m$ = 1 modes may be prograde whereas when  the
disk self-gravity and pressure forces are stronger, 
$m$ = 1 modes are retrograde. This suggests that for  
large $M_*/M_d$ and small $r_-/r_+$, self-gravity dominates pressure forces
causing the $m$ = 1 mode to be prograde. For lower $M_*/M_d$, pressure forces 
dominate self-gravity leading to retrograde $m$ = 1 modes.

\subsection{ Excitation of Disk Modes }  \label{sec_excite}

The excitation and growth of $m$ = 1 modes in star/disk systems 
could couple the star and disk and drive otherwise stable
nonaxisymmetric modes in the disk unstable. To see this, recall
how we incorporated the effects of the indirect potential in our 
calculations ($\S\ref{sec_in_lineq}$). We 
considered the indirect potential as an 
expansion of the point mass potential where the position of the star
was treated as a perturbation in that the magnitude of the 
perturbed stellar position, $|\delta_*|$, was assumed to be much
smaller than the radius of the inner edge of the disk, $r_-$. In 
our first-order calculation, the orbiting star interacted with the
unperturbed disk and the perturbed disk interacted with the unperturbed
star. In this way, we self-consistently followed linear growth of $m$ = 1
disk modes. For disk modes with $m$ $\ge$ 2, no coupling occurs. 
However, once the $m$ = 1 mode reaches nonlinear amplitude, the 
indirect potential may then couple strongly to stable disk modes and
drive $m$ $\ge$ 2 instabilities in the disk. This could have implications
for the nature of the angular momentum transport that arises from 
$m$ = 1 modes. If high-$m$ modes are driven, then the induced transport
in the disk may act as a local transport rather than a global mechanism
as would be expected from our simulations. A similar problem has been
studied in the context of migration of planetary cores in protoplanetary
disks ({\it e.g.,} see Papaloizou \& Terquem 2006).

Another mode of interaction between the star and disk could come from 
coupling 
of internal stellar modes and disk modes. Such interactions have 
not been studied except in a  handful of investigations, and,
even then, the studies were in the context 
of star/disk coupling which drove secular instability
in the central star and not instability in the surrounding disk. 
For example, Imamura {\it et al.} (1995) considered the coupling of 
$m$ $\ge$ 2 
$f$-modes in secularly unstable polytropic stars with a surrounding 
disk to drive instability in the central star 
(see also Yuan \& Cassen 1985) and 
Lai (2001) considered 
coupling of $m$ $\ge$ 2 $f$-modes in secularly unstable neutron stars and 
surrounding accretion disks to drive instability in the neutron star.
Imamura {\it et al.} (1995) were interested in how rapidly rotating
protostars could shed angular momentum and how the inner regions of
protostellar disks were truncated while Lai (2001) was
interested in whether $f$-mode instabilities could be driven
in rapidly rotating neutron stars and so limit the spin rate of 
{\it milllisecond} pulsars.
In both Imamura {\it et al.} and Lai, internal stellar modes were 
calculated but the disk modes were not. It was assumed the time-varying
nonaxisymmetric gravitational potential arsing from the internal stellar
modes drove spiral waves in the disks. The coupling torque was based on the 
approximation that the disk was thin and the coupling arose from the
excitation of spiral waves at the 
the Lindblad and corotation resonances in the disk (see
Papaloizou \& Terquem 2006). 

More recently Lin, Krumholz and Kratter (2011) modeled the collapse of 
an isothermal sphere of gas into a star-disk system, and investigated the 
effect of m = 1 and m = 2 disk modes on the evolution of the stars
 spin rate. In contrast to Lai and Imamura {\it et al.}, it was the disk modes 
that were calculated while the stellar modes were not.
Lin, Krumholz and Kratter found that the m = 1 dominant disk drove
orbital motion of the star's center of mass, inhibiting spin 
evolution, while the m = 2 dominant disk   
provided long-term gravitational spin-down torques.

\subsection{ AB Aurigae and Protostellar Disks }  \label{sec_AB}

AB Aurigae (AB Aur) is a well-observed Herbig-Haro Ae star, mass
$M_*$ = 2.4 $\pm$ 0.2 $M_{\odot}$
and age 4.0 $\pm$ 0.1 My (de Warf {\it et al.} 2003),
surrounded by a large, circumstellar disk, mass estimated as 
$M_d$ = 0.02 to 0.1 $M_{\odot}$ (Henning {\it et al.} 1998). 
{\it Subaru} (Takami {\it et al.} 2002) observations show that
the disk in AB Aur has structure from its inner regions $<$
22 astronomical units (A.U.) (Hashimoto {\it et al.} 2011) to its outer 
regions 554 A.U. (Fukugawa {\it et al.} 2004, Hashimoto {\it et al.} 
2011) so that the disk in AB Aur is wide $r_-/r_+$ $<$ 0.04.
The characteristic time scale for the disk in AB Aur is 
$\tau_{\circ}$ $\sim$ $2\pi r/v_{orb}(r)$ $\sim$ 10$^3$
$(r/10^2 A.U.)^{1.5}
(M_*/M_{\odot})^{-0.5}$ y under the assumption the disk is in Keplerian
motion. For the inner disk region, 
$\tau_{\circ}$ $\sim$ 600 y and for the outer disk region 
$\tau_{\circ}$ $\sim$ 2,000 y. Although these characteristic times
are only a small fraction of the star formation time scale and 
AB Aur's estimated age, the times 
are not negligible and features associated with them 
could plausibly be observed. Consequently, features 
attributed to physical processes that
operate on disk dynamical time scales, processes 
such as gravitational instability, could play roles in the
formation of the structure observed in AB Aur. We
consider this possibility here.

High-resolution observations of AB Aur have revealed 
two elliptical rings with
major axes $\sim$ 92 A.U.and 210 A.U.,
separated by an elliptical gap with major axis 170 A.U.
in AB Aur's disk (Fukugawa {\it et al.} 2004,
Hashimoto {\it et al.} 2011). The rings have small eccentricity
on the sky but are thought to be 
circular in shape; the eccentricity arising from 
inclination of the AB Aur system to the plane of the sky
(Hashimoto {\it et al.} 2011). 
The ring/gap structure is seen in optically thick IR emission
(Fukugawa {\it et al}. 2004) 
and optically thin sub-millimeter emission 
(Henning {\it et al.} 1998) suggesting that the
features extend to the midplane of the disk and are more than
corrugations in the surface density. The polarimetric high-resolution
images obtained by the {\it Subaru} telescope have also revealed the  
presence of several dips in the intensity of the rings interpreted as spiral 
structures (Hashimoto {\it et al.} 2011). Hashimoto {\it et al.} (2011)
found seven narrow dips with the most prominent, Dip A at $\sim$ 10$^2$ A.U..
Hashimoto {\it et al.} (2011) did not detect point-like structures in Dip A,
as would be expected if the gap was cleared by a massive protoplanet (however,
see Oppenheimer {\it et al.} 2008).
Hashimoto {\it et al.} noted that the appearance of 
the multi-armed spiral structure of AB Aur's disk was
consistent with expectations of gravitational instability 
(GI). However, they argued that GIs were not likely to be 
the explanation for the structure because of the large 
Toomre $Q$-parameter estimated for the disk, $Q$ $\sim$ 10 
(Pietu, Guilloteau, \& Dutrey 2005). Other explanations such as 
gaps cleared by massive planetary cores ({\it e.g.}, Papaloizou \& Terquem 
2006) and spiral patterns excited by low-mass planetary cores ({\it e.g.,}
Tanaka, Takeuchi, \& Ward 2002) were proposed, but such scenarios 
may suffer from the lack of detected point-like
sources (especially in in Dip A). 

Our simulations find disk systems with 
$M_*/M_d$ $>$ 24 and small disk aspect
ratios, $r_-/r_+$ $<$ 0.1, are unstable to 
low-$m$ tightly wound spiral modes in disks with $q$ 
$\sim$ 2. Low $m$ edge modes would appear to show multiple 
dips when tightly wound. For an adiabatic disk with
$q$ = 2, $r_-/r_+$ = 0.1, and $M_*/M_d$ = 25, the Toomre
$Q$ parameter $Q$ $>$ 10 everywhere in the disk.
However, the disk is unstable to 
tightly wound nonaxisymmetric instabilities (E modes).
See Figure \ref{Edgemode_plots} for the $m$ = 2 mode. The 
mode has eigenvalues $(y_1,y_2)$ = (0.48,0.056) with
disk torques, $\Gamma_g$ and $\Gamma_R$, shown in
Figure \ref{edge_m2_plot}. It shows a small inner bar, forward $\pi/2$ phase jump,
followed by trailing arms with very large winding number. At an $M_*/M_d=25$ self 
gravity is a relatively small factor, with the self-gravity parameter, $p$ = 0.110.

AB Aur may be a transitional
disk with the dust distribution a vestige of the original 
gas disk structure.

\section{SUMMARY}  \label{sec_summary}

We studied the 
I, J, P, and edge
modes of hot, isentropic disks including self-gravity. We considered 
disk material with polytropic index $n$ = 1.5, and disks with
power law angular velocity distributions, exponents
$q$ = 1.5, 1.75, and 2, star-to-disk mass ratios, $M_*/M_d$ = 0 to 10$^{3}$
(from purely self-gravitating to almost non-self-gravitating), 
and inner-to-outer disk radii, $r_-/r_+$ = 0.05 to 0.75.
The parameter space covers that occupied
by protoplanetary and protostellar disks. 

We provide a brief summary of mode types, and then discuss our findings.
Broadly, our equilibrium models had $p \gtrsim$ 7.5 for models 
identified as J modes, 3 $\lesssim$ $p \lesssim$ 7.5 
for models identified as I modes and $p \lesssim$ 3 for models
identified with P and edge modes (the P/I threshold being slightly
higher for higher $m$). The I$^-$/I$^+$ boundary, depending on $q$, $m$, is
typically from $r_-/r_+$ of .5 to .7, increasing with both parameters as
well as $M_*/M_d$.

J modes are strongly self-gravity driven, I modes are driven by coupling of
inertial waves and self-gravity, while the P and
edge modes are driven by coupling of inertial waves across corotation.
For the J , P and edge modes, corotation was usually located at or around the 
density maximum, while for the I+ (I-) modes corotation was found 
exclusively well outside (inside) of density maximum.
J modes had barlike perturbed density eigenfunctions inside of 
the density max, and trailing spiral arms outside. 
I modes had barlike perturbed density eigenfunctions inside and 
outside of $r_{\circ}$, coupled by a trailing or leading spiral
arm. P modes had bars near the inner edge of the disk which switched to
both short and long trailing spiral arms outside of $r_{\circ}$. The instabilites
with a large number of windings are referred to as edge modes.

We find that disks may be unstable to nonaxisymmetric modes, however high
the Toomre $Q$ parameter may be. For $q$ = 2 disks which formally have $Q=0$
disk systems are generally unstable dominated by  
$m$ = 1 modes over most of parameter space. Multi-armed modes,
$m$ $\ge$ 2 modes dominate only at large $r_-/r_+$ 
for given $M_*/M_d$ with only few exceptions. For small $M_*/M_d$ and
small $r_-/r_+$ I modes dominate giving way to J modes at high $r_-/r_+$. 
At large $M_*/M_d$, I and then P modes dominate.
Similar behavior is found for $q$ = 1.75 and 1.5 disks which can 
show Q $>$ 2 everywhere. We find, however, that the range over which 
$m$ $\ge$ 2 modes dominate covers a larger portion of the 
parameter space. For $q$ = 1.5, the region stretches to
$r_-/r_+$ $\sim$ 0.1 for the disk mass range 
$M_*/M_d$ $\approx$ 1 to 20. 
For protostellar and protoplanetary disks with near ${\it Keplerian}$ 
rotation, disks with 
$M_*/M_d$ $\approx$ 1 to 20, multi-armed 
modes, $m$ = 2, 3, and 4 modes, dominate instability
in contrast to $q$ $\sim$ 2 disks which are dominated by 
one-armed modes.

The instability regimes of multi-armed modes in disks track 
the strength of self-gravity as measured by parameter $p$. 
The fastest growing instabilities, the J modes, are in 
the upper left hand corner of $(r_-/r_+,M_*/M_d)$ space 
where $p$ is largest. Growth rates 
decrease away from this corner, then increase and 
decrease forming a bulls eye pattern. Two stable regions where
$y_2$ goes to 0 are found. There is a short arc sweeping from 
$r_-/r_+$ $\approx$ 0.425, $M_*/M_d$ = 0.01 to $r_-/r_+$ $\approx$ 0.475,
$M_*/M_d$ = 0.05,
and a long arc sweeping through parameter space from 0.1 $\le$
$r_-/r_+$ $\le$ 0.20, 0.01 $\le$ $M_*/M_d$ $\le$ 0.1 to 
$r_-/r_+$ $\approx$ 0.70, 17.5 
$\le$ $M_*/M_d$ = 50. The two arcs roughly follow $p$ $\approx$ 
7.5 and 3 breaking the parameter space into regions dominated by 
I modes and P/edge modes, respectively.
$y_1$ changes discontinuously between mode types, while $y_2$ changes
in a continuous manner.

We modeled torques driven by the Reynolds stress and 
gravitational stress that result from the development 
of nonaxisymmetric instability into the nonlinear regime 
using our linear eigenfunctions for 
comparison to the early stages of numerical simulations. 
For Jeans-like J modes, which dominate in the region
of low star mass to disk mass ratios, $M_*/M_d$ $<$ 0.1-0.5, and wide
disks, $r_-/r_+$ $>$ 0.6-0.7, the disk torque is driven by 
the gravitational stress. For larger $M_*/M_d$ and smaller $r_-/r_+$ 
J modes give way first to I modes, where disk self-gravity and 
inertial effects are comparable with disk torques driven by 
both the gravitational and Reynolds stresses, and then
to P and edge modes. These are dominated by the effects of shear except for
one-armed P modes, where the star/disk coupling was sometimes dominant.
For P and edge modes, disk torques are driven by the Reynolds stress 
except $m$ = 1 P modes.

Although illustrative, our work is incomplete 
as we did not investigate the saturation of the unstable modes in the 
nonlinear regime.  We are currently pursuing this question 
through nonlinear simulations of unstable disk systems which include
thermal effects such as entropy generation in shock waves and 
entropy loss through radiative cooling.

\acknowledgments{
The authors thank the National Science Foundation and the National Aeronautics
and Space Administration for support. The computations were
supported by a Major Research Instrumentation grant
from the National Science Foundation, Office of Cyber Infrastructure,
"MRI-R2: Acquisition of an Applied Computational Instrument for
Scientific Synthesis (ACISS)," Grant Number OCI-0960534.
The authors would like to thank an anonymous referee whose careful
reading, and thoughtful comments and criticisms, greatly improved
our manuscript.
JNI thanks Kobe University and host, Dr. Masayuki Itoh, for support and
hospitality during which a portion of this research was carried out.}

\end{document}